\documentclass[aip,jcp,amsmath,amssymb,reprint]{revtex4-2}
\usepackage{mathtools}%
\usepackage[dvipdfmx]{graphicx}%
\usepackage[caption=false]{subfig}
\usepackage{bmpsize}
\usepackage{dcolumn}%
\usepackage{bm}%
\usepackage{bbold}%
\usepackage[hidelinks]{hyperref}%
\usepackage{comment}
\usepackage{braket} %
\usepackage{color}
\usepackage{diagbox}
\usepackage{siunitx}
\usepackage{float}
\usepackage[shortcuts]{extdash}

\graphicspath{{./}{./graphics/plots/}}

\DeclareMathOperator{\Tr}{Tr}

\DeclareMathOperator{\Imag}{Im}
\DeclareMathOperator{\Real}{Re}
\DeclareMathOperator{\csch}{csch}
\DeclareMathOperator{\sech}{sech}
\renewcommand{\Re}{\mathord{\Real}\mkern1mu}
\renewcommand{\Im}{\mathord{\Imag}\mkern1mu}

\newcommand{\trsp}[1][0]{^{\mkern#1mu\top}} %transpose
\newcommand{\by}{\mathord{\times}}

\newcommand{\bessel}[1]{K_{#1}}
\newcommand{\rbessel}[1]{\widetilde{K}_{#1}}
\newcommand{\dawson}{\mathrm{daw}}
\newcommand{\ImF}{\ensuremath{\Im\mkern2mu F}}
\newcommand{\latin}[1]{\emph{#1}}

\newcommand{\iu}{\ensuremath{\mathrm{i}}}
\newcommand{\eu}{\mathrm{e}^}
\newcommand{\ud}{\, \mathrm{d}}
\newcommand{\rmd}{\mathrm{d}}

\newcommand{\eqn}[1]{Eq.~\eqref{#1}}
\newcommand{\Eqn}[1]{Equation~\eqref{#1}}
\newcommand{\eqs}[2]{Eqs.~\eqref{#1} and~\eqref{#2}}

\newcommand{\eqss}[3]{Eqs.~\eqref{#1}, \eqref{#2} and~\eqref{#3}}
\newcommand{\eqt}[2]{Eqs.~\eqref{#1}--\eqref{#2}}
\newcommand{\fig}[1]{Fig.~\ref{#1}}
\renewcommand{\sec}[1]{Sec.~\ref{#1}}

\newcommand{\secs}[2]{Secs.~\ref{#1} and \ref{#2}}

\newcommand{\app}[1]{Appendix~\ref{#1}}
\newcommand{\Refx}[1]{Ref.~\onlinecite{#1}}
\newcommand{\Refs}[1]{Refs.~\onlinecite{#1}}
\newcommand{\tab}[1]{Table~\ref{#1}}

\newcommand{\der}[3][]{\frac{\rmd^{#1}{#2}}{\rmd{#3}^{#1}}}
\newcommand{\pder}[3][]{\frac{\partial^{#1}{#2}}{\partial{#3}^{#1}}}
\newcommand{\pders}[3]{\frac{\partial^2{#1}}{\partial{#2}\partial{#3}}}
\newcommand{\grad}{\nabla}

\newcommand{\ketbra}[2]{\ket{#1}\!\bra{#2}}
\newcommand{\op}{\hat}
\newcommand{\Hop}{\op{H}}
\newcommand{\Hn}{\op{H}_n}
\newcommand{\DRP}{\op{\Delta}}
\newcommand{\Gop}{\op{\mathcal{G}}}
\newcommand{\Gn}{\Gop_n}
\newcommand{\GR}{\Gop_{0}}
\newcommand{\GP}{\Gop_{1}}
\newcommand{\IGop}{\Im\Gop}
\newcommand{\IGn}{\Im\Gn}
\newcommand{\IGR}{\Im\GR}
\newcommand{\IGP}{\Im\GP}
\newcommand{\RGn}{\Re\Gn}
\newcommand{\RGR}{\Re\GR}
\newcommand{\RGP}{\Re\GP}
\newcommand{\HR}{\op{H}_0}
\newcommand{\HP}{\op{H}_1}

\newcommand{\ZR}{Z_{\mathrm{r}}}
\newcommand{\ZRred}{\widetilde{Z}_{\mathrm{r}}}
\newcommand{\ZRn}[1]{Z_{\mathrm{r},#1}}
\newcommand{\ZPn}[1]{Z_{\mathrm{p},#1}}
\newcommand{\ZRredn}[1]{\widetilde{Z}_{\mathrm{r},#1}}

\newcommand{\ZP}{Z_{\mathrm{p}}}
\newcommand{\popR}{\chi_{\mathrm{r}}}
\newcommand{\popRn}[1]{\chi_{\mathrm{r},#1}}

\newcommand{\rhoR}{\hat{\rho}_{\mathrm{r}}}

\newcommand{\Vdd}{V^{\ddagger}}
\newcommand{\kappaR}{\kappa_0}
\newcommand{\kappaP}{\kappa_1}
\newcommand{\Kop}{\op{\mathcal{K}}}
\newcommand{\Kn}{\Kop_{n}}
\newcommand{\KR}{\Kop_0}
\newcommand{\KP}{\Kop_1}
\newcommand{\Ksc}{\mathcal{K}^{\mathrm{sc}}}
\newcommand{\Knsc}{\Ksc_n}

\newcommand{\tpl}{\ensuremath{t_{\mathrm{pl}}}}
\newcommand{\boltz}{\ensuremath{k_{\mathrm{B}}}}
\newcommand{\fftcf}{\ensuremath{C_{FF}}}

\newcommand{\SC}{\mathrm{sc}}
\newcommand{\Pf}{P_{4}}

\newcommand{\fa}{{4\mathrm{A}}}
\newcommand{\fb}{{4\mathrm{B}}}
\newcommand{\Pfa}{P_{\fa}}
\newcommand{\Pfb}{P_{\fb}}
\newcommand{\kred}{\tilde{k}}
\newcommand{\cgr}{c_{2}}
\newcommand{\kgr}{k_{2}}
\newcommand{\kgrred}{\kred_{2}}
\newcommand{\kgrsc}{k_{2}^{\SC}}
\newcommand{\kgrredsc}{\kred_{2}^{\SC}}
\newcommand{\kf}{k_{4}}
\newcommand{\kfsc}{k_{4}^{\SC}}
\newcommand{\kfred}{\kred_{4}}

\newcommand{\cfa}{c_{\fa}}
\newcommand{\kfa}{k_{\fa}}
\newcommand{\kfreda}{\kred_{\fa}}

\newcommand{\kfredasc}{\kred_{\fa}^{\SC}}
\newcommand{\kfredaht}{\kred_{\fa}^{\mathrm{ht}}}
\newcommand{\kfredaltl}{\kred_{\fa}^{\mathrm{lt,lin}}}
\newcommand{\cfb}{c_{\fb}}
\newcommand{\kfb}{k_{\fb}}
\newcommand{\kfredb}{\kred_{\fb}}
\newcommand{\kfbsc}{k_{\fb}^{\SC}}
\newcommand{\kfredbsc}{\kred_{\fb}^{\SC}}
\newcommand{\kfpade}[1][]{\ensuremath{\left[k_{2+4}^{#1}\right]_{\text{Pad\'{e}}}}}
\newcommand{\tauR}{\tau_{\mathrm{r}}}
\newcommand{\tauP}{\tau_{\mathrm{p}}}
\newcommand{\tauGR}{\tau_{\mathrm{g}}}
\newcommand{\zR}{z_{\mathrm{r}}}
\newcommand{\zP}{z_{\mathrm{p}}}
\newcommand{\xin}{\vec{x}_{\mathrm{i}}}
\newcommand{\xfin}{\vec{x}_{\mathrm{f}}}

\newcommand{\margfa}{L_{\fa}} % marginalised c4B integral
\newcommand{\margfb}{L_{\fb}} % marginalised c4B integral
\newcommand{\bred}{\widetilde{\beta}} %reduced beta for linear systems
\newcommand{\kZUS}{\ensuremath{k_{\mathrm{Zus}}}}
\newcommand{\kad}[1]{\ensuremath{k_{\mathrm{ad}{#1}}}}
\newcommand{\kgrht}{\ensuremath{\kgr^{\mathrm{ht}}}}
\newcommand{\PH}{\ensuremath{P_{\mathrm{H}}}}
\newcommand{\PLZ}{\ensuremath{P_{\mathrm{LZ}}}}

\newcommand{\mat}[1]{\mathsf{#1}}
\renewcommand{\vec}[1]{\mathsf{#1}}
\newcommand{\gvec}[1]{\bm{#1}} %Greek vector
\newcommand{\cvec}[1]{\bm{#1}} %concatenation of vectors/scalars

\newcolumntype{d}[1]{D{.}{.}{#1}} %
\newcolumntype{L}[1]{>{\raggedright\arraybackslash$}p{#1}<{$}}
\newcolumntype{R}[1]{>{\raggedleft\arraybackslash$}p{#1}<{$}}
\newcolumntype{C}[1]{>{\centering\arraybackslash$}p{#1}<{$}}

\makeatletter
\renewcommand*\env@matrix[1][\arraystretch]{%
	\edef\arraystretch{#1}%
	\hskip -\arraycolsep
	\let\@ifnextchar\new@ifnextchar
	\array{*\c@MaxMatrixCols c}}
\makeatother

\newlength{\mycolw}

\begin{document}

\preprint{AIP/123-QED}

\title{Nonadiabatic instanton rate theory beyond the golden-rule limit}

\author{George Trenins}
\email{georgijs.trenins@phys.chem.ethz.ch}
\author{Jeremy O. Richardson}
\email{jeremy.richardson@phys.chem.ethz.ch}
\affiliation{Laboratory of Physical Chemistry, ETH Z\"{u}rich, 8093 Z\"{u}rich, Switzerland}

\date{\today}

\begin{abstract}
Fermi's golden rule describes the leading-order behaviour of the reaction rate as a function of the diabatic coupling. Its asymptotic (\parbox[b][0pt]{\widthof{$\hbar \to 0$}}{$\hbar \to 0$}) limit is the semiclassical golden-rule instanton rate theory, which rigorously approximates nuclear quantum effects, lends itself to
efficient numerical computation and gives physical insight into reaction mechanisms. However the golden rule by itself becomes insufficient as the strength
of the diabatic coupling increases, so higher-order terms must be additionally considered.
In this work we give a first-principles derivation of the next-order term
beyond the golden rule, represented as a sum of three components. Two of them lead to new instanton pathways that extend the golden-rule case and, among
other factors, account for effects of recrossing on the full rate.
The remaining component
derives from the equilibrium partition function and accounts for changes in
potential energy around the reactant and product wells due to diabatic coupling. The new semiclassical theory demands little computational effort beyond a golden-rule instanton calculation. It
makes it possible to rigorously assess the accuracy of the golden-rule approximation and sets the stage for future work on general semiclassical nonadiabatic rate theories.
\end{abstract}

\maketitle

\section{\label{sec:intro}%
Introduction}

Semiclassical instanton rate theory\cite{Miller1975semiclassical,Chapman1975rates,%
Andersson2009Hmethane,RPInst,Rommel2011locating,AdiabaticGreens,Perspective,InstReview} is becoming a well-established method for describing chemical reactions on a single Born--Oppenheimer potential energy surface (PES).\cite{dimersurf,porphycene,Hgraphene,Asgeirsson2018instanton,Rommel2012enzyme,HCH4,Kryvohuz2014KIE,GPR,Meisner2016water}
The theory accounts for nuclear quantum effects such as tunnelling in an approximate but rigorous manner and scales favourably with
system size. Hence it can be readily applied to full-dimensional \latin{ab~initio}
simulations of chemical transformations. The favourable scaling arises because
(in the simplest case) the instanton rate derives from a single
classical trajectory. This not only simplifies the calculation compared to the full quantum treatment, but also offers direct
mechanistic insight, as this trajectory defines the dominant tunnelling pathway,
which can easily be visualised.

However, a description in terms of a single Born--Oppenheimer PES is only valid
for systems with strong diabatic coupling, $\Delta$. Away from this
limit the Born--Oppenheimer approximation breaks down, and
the chemical process is said to be nonadiabatic. For weak diabatic coupling,
nonadiabaticity can be effectively described with
perturbation theory, taking the uncoupled
diabatic Hamiltonian\cite{ChandlerET} as the reference. In the case of reaction rates,
the leading-order perturbation term (second order in $\Delta$) is the well-known Fermi's golden rule (GR).\cite{Dirac1927radiation,Wentzel1927golden}
This can be formally expressed in terms of path integrals, whose
evaluation by steepest descent leads to semiclassical golden-rule instanton rate theory.\cite{GoldenGreens,GoldenRPI,AsymSysBath,MLJ,Ansari2021} Like its Born--Oppenheimer counterpart, this semiclassical theory rigorously approximates the quantum rate using information
from (typically) a single classical trajectory and hence enjoys the same computational advantages. The theory has been extended\cite{inverted,Ansari2021} into the Marcus inverted regime\cite{Marcus1960ET,Marcus1985review}---a considerable methodological challenge for most path-integral methods---and has been used
for \emph{ab~initio} calculations of spin-crossover rates for thiophosgene, showing unprecedented accuracy
when compared with experiment.\cite{Heller2021}

That said, Fermi's golden rule is only accurate if
the next term in the perturbation expansion of the full nonadiabatic rate is small by comparison.
In other words, the GR expression should only be used if the diabatic coupling is sufficiently weak. This assumption is often true for electron-transfer\cite{Marcus1964review,Marcus1985review,Marcus1993review,HammesSchiffer2008PCET,UlstrupBook,ChandlerET} and spin-crossover\cite{Harvey2007review,Lykhin2016NATST} reactions,
although its validity is by no means guaranteed. One expects to also find
systems with intermediate values of the diabatic coupling, lying in a ``grey area'' where neither Born--Oppenheimer nor golden-rule rate theory are valid.\cite{Lomont2012,Sousa2013,Daniel2015,Fang2019,Valentine2019,Valentine2022} This intermediate regime has always been of great interest,
and the search for a corresponding practical nonadiabatic rate theory is ongoing.\cite{Althorpe2016a}

Considerable progress \cite{Zusman1980,Rips1995ET,Rips1996ET,Sparpaglione1988,Cao2000,Gladkikh2005} has been made for dissipative
systems that can be mapped onto a spin--boson model,\cite{Leggett1987spinboson}
which is typically used to represent electron transfer in solution.\cite{Nitzan}
Our goal, however, is to develop a full-dimensional theory of molecular
reactions, for which instanton methods are particularly well suited.
In contrast to studies of electron transfer in solution, here we do not focus on solvent effects or chemical processes
dominated by diffusive motion along the reaction coordinate.
Instead the key requirement for our theory is that it be applicable to
multidimensional anharmonic potentials, such as those found in \latin{ab~initio}
simulations of gas-phase reactions.

Early work that paved the way for such developments includes Landau--Zener (LZ) theory,\cite{Landau1932LZ,Zener1932LZ} which gives the correct classical (high-temperature) rate constant for a one-dimensional linear system
in the GR limit\cite{PetersBook,NikitinBook,UlstrupBook} and which has been used in uniform rate expressions for reduced models
of electron transfer.\cite{Rips1995ET,Rips1996ET,Nitzan}
Zhu--Nakamura theory\cite{Zhu1994ZN,Zhu1994linear,Zhu1995ZN,NakamuraNonadiabatic}
is a generalisation of LZ that accounts for nuclear quantum effects
and hence yields an approximation to the reaction rate, applicable at any temperature
or strength of diabatic coupling.
Both theories are, however, limited to one-dimensional systems, and are only strictly rigorous for linear potentials.\cite{NikitinBook,NakamuraNonadiabatic}
Their other limitation is conceptual, as neither provides the same mechanistic
insight as semiclassical instantons, since no stationary-action pathways are
computed.

An alternative approach was proposed
in a recent publication by Lawrence~\emph{et~al.},\cite{Lawrence2019ET} who
introduce an interpolation formula connecting the golden-rule and the
adiabatic (Born--Oppenheimer) limits. The formula is inspired by the Zusman equation,
originally derived for the rate of electron transfer in the classical, high-friction limit.\cite{Zusman1980,Garg1985spinboson,Gladkikh2005} In contrast to the Zusman equation, the interpolation
formula can account for nuclear tunnelling and zero-point energy effects, as it %
inherits ``quantumness'' from its inputs of a golden-rule and an adiabatic rate theory (originally
Wolynes theory\cite{Wolynes1987nonadiabatic} and
ring-polymer molecular dynamics (RPMD),\cite{RPMDrate,RPMDrefinedRate,Habershon2013RPMDreview}
but could equally be the closely related instanton rate theories\cite{RPInst,GoldenGreens}).
Unlike Landau--Zener and Zhu--Nakamura theories, the interpolation formula
is obtained following a more heuristic approach, wherein lie both its strength
and its weakness. On the one hand,
the method is readily applicable to
multidimensional systems. On the other hand,
because it is not derived from a rigorous
nonadiabatic rate theory, there is no
systematic way of improving it.
It is also not obvious that the interpolation
formula should be
effective in cases where
golden-rule and adiabatic reaction mechanisms are qualitatively different.
An extreme example of this is provided by the Marcus inverted regime,
for which the adiabatic rate is undefined.\cite{Lawrence2019ET} In light of this
we think that approaches such as the interpolation formula can be complemented by a first-principles
rate theory tackling the intermediate nonadiabatic regime.

A general nonadiabatic instanton solution for arbitrary diabatic coupling
strengths has been proposed
by Voth and co-workers.\cite{Cao1995nonadiabatic,Cao1997nonadiabatic,%
Schwieters1998diabatic,Schwieters1999diabatic} It is based on the assumption that the reaction
rate is related to the imaginary part of the barrier partition function. This
approach, known as the \ImF~premise,\cite{Langer1967ImF,Langer1969ImF,Uses_of_Instantons,Affleck1981ImF,Cao1996QTST}
has been validated for both adiabatic\cite{Althorpe2011ImF,AdiabaticGreens,InstReview} and golden-rule instantons\cite{Cao1997nonadiabatic,GoldenGreens} by demonstrating that the \ImF\ expressions recover the semiclassical limits of the corresponding
quantum
rate theories based on the flux-correlation formalism.\cite{Miller1983rate}
The same has not yet been accomplished in the general
nonadiabatic case, implying that the definitions of (imaginary) barrier partition functions used by
Voth and co-workers are not rigorously justified.
The first formulation to be suggested\cite{Cao1995nonadiabatic,Cao1997nonadiabatic} was
later determined to be incomplete,
as it fails to recover adiabatic rates.\cite{Schwieters1998diabatic}
It was superseded by the approach in \Refx{Schwieters1998diabatic},
later given the name ``mean-field ring-polymer instanton''.\cite{Ranya2020instanton}
Although this has the correct behaviour in the adiabatic limit, like all mean-field path-integral
methods,
it fails to recover the classical golden-rule limit without \emph{ad~hoc} corrections.\cite{Schwieters1999diabatic}
Furthermore, like with
the interpolation formula, it is not obvious how it can be generalised to describe
the Marcus inverted regime.

In light of this, the search is still out for a rigorous nonadiabatic instanton theory. This
would be of interest not only as a numerical method for predicting rates, but also
as a key to understanding the interplay of nuclear tunnelling
and electronic nonadiabaticity in chemical reactions.
Moreover, instanton theories are a powerful tool for the
design and/or theoretical justification of path-integral sampling
and dynamics approaches that go beyond steepest-descent integration.
Such a connection to semiclassical instantons has been made both
in the adiabatic and weak-coupling limits for RPMD,\cite{RPMDrate,RPMDrefinedRate,RPInst,Habershon2013RPMDreview} quantum transition-state theory, \cite{RPInst,Mills1997QTST,Hele2013QTST} quantum instanton\cite{Miller2003QI,Vanicek2005QI,QInst} and golden-rule quantum transition-state theory (GR-QTST).\cite{GRQTST,GRQTST2} Approaches proposed for intermediate
coupling strengths\cite{Shushkov2012RPSH,mapping,Ananth2013MVRPMD,Duke2015MVRPMD,Chowdhury2017CSRPMD,Menzeleev2014kinetic,Kretchmer2016KCRPMD,Kretchmer2018KCRPMD,Tao2018isomorphic,Tao2019RPSH,Lawrence2020NQI,Schwieters1999diabatic} currently lack this rigorous connection,\cite{Shushkov2013instanton,GRQTST,GRQTST2,Lawrence2019isoRPMD}
and their future development may be inspired and
aided by a first-principles semiclassical
theory.

In this paper we develop
a semiclassical instanton theory for the second term in the perturbation expansion of the exact nonadiabatic reaction
rate.
The term appears in the series immediately after
the golden-rule expression and is fourth order in the diabatic coupling $\Delta$. In \sec{sec:background} we
formally define the rate,
develop a perturbation series in the diabatic representation and give a brief summary of how the semiclassical
limit of the leading-order term (the GR instanton) can be
derived. In~\sec{sec:new-theory} we derive the quantum-mechanical fourth-order term, casting it as a sum of three components that are individually amenable to approximation by steepest descent. Applying the approximation
in \sec{sec:k4-sc} we arrive at a semiclassical formula for the new term,
whose numerical accuracy is tested on a one-dimensional predissociation
model and a multidimensional spin--boson system in \sec{sec:models}.
Our results and outlook on future work are summarised in \sec{sec:discuss}.

Because the fourth-order rate comprises three components, and
the asymptotic behaviour of each has to be analysed individually,
the amount of mathematical manipulation in \secs{sec:new-theory}{sec:k4-sc} is considerable.
The salient point is
that the semiclassical instanton expressions in this work
derive from a rigorous flux-correlation formulation of the quantum rate.
For reference, the final results are given by \eqt{eq:d0d1-defn}{eq:k4asc},
\eqt{eq:k4b-ag}{eq:k4b-sc-uni} and \eqt{eq:z2red-ints}{eq:zrred2-bound}, which define the three
components of the fourth-order rate constant. Despite the lengthy derivations, these expressions
are all easily evaluated and require little computational effort
beyond a golden-rule instanton calculation.

\section{\label{sec:background}%
Theoretical background}

We consider a system that comprises two diabatic states,\cite{Nitzan}
$\ket{0}$ and $\ket{1}$, with
corresponding nuclear Hamiltonian operators
\begin{equation}
\Hn = \sum_{j=1}^{f}
\frac{\op{p}_j^2}{2 m} + V_n(\hat{\vec{x}}),
\end{equation}
where $n \in \{0,1\} $. Here $f$ is the number of nuclear coordinates, which
have been mass-weighted so that each has the same mass $m$, and $V_n$
are the diabatic potential energy surfaces (PESs). We assume that the
system is in the Marcus normal regime, implying that the diabatic PESs
intersect along a seam for which
$\grad V_0 \cdot \grad V_1 < 0$. The states interact via
the diabatic electronic coupling operator, $\DRP = \Delta(\op{\vec{x}})$,
which is assumed to vary slowly with respect to $\vec{x}$.
The total Hamiltonian expressed in the diabatic basis is then
\begin{equation}
\label{eq:htot}
\Hop = \Hop^{(0)} + \lambda \Hop^{(1)}
= \left(\!
\begin{array}{cc}
\HR & 0 \\
0 & \HP
\end{array}\!
\right)
+
\lambda \left( \!
\begin{array}{cc}
0 \, & \, \DRP \\
\DRP \, & \, 0
\end{array} \!
\right)\!,
\end{equation}
where $\lambda$ is a dimensionless parameter used to track the order of the
perturbative terms. At the end of the derivation we set
$\lambda = 1$.
Identifying $\ket{0}$ and $\ket{1}$ as the reactant and
product states respectively, we introduce the operator $\op{h} = \ketbra{1}{1}$
that projects onto the product state. Its time derivative is the flux operator,
\begin{equation}
\op{F} = \frac{1}{\iu \hbar} [\op{h}, \op{H}] = \frac{\lambda}{\iu \hbar}
\left(
\begin{array}{cc}
0 & \!-\op{\Delta} \\
\op{\Delta} & \!\hphantom{-}0
\end{array}
\right).
\end{equation}
In what follows we give the formally exact expression for the thermal rate constant
associated with the transition from $\ket{0}$ to $\ket{1}$ and outline the derivation
of its \mbox{$\lambda \to 0$} limit (\latin{i.e.}, Fermi's golden rule) and the corresponding
semiclassical approximation.

\subsection{\label{ssec:greens-tcf}%
The flux-correlation and Green's function formalisms}

Following \Refx{Craig2007condensed}, we define
the reactant and product partition functions
\begin{equation}
\ZR = \Tr\!\big[ \eu{-\beta \Hop} (1 - \hat{h}) \big],
\quad
\ZP = \Tr\!\big[ \eu{-\beta \Hop} \hat{h} \big],
\end{equation}
and the initial reactant population
\begin{equation}
\popR = \frac{1}{\ZR}\Tr \big[ \rhoR (1-\hat{h}) \big],
\end{equation}
where $\beta = 1 / \boltz T$ and $\ZR = \Tr[\rhoR].$ Here
$\rhoR$ is the symmetric thermal density operator\cite{Schofield1960,Miller1983rate}
\begin{equation}
\rhoR = \eu{-\beta \Hop /2 } (1 - \hat{h}) \, \eu{-\beta \Hop /2 },
\end{equation}
which models the system at thermal equilibrium
in the reactant well. Depending on the system, the initial reactant population may be
exactly equal, or very close to, one.
In our theory the deviation of $\popR$ from unity cannot always be neglected,
and so we use Eq.~(2.21) of \Refx{Craig2007condensed}
to define the rate $k(\beta) $ as
\begin{subequations}
\label{eq:k-craig}
\begin{align}
\label{eq:k-plateau}
k(\beta) & =
 \frac{\kred(\beta)}{\ZRred} =
 \frac{1}{\ZRred} \int_0^{\mathrlap{\tpl}} \ \, \fftcf(t) \, \rmd t, \\
\fftcf(t) & = \frac{1}{\iu \hbar }\Tr\left\{
[\Hop, \rhoR] \, \eu{\iu \Hop t/\hbar} \op{F} \eu{-\iu \Hop t/\hbar}
\right\} \\
\label{eq:zrred}
\rule{0pt}{1.4em}%
\ZRred & = \ZR \left( \popR - [1-\popR]\ZR / \ZP  \right),
\end{align}
\end{subequations}
where $\fftcf(t)$ is the flux-correlation function and $\kred(\beta)$ is
a ``reduced'' thermal rate introduced for
notational convenience. The expression assumes a separation of time scales,
such that the flux-correlation function decays to zero for
$\tpl < t \ll t_{\mathrm{rxn}}$. Here $\tpl$ is some ``plateau time'' that is
significantly shorter than the reaction time scale $t_{\mathrm{rxn}}$.\cite{ChandlerET,ChandlerGreen}
In this work we consider the limit of small diabatic
coupling, for which $t_{\mathrm{rxn}} \to \infty$, enabling us to
also take the limit $\tpl \to \infty$.

In cases when $\popR = 1$
(\latin{e.g.}, bimolecular scattering reactions),
this rate expression reduces to the well-known
form\cite{Miller1983rate,Miller1998rate} with
$\ZR$ in place of $\ZRred$.
However, if $\popR \neq 1$ (\latin{e.g.}, in the condensed phase),
there exist other reasonable choices for $\rhoR$, not necessarily leading to equivalent rate constants
(see Appendix of \Refx{Craig2007condensed}).
In all of the systems and regimes
considered here (see \sec{sec:models})
this effect is essentially negligible,
as can be explained with the help of
the asymptotic analysis presented in this work.\footnote{
Our expressions are readily generalisable to $\rhoR = \tfrac{1}{2}
\eu{-\Hop (\beta \hbar - \sigma) / \hbar } {(1-\hat{h})}\, \eu{-\Hop \sigma  / \hbar } + \mathrm{h.c.}$, for $0 < \sigma < \beta \hbar$, which encompasses three of
the four possibilities presented in \Refx{Craig2007condensed}.
Under this definition, the only component of $\kfsc$ that
depends on $\sigma$ is $\tilde{Z}_{\mathrm{r}}^{\mathrm{sc}}$
[\eqn{eq:z2red-ints}].
The dependence is weak, as it is confined to terms proportional
to either $\eu{-\sigma \alpha_{\mathrm{s}}/ \hbar}$ or
$\eu{-(\beta \hbar -\sigma) \alpha_{\mathrm{s}}/ \hbar}$
 [\eqn{eq:zrred2-bound}], which all make a subdominant contribution.} The same
 analysis shows that
 significant differences between alternatives can in principle
emerge in extremely asymmetric and/or low-temperature systems.
Cases where this effect becomes noticeable
may arise for one of the
following reasons. Firstly, the assumption
of separation of time scales may break down,
so that the more general expression in Eq.~(2.13) of \Refx{Craig2007condensed}
must be used to define
the rate.\cite{Lawrence2019ET}
Secondly, the ambiguity may be due to how we determine if a
chemical species is a reactant or a product (\latin{i.e.}, the definition of the projection operator
$\hat{h}$).\cite{ChandlerGreen} Lastly,
the phenomenological rate constant
may depend on the initial
state of the system.\cite{Craig2007condensed}
The last two possibilities indicate that
a ``unique'' thermal rate constant
can occasionally be an ill-defined quantity,
even within exact quantum theory and experiment.
However, in many cases (such as those we consider in this paper) the rate constant
is well behaved and essentially independent of these choices.

With the terms in \eqn{eq:htot} treated as
a reference Hamiltonian, $\Hop^{(0)}$, and a perturbation, $\Hop^{(1)}$,
we can expand the rate
constant $k(\beta)$ as a power series in $\lambda$, namely
\begin{equation}
\label{eq:k-series}
k(\beta) = \sum_{\nu = 1}^{\infty} \lambda^{2 \nu} k_{2 \nu}(\beta).
\end{equation}
Note that in the above expression, coefficients of odd powers
of $\lambda$ are identically zero.
The leading (second-order) term is obtained by noting that
$\ZRred \sim \ZRn{0} = \Tr \! \big[ \eu{-\beta \HR}\big]$
and $\eu{-\Hop z / \hbar} \sim \KR(z) \ketbra{0}{0} + \KP(z) \ketbra{1}{1}$ as
$\lambda \to 0$, where we define the propagator $\Kn(z) = \eu{-\Hn z / \hbar}$
and complex time $z \equiv \tau + \iu t$. Here $\Re(z) \equiv \tau$ corresponds to
imaginary time, and $\Im(z) \equiv t$ corresponds to real time.
To second order, the flux correlation is $\fftcf(t) \sim \lambda^2 [\cgr(z) +
\cgr(\beta \hbar - z)]$, with
\begin{equation}
\label{eq:cgr-exact}
c_2(z) = \Tr\!\left[
\KR\big(\beta \hbar - z\big) \tfrac{\DRP}{\hbar} \mkern1mu
\KP\big(z\big) \tfrac{\DRP}{\hbar}
\right].
\end{equation}%
Substituting this into \eqn{eq:k-plateau} with
$\tpl \to \infty$ yields upon simplification\cite{ChandlerET,Wolynes1987nonadiabatic,Bader1990golden,InstReview}
\begin{equation}
k_2(\beta) = \frac{1}{\ZRn{0}} \int_{-\infty}^{\infty} \! c_2(z)
 \ud t, \label{eq:kgr-exact}
\end{equation}
which follows because $\Re[\cgr(z)]$ is even in $t$, allowing us to first replace the original
integral with $\tfrac{1}{2}\int_{-\infty}^{\infty} \rmd t$. Then, since $\cgr(z)$
is an analytic function, we can use Cauchy's integral theorem\cite{ComplexVariables}
to show that the integrals over $\cgr(z)$ and $\cgr(\beta \hbar - z)$ are equal, resulting in \eqn{eq:kgr-exact}. This also
shows that $k_2$ does not depend on $\tau$, since both $\ZRn{0}$
and the integral are $\tau$-independent. An important
consequence of this is that
$\tau$ can be freely chosen in a way that simplifies the evaluation of the integral,
as discussed later.

\Eqn{eq:kgr-exact} is precisely the GR limit
mentioned previously. To go beyond it to the next-order
contribution, $\kf(\beta)$,
one could proceed via the flux-correlation formalism,
expanding the propagator in a time-dependent perturbation series.\cite{feynman2010quantum,KuehnBook}
Doing so results in a $\kf$ expressed in terms of components that
have similar magnitudes but opposite signs. Apart from the numerical
difficulties associated with evaluating such an expression accurately,
it is not obvious how it can be effectively approximated using
integration by steepest descent. We will show that a better starting point is afforded by the Green's function formalism.\cite{Miller1997,GoldenGreens,Nitzan}
Central to this is the Green's function operator, which is the Laplace transform
of the full propagator,
\begin{equation}
\label{eq:Gint}
\begin{aligned}
  \Gop(E)  & = \lim_{\eta \to 0^{+}} -\frac{\iu}{\hbar} \int_0^{\infty}
  		     \eu{-\iu \Hop t/\hbar} \,
  		     \eu{\iu (E + \iu \eta)t/\hbar} \ud t \\
  		  & =  \lim_{\eta \to 0^{+}} \frac{1}{E + \iu \eta - \Hop}.
\end{aligned}
\end{equation}
Its imaginary part can be more simply written as the Fourier transform of the propagator, and is thus related to the density of states,
\begin{equation}
 \label{eq:ImGint}
\begin{aligned}
   \IGop(E) & = -\frac{1}{2 \hbar} \int_{-\infty}^{\infty}
   \eu{-\iu \Hop t/\hbar} \, \eu{\iu E t/\hbar} \ud t \\
            & = -\pi  \delta(E-\Hop).
\end{aligned}
\end{equation}
This can be used to calculate the cumulative reaction probability at energy $E$,\cite{Miller1983rate,Miller1997}
\begin{equation}
\label{eq:micro-rate}
P(E) = 2 \hbar^2 \Tr \left[
 \IGop(E) \op{F} \IGop(E) \op{F}
\right],
\end{equation}
which in turn is related to the thermal rate constant in \eqn{eq:k-plateau}
with $\tpl \to \infty$ via
\begin{equation}
\label{eq:k-therm}
\kred(\beta) = \int_0^{\infty} \!\! \fftcf(t) \, \rmd t =
 \frac{1}{2 \pi \hbar} \int_{-\infty}^{\infty} \!\eu{-\beta E} P(E) \ud E.
\end{equation}
One advantage of this formalism is that the perturbation series for the Green's
function operator, readily obtained from the Dyson equation,\cite{Nitzan}
has a particularly simple form,
\begin{equation}
\label{eq:dyson}
\Gop = \sum_{\nu = 0}^{\infty} \lambda^{\nu} \left[ \Gop^{(0)} \Hop^{(1)} \right]^{\!\nu} \! \Gop^{(0)},
\end{equation}
where $\Gop^{(0)} = \GR \ketbra{0}{0} + \GP \ketbra{1}{1}$
is the Green's function operator for the unperturbed problem, with $\Gop_{n}$ given by \eqn{eq:Gint} with Hamiltonian $\Hop_{n}$. Substituting this
into \eqs{eq:micro-rate}{eq:k-therm} leads to a perturbation series for $\kred$ that is analogous to \eqn{eq:k-series}.
The crucial difference is that this
formalism leads to expressions for
rate constants that are naturally separated into terms amenable to semiclassical approximation.

To illustrate how the two alternative formalisms connect, let us re-derive \eqn{eq:kgr-exact} starting
from the perturbation series
\begin{equation}
\label{eq:p-series}
P(E) = \sum_{\nu=1}^{\infty} \lambda^{2 \nu} P_{2 \nu}(E)
\end{equation}
and considering its $\nu=1$ term,
\begin{equation}
\label{eq:pgr}
P_2(E) = 4 \Tr \left[\IGR(E) \DRP \IGP(E) \DRP\right].
\end{equation}
Taking the Boltzmann average and using the relation in \eqn{eq:ImGint} gives
the first term in the perturbation series for~$\kred$,
\begin{align}
\kgrred(\beta) & = \frac{1}{2 \pi \hbar} \int_{-\infty}^{\infty} \eu{-\beta E} P_2(E) \, \rmd E
\nonumber \\
         & =
             \int_{-\infty}^{\infty} \frac{\rmd E}{2 \pi \hbar}
             \int_{-\infty}^{\infty} \!\! \rmd u_1
             \int_{-\infty}^{\infty} \!\! \rmd u_2  \\
         & \quad \quad
         \eu{-E(\beta \hbar - \iu (u_1+u_2))/\hbar}
         \Tr\left[
         \KR(\iu u_2) \tfrac{\DRP}{\hbar} \mkern1mu \KP(\iu u_1)
          \tfrac{\DRP}{\hbar}
         \right]. \nonumber
\end{align}
The integral over energy can be simplified
by the variable transformation $\bar{t} = u_1 + u_2$, $t = (u_2 - u_1)/2$,
and the integration contour for the new variable $\bar{t}$ can be shifted along the
imaginary-time axis, $\bar{t} \to \bar{t} - \iu \beta \hbar$,
so that the energy integral becomes
\begin{equation}
 \frac{1}{2 \pi \hbar} \int_{-\infty}^{\infty} \eu{\iu E \bar{t} / \hbar} \, \rmd E = \delta(\bar{t}).
\end{equation}
Integrating over $\bar{t}$ leaves
\begin{equation}
\kgrred =  \!
             \int_{-\infty}^{\infty} \!
         \Tr\!\left[
         \KR\big(\tfrac{\beta \hbar}{2} - \iu t\big) \tfrac{\DRP}{\hbar} \mkern1mu
         \KP\big(\tfrac{\beta \hbar}{2} + \iu t\big)
		 \tfrac{\DRP}{\hbar}
         \right]\rmd t.
\end{equation}
At this point we recognise that the remaining integration variable can also be shifted along the imaginary-time axis, and use
$k_2 = \kgrred / \ZRn{0}$ to recover
\eqn{eq:kgr-exact}. This approach is generalised in \sec{sec:new-theory} to derive the next
term in the series,~$\kf$.

\subsection{\label{ssec:semiclassical}%
Semiclassical approximation}

We now summarise previous derivations of the semiclassical approximation to $\kgr(\beta)$.\cite{InstReview,Ansari2021}
Calculating the quantum correlation
function is only computationally feasible for low-dimensional or particularly
simple model systems. Applications
to realistic chemical reactions call for approximations to \eqn{eq:kgr-exact}
that make the calculation computationally tractable. The approach that we use
is to expand the trace in terms of position eigenstates and replace the exact
quantum propagator with its semiclassical counterpart,\cite{vanVleck1928correspondence,Miller1971density,GutzwillerBook}
\begin{equation}
\label{eq:sc-propa}
\braket{\xfin | \Kn(\tau_n) | \xin} \sim
\Knsc(\xfin,\xin,\tau_n) = \sum_{\text{traj.}} \sqrt{\frac{C_n}{(2 \pi \hbar)^f}} \,
\eu{-S_n / \hbar},
\end{equation}
where  `$\mathord{\sim}$' denotes an asymptotic
relationship\cite{BenderBook} and the sum is over all classical trajectories that travel from $\xin$ to $\xfin$ in imaginary time $\tau_n$
(typically only one of which dominates). $S_n \equiv S_n(\xfin,\xin,\tau_n)$ is
the corresponding stationary Euclidean action,\
\begin{equation}
\label{eq:action}
S_n[\vec{x}(\mathord{\cdot})] = \int_{0}^{\tau_n} \left[
\frac{m}{2} \left\lVert \der[]{\vec{x}}{u} \right\rVert^2 + V_n\big(\vec{x}(u)\big)
\right] \rmd u,
\end{equation}
and
\begin{equation}
C_n = \left \rvert
-\pders{S_n}{\xin}{\xfin}
\right \rvert.
\end{equation}
Hence we can write
\begin{align}
\label{eq:kgr-sc}
\kgrred \sim \iiint_{-\infty}^{\infty} \!\! \rmd \vec{x}' \rmd \vec{x}'' \rmd t
\, \sqrt{\frac{C_0 C_1}{(2 \pi \hbar)^{2f}}} \frac{\Delta(\vec{x}')
 \Delta(\vec{x}'')}{\hbar^2} \, \eu{-S_2 / \hbar},
\end{align}
where $S_2(\vec{x}', \vec{x}'', z) = S_0(\vec{x}', \vec{x}'', \beta \hbar -  z) + S_1(\vec{x}'', \vec{x}', z) $. The semiclassical approximation to the exact
propagator in \eqn{eq:sc-propa} is obtained
by steepest-descent integration of its path-integral representation.\cite{Kleinert}
The same technique can be applied to approximate the remaining integrals
in \eqn{eq:kgr-sc}. For a one-dimensional example this typically reads
\begin{align}
\label{eq:sd-1d}
&\int_a^b g(x) \mkern1mu \eu{-\phi(x) / \hbar} \, \rmd x \\
& \qquad {} \sim
g(\bar{x}) \mkern1mu \eu{-\phi(\bar{x}) / \hbar} \! \int_{-\infty}^{\infty} \! \exp \!\left[
- \der[2]{\phi}{x} \frac{(x-\bar{x})^2}{2 \hbar}
\right] \rmd x \nonumber \\
& \qquad {} = \sqrt{2 \pi \hbar} \, \left(
\der[2]{\phi}{x}
\right)^{\mathrlap{\!\!-1/2}} \
\, g(\bar{x}) \mkern1mu \eu{-\phi(\bar{x}) / \hbar}  \nonumber
\end{align}
as $\hbar \to 0$, where the double derivative
$\rmd^2 \phi / \rmd x^2 > 0$ is evaluated at the point
$\bar{x}$ where $\phi(x)$ reaches its minimum value
on the interval $(a, b)$, with the corresponding
first derivative satisfying $\rmd \phi / \rmd x = 0$.
In effect, the
steepest-descent approximation consists in replacing the exponentiated
function with its Taylor series expansion about $\bar{x}$, truncated at the second-order term. All remaining factors are replaced with the
leading-order terms in their series expansions about the
same point, and the integration bounds are extended to $\pm \infty$.
The resulting Gaussian integral can then be evaluated analytically.
The relative error associated with this approximation
becomes vanishingly small as $\hbar \to 0$, provided certain conditions
are satisfied. In particular, we can
ignore higher-order terms in the Taylor series expansion of $\phi(x)$
as long as $\rmd^2 \phi / \rmd x^2$
is not itself vanishingly small. Similarly,
we can extend the integration range to span the entire real line provided
the stationary point $\bar{x}$ does not lie infinitesimally close to either
of the bounds. Finally, we can replace $g(x)$ with its value at the
stationary point, provided $g(x)$ does not vary rapidly in its vicinity.
Later in the paper we encounter cases where some of these conditions are not satisfied, at which point the procedure is modified accordingly.

For completeness we note that all of the above also
applies to functions with multiple minima within the integration domain, provided these are well separated from each other. If this
condition is satisfied, one may sum over the contributions
from all such minima, as in \eqn{eq:sc-propa}, which ultimately leads to a sum over contributions from competing reaction mechanisms.
However, no general steepest-descent prescription is available for when
this condition is not satisfied, such as is typically
encountered when calculating reaction rates in liquid systems,
to which the instanton
approach is not directly applicable.\cite{InstReview} Tackling such
systems would require the use of path-integral sampling methods.
From here on we assume that our system
is well behaved such that the relevant minima
are isolated from each other, and focus
on one minimum at a time.

The asymptotic limit of
\eqn{eq:kgr-sc} can then be obtained by generalising \eqn{eq:sd-1d}
to multiple dimensions. The stationary point
$(\vec{x}', \vec{x}'\mathrlap{'}, \tau)$ satisfies\cite{GoldenGreens,InstReview,Ansari2021}
\begin{subequations}
\label{eq:gr-stat}
\begin{align}
\pder{S_2}{\vec{x}'} & = \vec{p}_0' - \vec{p}_1' = \vec{0} \\
\pder{S_2}{\vec{x}''} & = \vec{p}_1'' - \vec{p}_0'' = \vec{0} \\
\pder{S_2}{\tau} & = E_1 - E_0 = 0
\end{align}
\end{subequations}
where $E_n$ are the energies of a classical trajectory comprised
of two parts ($n \in \{0,1\}$). The first part
corresponds to travelling on $V_0$ from $\vec{x}''$ to $\vec{x}'$
over an imaginary time
$\beta \hbar - \tau$, with initial and final momenta $\vec{p}_0''$ and $\vec{p}'_0$. The second part corresponds to travelling
on $V_1$ from $\vec{x}'$ to $\vec{x}''$ over an imaginary time $\tau$.
The conditions in \eqn{eq:gr-stat} follow directly from the definition of $S_2$ and
the relations
\begin{equation}
\label{eq:dsdstuff}
\pder{S_n}{\vec{x}_{\mathrm{i}}} = -\vec{p}_{\mathrm{i}}, \quad
\pder{S_n}{\vec{x}_{\mathrm{f}}} = \vec{p}_{\mathrm{f}}, \quad
\pder{S_n}{\tau_n} = E_n.
\end{equation}
The path that makes the combined $S_2$ action stationary is therefore
a periodic classical trajectory in imaginary time with period $\beta \hbar$,
comprised of
a reactant and a product segment. Conservation of energy and momentum
imposed by \eqn{eq:gr-stat} implies that at the hopping points $\vec{x}', \, \vec{x}''$, where the two segments join, the trajectory is continuous. Together with time-reversal
symmetry this typically imposes $\vec{x}' = \vec{x}'' = \vec{x}^{\ddagger}$, where
$\vec{x}^{\ddagger}$ is a point on the seam along which the reactant
and product potential energies are equal. This stationary path is known as the golden-rule instanton.

In the Marcus normal regime, which we assume throughout this work,
the value of $\tau$ that extremises
the action (which we will call $\tauGR$) is in the range \mbox{$0 < \tauGR < \beta \hbar$}. The instanton is a first-order saddle point
of the action and can be found by discretising the trajectory (\latin{i.e.}, representing it as a ring polymer) and performing multidimensional optimisation
of the resulting extended classical
system.\cite{GoldenGreens,GoldenRPI,InstReview,Ansari2021} This is closely
related to the analogous procedure for adiabatic instantons,\cite{Andersson2009Hmethane,RPInst,Rommel2011locating} except now one
must optimise the action in $\tau$ as well as $\vec{x}'$ and $\vec{x}''$. At the end of the
optimisation one calculates the action $S_2$ and its Hessian
\begin{equation}
\label{eq:sigma2}
\mat{\Sigma}_2 =
\left[
 \begin{array}{*{3}{>{\rule[-0.5em]{0pt}{1.75em}}C{2.5em}}}
    \pders{S_2}{\vec{x}'}{\vec{x}'} &  \pders{ S_2}{\vec{x}'}{\vec{x}''} & \pders{S_2}{\vec{x}'}{\tau} \\
    \pders{S_2}{\vec{x}''}{\vec{x}'} &  \pders{S_2}{\vec{x}''}{\vec{x}''} & \pders{S_2}{\vec{x}''}{\tau} \\
    \pders{S_2}{\tau}{\vec{x}'} &  \pders{S_2}{\tau}{\vec{x}''} & \pder[2]{S_2}{\tau}
    \end{array}
    \right]
\end{equation}
at the stationary point $(\vec{x}^{\ddagger}, \vec{x}^{\ddagger}, \tauGR)$. The
multidimensional generalisation of \eqn{eq:sd-1d}, along with the Cauchy--Riemann equations\cite{ComplexVariables}
relating the partial derivatives with respect to $\tau$ and $t$, gives
\begin{align}
\label{eq:kgrred-sc}
\kgrred \sim \kgrredsc = \sqrt{2 \pi \hbar} \mkern1mu
 \frac{\Delta^2}{\hbar^2}
\sqrt{
\frac{C_0 C_1}{-\lvert \mat{\Sigma}_2 \rvert}
} \mkern1mu \eu{-S_2 / \hbar}.
\end{align}
To complete the derivation, the semiclassical approximation to the reactant partition function $\ZRn{0}$ is evaluated following the same approach. This time the
trajectory that makes the action stationary is collapsed at the
bottom of the reactant well, and in the absence of translational or rotational degrees
of freedom the expression reduces to\cite{InstReview,Kleinert}
\begin{equation}
\label{eq:zreact-sc}
\ZRn{0} \sim \ZRn{0}^{\SC} = \eu{-\beta \epsilon_{\mathrm{r}} }
\prod_{j=1}^{f} \frac{1}{2} \csch\!\left( \frac{\beta \hbar\mkern1mu \omega_{\mathrm{r},j}}{2}\right),
\end{equation}
where $\epsilon_{\mathrm{r}}$ is the energy at the minimum of the reactant well, and
$ \omega_{\mathrm{r},j} $ is the frequency of its $j$-th vibrational normal mode.
If present, translational and rotational modes can also be accounted for.\cite{Ansari2021} Combining \eqs{eq:kgrred-sc}{eq:zreact-sc}
gives the final result, $\kgr \sim \kgrsc = \kgrredsc / \ZRn{0}^{\SC}$.

By representing the instanton trajectory as a
ring polymer,\cite{GoldenRPI,InstReview,Ansari2021}
the formula can be readily applied to realistic molecular potentials,\cite{Heller2021}
for which the analytic form of the stationary action is not known.
In this representation finding the stationary action is equivalent to
a multidimensional optimisation problem, which can be solved
efficiently with well-established numerical techniques.

Semiclassical instanton theory is however not limited to the golden-rule term.
In other work, we have already shown\cite{Ansari2021} how to generalise the approach to tackle the
breakdown of GR in multistate systems
reacting via the superexchange mechanism.\cite{Anderson1950Superex,KuehnBook,Jortner2002DNASuperex,Franzen1993SupexET,Jang2001nonadiabatic}
Specifically, a three-state system that reacts via this mechanism
has a rate constant with a \emph{leading} fourth-order
dependence on $\Delta$. We have shown how
instanton theory can be applied to such systems,
enabling semiclassical calculations of bridge-mediated
electron-transfer rates.\cite{Ansari2021}
In what follows, we develop another kind
of fourth-order rate theory, one that
describes contributions to the nonadiabatic rate \emph{beyond} leading order. The
underlying instantons (see \sec{sec:k4-sc})
share some similarities with those presented
in \Refx{Ansari2021}, but at the same time exhibit a set of new features that stem from nuclear tunnelling and nonadiabaticity  combining to influence the reaction mechanism.

\section{\label{sec:new-theory}%
Exact fourth-order rate expression}

Substituting \eqn{eq:dyson} into \eqn{eq:micro-rate} shows that the fourth-order
contribution to the cumulative reaction probability is
\begin{equation}
\Pf(E) = -\Pfa(E) + \Pfb(E),
\end{equation}
where
\begin{subequations}
 \label{eq:pf-tr}
	\begin{align}
	    \label{eq:pfa-tr}
		\Pfa(E) & = 8 \Tr \left[
			\IGR \DRP \IGP \DRP \IGR \DRP \IGP \DRP
		\right], \\
        \label{eq:pfb-tr}
		\Pfb(E) & = 8 \Tr \left[
			\IGR \DRP \RGP \DRP \RGR \DRP \IGP \DRP
		\right].
	\end{align}
\end{subequations}
Both terms are comprised of four $\Gn$ factors,
alternating between the reactant ($n=0$) and product ($n=1$)
diabats. This corresponds to a total of four state changes
in a single trace, as opposed to the two changes in the golden-rule
expression [\eqn{eq:pgr}]. One could directly approximate the terms in \eqn{eq:pf-tr} with
semiclassical methods, arriving at expressions
that have simple physical interpretations in terms of instanton
trajectories. As discussed in \Refx{GoldenGreens}, in the forbidden regime the dominant contribution to
$\braket{\xfin|\RGn| \xin}$
comes from a trajectory that goes directly from $\xin$ to
$\xfin$. On the other hand, the dominant contribution to
$\braket{\xfin|\IGn|\xin}$
comes from a trajectory that reaches a turning point where
$V_n(\vec{x}) = E$. This feature is known as a ``bounce''.
The GR probability  in \eqn{eq:pgr}
corresponds therefore to an instanton trajectory that bounces
a total of two times: once off the reactant and once off the
product diabat, as shown schematically in \fig{fig:k4a-instanton}(c).
The $\Pfa$ term corresponds to an instanton with four bounces, as
in \fig{fig:k4a-instanton}(b), and the $\Pfb$ term to an instanton
with two bounces and two consecutive direct segments on alternating
diabats, as in \fig{fig:k4b-blip}.

Despite the ease of physical interpretation, the semiclassical
Green's function is not as well behaved
as the imaginary-time propagator.\cite{GoldenGreens,Carlitz1985semiclassical,Faraday}
For this reason we convert the microcanonical
reaction probabilities into thermal
rate constants, expressed in terms of imaginary-time propagators.
Throughout the derivation it is assumed that we can take the
limit $\tpl \to \infty$, just as in the case of the GR rate constant.
The derivation then amounts to finding the
$\mathcal{O}(\lambda^4)$ term $\kfred = \kfredb - \kfreda$ in the expansion of $\kred(\beta)$ [\eqn{eq:k-therm}] and the
$\mathcal{O}(\lambda^2)$ term $\ZRredn{2}$ of $\ZRred$ [\eqn{eq:zrred}].
Together these give the total fourth-order rate constant
\begin{equation}
\label{eq:k4-therm}
k_4 = \kgr \left(
  - \frac{\kfreda}{\kgrred} + \frac{\kfredb}{\kgrred} -
\frac{\ZRredn{2}}{\ZRn{0}}\right).
\end{equation}
Continuing in the same fashion one can obtain analogous expressions for
$k_{6},\,k_{8},$ etc., which we intend to pursue in future work.

\subsection{\label{ssec:exact-ka}%
\emph{A}-type term}

Following a similar approach to \sec{ssec:greens-tcf}, we
use the Fourier transform representation of the imaginary Green's function operator to yield
\begin{equation}
\begin{aligned}
\kfreda = \frac{1}{2 \hbar^4} \int_{-\infty}^{\infty} \frac{\rmd E}{2 \pi \hbar}
\int \rmd^4 \cvec{u} \, \eu{
-E \left(
\beta \hbar - \iu  \sum u_i
\right)/\hbar
} & \\
 {} \times  \Tr \left[
\KR(\iu u_4) \DRP \KP(\iu u_3) \DRP \KR(\iu u_2)  \DRP \KP(\iu u_1)  \DRP
\right], &
\end{aligned}
\end{equation}
where $\int \rmd^4 \cvec{u} \equiv \int_{-\infty}^{\infty} \rmd u_1 \cdots \int_{-\infty}^{\infty} \rmd u_4$. Applying the variable transformation
\begin{align}
\label{eq:trans-4a}
  \begin{bmatrix}[1.5]
     u_1 \\
     u_2 \\
     u_3 \\
     u_4
  \end{bmatrix} & =
  \begin{bmatrix}[1.5]
      0 & 0 & -1 & \frac{1}{2} \\
      \frac{1}{2} & -1 & 0 & -\frac{1}{2} \\
      0 & 0 & 1 & \frac{1}{2} \\
      \frac{1}{2} & 1 & 0 & -\frac{1}{2}
  \end{bmatrix}
    \begin{bmatrix}[1.5]
      \bar{t} \\
      t_0 \\
      t_1 \\
      t
    \end{bmatrix},
\end{align}
allows us to integrate over $E$ and $\bar{t}$ in the same fashion as before. The
remaining integration variables are shifted into the complex plane, so that
$\iu t_0 \to z_0$,  $\iu t_1 \to z_1$ and $\iu t \to z$,
where $z_n \equiv \tau_n + \iu t_n$. This leads to the final expression
\begin{subequations}
\label{eq:k4a-exact}
\begin{align}
\label{eq:k4a-exact-rate}
&\kfreda = \frac{1}{2}
        \int_{-\infty}^{\infty} \!\! \rmd t_0
        \int_{-\infty}^{\infty} \!\! \rmd t_1
        \int_{-\infty}^{\infty} \!\! \rmd t \ \cfa(z_0, z_1, z), \\
\label{eq:k4a-exact-tcf}
& \cfa(z_0, z_1, z) = \Tr \! \bigg[
      \KR\Big(\tfrac{\zR}{2} + z_0\Big)
      \tfrac{\DRP}{\hbar} \mkern1mu
      \KP\Big(\tfrac{\zP}{2} + z_1\Big)
      \tfrac{\DRP}{\hbar} \nonumber \\
& \qquad \qquad \quad
{} \times  \KR\Big(\tfrac{\zR}{2} - z_0\Big)
           \tfrac{\DRP}{\hbar} \mkern1mu
           \KP\Big(\tfrac{\zP}{2} - z_1\Big)
           \tfrac{\DRP}{\hbar}
    \bigg],
\end{align}
\end{subequations}
where we introduce the notation $\zR \equiv \beta \hbar - z$ and $\zP \equiv z$.
The expression defines the quantum $\kfreda$ and will be given a semiclassical
treatment in \sec{ssec:ka}.

\subsection{\label{ssec:exact-kb}%
\emph{B}-type term}
We rewrite the second component of the fourth-order reaction probability as
\begin{equation}
\Pfb = \Pfa + 8 \Real \left\{
\Tr \left[
    \IGR \mkern1mu \DRP \mkern1mu
    \GP  \DRP \mkern1mu
    \GR  \mkern1mu \DRP \mkern1mu
    \IGP \DRP
\right]
\right\}
\end{equation}
and recast the Green's function operators as integral transforms according to \eqs{eq:Gint}{eq:ImGint}, to yield
\begin{align}
    & \kfredb  = \kfreda + \Real\bigg\{\!{-\frac{2}{\hbar^4}} \! \int_{-\infty}^{\infty} \frac{\rmd E}{2 \pi \hbar}
    \int \! \rmd^4 \cvec{u} \
    \eu{
        -E \left(
        \beta \hbar - \iu  \sum u_i
        \right)/\hbar
        } \nonumber \\
    & \ {} \times  \Tr \left[
    \KR(\iu u_4) \DRP \KP(\iu u_3) \DRP
    \KR(\iu u_2) \DRP \KP(\iu u_1) \DRP
    \right]\!\bigg\},
\end{align}
where the integration ranges are now
\begin{equation}
\int \rmd^4 \cvec{u} \equiv
    \int_{-\infty}^{\infty} \!\! \rmd u_1
    \int_{0}^{\infty} \! \rmd u_2
    \int_{0}^{\infty} \! \rmd u_3
    \int_{-\infty}^{\infty} \!\! \rmd u_4.
\end{equation}
Under the variable transformation
\begin{align}
\label{eq:trans-4b}
  \begin{bmatrix}[1.5]
     u_1 \\
     u_2 \\
     u_3 \\
     u_4
  \end{bmatrix} & =
  \begin{bmatrix}[1.5]
      0 & 0 & -1 & 1 \\
      0 & 1 & 0 & 0 \\
      0 & 0 & 1 & 0 \\
      1 & -1 & 0 & -1
  \end{bmatrix}
    \begin{bmatrix}[1.5]
      \bar{t} \\
      t_0 \\
      t_1 \\
      t
    \end{bmatrix}
\end{align}
the expression simplifies and can be integrated over $E$ and $\bar{t}$ as before, so that
\begin{subequations}
\label{eq:k4b-exact}
\begin{align}
    & \kfredb  = \kfreda + 2 \Real\bigg\{
    \int_{0}^{\mathrlap{\iu \infty}} \ \rmd z_0 \!
    \int_{0}^{\mathrlap{\iu \infty}} \ \rmd z_1 \!
    \int_{\mathrlap{-\infty}}^{\mathrlap{\infty}} \ \rmd t \
    \cfb  \bigg\} \\
    & \cfb(z_0, z_1, z) =
    \begin{aligned}[t]
    \Tr \Big[ & \KR(\zR - z_0)
    \tfrac{\DRP}{\hbar} \mkern1mu
    \KP(z_1) \tfrac{\DRP}{\hbar}  \\
    {} \times {} & \KR(z_0) \tfrac{\DRP}{\hbar} \mkern1mu
    \KP(\zP - z_1) \tfrac{\DRP}{\hbar} \Big].
    \end{aligned}
\end{align}
\end{subequations}
The integration contours for $z_{n}$ can be deformed to simplify evaluation. Introducing
\begin{equation}
\label{eq:J-def}
\margfb(z_0, z_1) = \int_{-\infty}^{\infty} \! \cfb(z_0, z_1, z) \, \rmd t,
\end{equation}
we apply Cauchy's integral theorem\cite{ComplexVariables} iteratively to get
\begin{align}
\label{eq:z0z1-int}
&\int_0^{\mathrlap{\iu \infty}} \ \,
 \int_0^{\mathrlap{\iu \infty}} \ \ \margfb(z_0, z_1) \, \rmd z_0
 \rmd z_1 = {} \\
& \left[
\int\limits_0^{\mathrlap{\tau_0^{*}}}
\int\limits_0^{\mathrlap{\tau_1^{*}}} \, + \,
\int\limits_0^{\mathrlap{\tau_0^{*}}}
\int\limits_{\tau_1^{*}}^{\mathrlap{\sigma_1^{*}}}
\, + \,
\int\limits_{\tau_0^{*}}^{\mathrlap{\sigma_0^{*}}}
\int\limits_0^{\mathrlap{\tau_1^{*}}}
\, + \,
\int\limits_{\tau_0^{*}}^{\mathrlap{\sigma_0^{*}}}
\int\limits_{\tau_1^{*}}^{\mathrlap{\sigma_1^{*}}} \ \
\right] \! \margfb(z_0, z_1) \, \rmd z_0 \rmd z_1, \nonumber
\end{align}
where $\sigma^{*}_n \equiv \tau^{*}_n + \iu \infty$. For future reference
we label the integration domains on the right-hand side of \eqn{eq:z0z1-int} as
$\mathcal{A}_{\tau\tau}$, $\mathcal{A}_{\tau t}$, $\mathcal{A}_{t\tau}$ and
$\mathcal{A}_{tt}$. A suitable choice of $\tau_n^{*}$ simplifies
the numerical evaluation of the integral by minimising the oscillations
in $\margfb(z_0, z_1)$. The final result of this section
is the formally exact definition of the quantum $\kfredb$, which will be
given a semiclassical treatment in \sec{ssec:kb}.

At this point it is worth re-emphasising that we have pursued this
particular route to $\kfred$ because the $\kfreda$ and
$\kfredb$ terms can be well approximated by semiclassical
techniques.
Both terms are expressed as integrals over three-time correlation
functions which quickly decay to zero
along appropriately chosen contours (see \sec{sec:k4-sc}).
Taking the route via time-dependent perturbation theory\cite{KuehnBook}
similarly results in a sum of integrals
over three-time correlation functions.
Although the corresponding value of $\kfred$ is identical to that obtained from the
Green's function approach,
the individual correlation
functions are not
straightforwardly related to $\cfa$ and $\cfb$.
The various transforms leading to the latter mean that the $\Im z \equiv t$ in these expressions does not refer to the same physical quantity as in $\fftcf(t)$. Even among $\cfa$ and $\cfb$ the symbol $t$ assumes different meanings, which follows from the different transformations in \eqs{eq:trans-4a}{eq:trans-4b}.

Crucially, as far as
we can tell, the functions that emerge from time-dependent
perturbation theory cannot be made to decay quickly with time,
regardless of how the integration contours are deformed.
For this reason they are not readily amenable to steepest-descent integration,
which is why we employ the Green's function approach.

\subsection{Partition function%
\label{sec:exact-partf}}

To derive the $\mathcal{O}(\lambda^2)$ term in the
perturbation expansion of $\ZRred$, we use\cite{Weiss}
\begin{subequations}
\label{eq:partf-series}
\begin{align}
\ZR & =
\sum_{\nu = 0}^{\infty} \lambda^{2\nu} \ZRn{2\nu},\\
\ZRn{2\nu} & = \frac{1}{\hbar^{2\nu}} \int_{0}^{\beta \hbar} \!\! \rmd u_{2\nu}
\int_{0}^{u_{2\nu}} \!\! \rmd u_{2\nu-1} \cdots
\int_{0}^{u_{2}} \!\! \rmd u_{1} \\
& \quad \Tr \Bigg[
\eu{-\beta \HR} \! \prod_{\mu=0}^{\nu-1}
\KR(r_{\nu-\mu}) \DRP \mkern1mu \KP(s_{\nu-\mu}) \DRP
\Bigg], \nonumber
\end{align}
\end{subequations}
where $s_{\mu} \equiv u_{2\mu} - u_{2\mu-1}$, $r_{\mu} \equiv u_{2\mu+1} - u_{2\mu}$ and $u_{2\nu+1} \equiv u_1$. Explicitly, the first two terms are
\begin{subequations}
\begin{align}
\ZRn{0} & = \Tr \left[ \eu{-\beta \HR} \right], \\
\label{eq:z2}
\ZRn{2} & = \int_0^{\beta \hbar} \!\!
 [\beta \hbar - u] \,
 c_2(u) \, \rmd u.
\end{align}
\end{subequations}
The integrand in \eqn{eq:z2} contains a factor that is precisely the golden-rule correlation function in \eqn{eq:kgr-exact}, except it is now integrated over imaginary time instead of $t$. Analogous results for $\ZPn{2 \nu}$ can be obtained by exchanging
the diabatic state labels $0 \leftrightarrow 1$, \latin{e.g.}, $\ZPn{0} = \Tr \big[ \eu{-\beta \HP} \big]$.
We can write a similar perturbation series for the initial reactant
population,
\begin{equation}
\popR = \sum_{\nu=0}^{\infty} \lambda^{2 \nu} \popRn{2\nu}
\end{equation}
where $\popRn{0} = 1$ and
\begin{align}
    \popRn{2} = \frac{1}{\ZRn{0}} \bigg\{
    &\int_0^{\frac{\beta \hbar}{2}} \! [\beta \hbar - 2 u] \,
    c_2(u) \, \rmd u - \ZRn{2} \bigg\}.
\end{align}
It follows that
\begin{equation}
\label{eq:zrred2}
\ZRredn{2} = \ZRn{2} + \ZRn{0} \, \popRn{2} \left( 1 + \frac{\ZRn{0}}{\ZPn{0}} \right),
\end{equation}
which, together with our previous expressions for $\kfreda$ and $\kfredb$,
gives the overall $\kf$ rate.
This can then be
added to $\kgr$, producing a revised weak-coupling approximation to the full (non-perturbative) rate.
However one may rightly question whether a partial sum makes the
best use of the post-GR terms, since perturbation series
often have poor convergence properties and may even diverge. We now discuss
how this issue can be circumvented.

\subsection{\label{sec:pade}%
Pad\'{e} summation}

An introductory account of series acceleration
can be found in Chapter~8 of the textbook by \citeauthor{BenderBook}.\cite{BenderBook} Here we only
summarise some key points.
Slowly convergent and divergent
series are common features of asymptotic analysis
that typically arise when the quantity being described
is not an analytic function of the perturbation.
There nonetheless exist several techniques that
can yield high-accuracy approximations to the true values
of such functions given a limited number of terms.
These approaches work by assuming a representation
of the target function that, unlike
a Taylor series, can describe non-analytic features such as poles.
One widely used
approach of this type is Pad\'{e} summation, which represents the target function as a ratio
of two polynomials. The ratio is known as a
Pad\'{e} approximant and is characterised by the degrees of
the denominator and numerator polynomials, $M$ and $N$.

The advantages of Pad\'{e} summation can be seen by taking
as an example the geometric series $1 - x + x^2 - x^3 + \ldots,$ which
diverges for $|x|\ge1$. However its formal
sum $(1+x)^{-1}$ is defined everywhere except at $x = -1$, and
any Pad\'{e} approximant with denominator degree $M \geq 1$
will exactly recover the formal result. This particular example
of a divergent series may seem oversimplified, but it
does appears in physical contexts, \latin{e.g.}, in adiabatic
microcanonical instanton theory.\cite{Miller1975semiclassical}

Another example can be drawn from the perturbation expansion
of the nonadiabatic cumulative reaction probability,
\eqn{eq:p-series}. At present we do not have a useful
resummation formula for the entire series, however one can be derived for the subset of terms that only involve imaginary
parts of reactant and product Green's function operators (``all-bounce terms''), such
as $P_2$ in \eqn{eq:pgr} and $\Pfa$ in \eqn{eq:pfa-tr}. Denoting
the sum over that subset with $P_{\mathrm{A}}$, it can be shown that
\begin{equation}
\label{eq:pa}
P_{\mathrm{A}}(E) = -\! \sum_{s = 1}^{\infty} 4 s \left(
\! {-\frac{P_2(E)}{4}}
\right)^{\mathrlap{\!\!s}}  = \frac{P_2(E)}{\left[1 + \tfrac{1}{4}P_2(E)\right]^2},
\end{equation}
which converges for $\lvert P_2(E)\rvert < 4$. In this case
the exact result is recovered by any Pad\'{e} approximant with $M \geq 2$ and $N \geq 1$. Even
the approximant derived from just the first two terms in the series, $P_2(E) / \big[1 + \tfrac{1}{2} P_2(E)\big]$,
offers a substantial improvement over partial summation.

Of course these examples are series for which all
of the terms are known and can be summed without recourse to series acceleration. Here the effectiveness of Pad\'{e} summation is easy to prove since the resummed series is in each case precisely
a ratio of polynomials. However in cases when Pad\'{e} summation becomes practically useful one only knows the first few terms of
a series, as is the case for the thermal rate constant (even within the all-bounce subset).
To quote from \citeauthor{BenderBook}, ``Pad\'{e} approximants often work quite well, even beyond their proven range of
applicability.'' In the spirit of this quotation, we do not attempt to prove that
Pad\'{e} summation of the perturbation series for the thermal rate converges to the exact
result, and use it as a heuristic means of
extending the range of coupling strengths at which our new theory
gives reasonable predictions.
As with \eqn{eq:pa}, we only \emph{Pad\'{e}} the all-bounce
terms ($\kgrred$ and $\kfreda$), since they appear in the series with alternating signs. This ensures that the resulting
Pad\'{e} approximant does not have any artificial singularities for physically relevant values of the diabatic coupling.
The remaining contribution, $\kfredb / \ZRn{0} - \kgr \ZRredn{2}/ \ZRn{0}$, is usually positive, and we find it is best to add it
to the $\kfreda$ Pad\'{e} approximant in the usual manner,\footnote{
Quite possibly there are cases when it is better to also include  fourth-order components other than $\kfreda$
into the Pad\'{e} approximant, \latin{e.g.},
for systems in the Marcus inverted regime, where $\kfredb$ [\eqn{eq:k4b-exact}] likewise serves to
decrease the full nonadiabatic rate.}
yielding
\begin{equation}
\label{eq:k24-pade}
\kfpade = \kgr\Bigg(
 \bigg[
 1 + \frac{\kfreda}{\kgrred}
 \bigg]^{\!\mathrlap{-1}}\ + \frac{\kfredb}{\kgrred}
  - \frac{\ZRredn{2}}{\ZRn{0}}
\Bigg).
\end{equation}
The choice of terms included in the Pad\'{e} approximant
is not unique, and we do not undertake to explore all the possibilities at this stage.
A rigorous justification of the current approach (or indeed an alternative)
can only emerge once the theory of further high-order terms ($k_6, k_8,$
\latin{etc.}) is developed. For the present we note that: (i) \eqs{eq:k4-therm}%
{eq:k24-pade} by construction agree at small $\Delta$ up to fourth order;
(ii) \eqn{eq:k24-pade} makes remarkably accurate numerical predictions
at intermediate $\Delta$ for multidimensional spin--boson models
in \sec{ssec:spin-boson} (see \fig{fig:sb-rates}); and
(iii) our Pad\'{e}-summed term is analogous to the nonadiabatic rates derived in Refs.~\citenum{Sparpaglione1988} and~\citenum{Cao2000} for a reduced model of electron transfer in solution using complementary techniques.

The total rate defined either as in \eqn{eq:k24-pade} or as
a partial sum satisfies detailed balance for the forward and backward reactions to fourth order in the diabatic coupling.
This also holds when the constituent terms are replaced by their
semiclassical approximations, which we derive next.

\section{\label{sec:k4-sc}%
Fourth-order instanton rate theory}

Having split $\kf$ into three different terms, we now derive a semiclassical approximation for each of them in turn.
Derivations for $\kfredasc$ and $\kfredbsc$ follow the same pattern.
First, we consider their high-temperature limits, which happen to
be identical to the quantum rate constants
for a system of two one-dimensional linear diabatic potentials,
$V_n(x) = \Vdd + \kappa_n x$. Then we generalise this result to arbitrary temperatures
and potentials. The remaining term, which describes changes in the reactant partition
function, is obtained directly, without special consideration of the high-temperature
regime. To keep the derivations as simple as possible we assume here that there are no translational or rotational normal modes. If such zero-frequency modes are present,
they can be accounted for as described in \Refs{InstReview,Ansari2021}.

\subsection{\label{ssec:ka}%
Semiclassical \emph{A}-type contribution}

\begin{figure}[ht]
\includegraphics{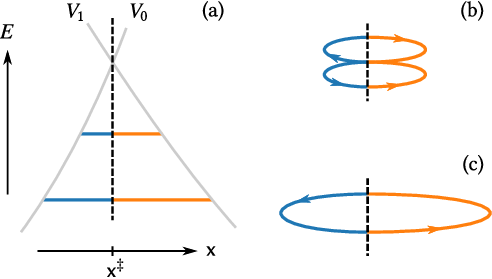}
\caption{\label{fig:k4a-instanton}
Instantons for a nonadiabatic reaction involving a pair of diabats $V_{n}(\vec{x})$.
The blue and orange
segments are classical imaginary-time trajectories on $V_{0}$ and $V_1$
that traverse a path starting and ending at the hopping point $\vec{x}^{\ddagger}$, bouncing once at the turning point.
(a)~shows a 4A (top) and a GR (bottom) instanton; (b) and (c) show
expanded diagrams of these two instantons, in the same order. Here the vertical dimension has no physical meaning
and is introduced for clarity. Each of the two loops of the 4A instanton in (b) is identical to a GR instanton at twice the temperature.
}
\end{figure}

In order to derive the semiclassical instanton formulation of $\kfreda$, we insert a set of four position-state resolutions of the identity into \eqn{eq:k4a-exact-tcf} and replace each instance of $\braket{\vec{x}_i | \Kn | \vec{x}_j}$
with the corresponding semiclassical propagator [\eqn{eq:sc-propa}]. This gives rise to the
combined action
\begin{align}
\label{eq:s4a-defn}
S_{\fa} & \equiv
     S_0\big(\vec{x}_1, \vec{x}_4, \tfrac{\zR}{2} + z_0\big)
   + S_1\big(\vec{x}_4, \vec{x}_3, \tfrac{\zP}{2} + z_1\big)  \nonumber \\
& {}
   + S_0\big(\vec{x}_3, \vec{x}_2, \tfrac{\zR}{2} - z_0\big)
   + S_1\big(\vec{x}_2, \vec{x}_1, \tfrac{\zP}{2} - z_1\big).
\end{align}
We then look for the path that makes the action stationary.
As discussed in \sec{sec:new-theory}, we expect this to be
a four-bounce trajectory comprised of two identical loops, as
depicted in \fig{fig:k4a-instanton}(b).

\paragraph{High-temperature limit.}

As $\beta \to 0$, the action in \eqn{eq:s4a-defn} becomes stationary for trajectories
in the immediate vicinity of the minimum-energy crossing
point (MECP) between the diabats.
For the sake of convenience, we position it at the origin of our coordinate system. The compactness of the high-temperature instanton allows us to approximate the diabatic PESs as\cite{GoldenGreens}
\begin{equation}
\label{eq:ho-pes}
V_n(\vec{x}) \sim V^{\ddagger} + \kappa_n q
    + \frac{1}{2} \vec{Q}\trsp[0] \mat{H}_n \vec{Q}.
\end{equation}
The gradients $\vec{g}_n = \grad V_n(\vec{0})$
of the two diabatic surfaces
are antiparallel at the MECP, and we use $\kappa_n$ to denote their signed norms,
$\kappaR = \lVert \vec{g}_0 \rVert,\, \kappaP = -\lVert \vec{g}_1 \rVert$. We call the Cartesian coordinate aligned with
the gradients $q$, and the $f-1$ orthogonal coordinates $\vec{Q}$.
Each appears in the expansion to its lowest order. The results
in this section can equally be derived using a more general
expansion, keeping \emph{all} second-order terms, but such a derivation
is more laborious and has the same $\beta \to 0$ limit. In this limit,
the terms appearing in the semiclassical propagator are
\begin{subequations}
\label{eq:ht-propa}
\begin{align}
& C_n(\vec{x}'', \vec{x}', z) \sim \left(
\frac{m}{2 \pi \hbar z}
\right)^{\!f/2}, \\
& S_n(\vec{x}'', \vec{x}', z) \sim S_n^{\mathrm{lin}}(q'', q', z)
                  + S_n^{\mathrm{vib}}(\vec{Q}'', \vec{Q}', z), \\
&
\label{eq:sn-lin}
S_n^{\mathrm{lin}}(q'', q', z) = V^{\ddagger} z +
\frac{m q_{-}^2}{2 z} + z \kappa_n q_{+} - \frac{\kappa_n^2 z^3}{24 m}, \\
&
S_n^{\mathrm{vib}}(\vec{Q}'', \vec{Q}', z) \sim
 \frac{m \lVert \vec{Q}_{-} \rVert^2}{2 z}
+ \frac{z}{2} \vec{Q}_{+}\trsp \mat{H}_n \vec{Q}_{+}.
\end{align}
\end{subequations}
Here $q_{-} = q''-q'$ and $q_{+} = (q'+q'')/2$, with analogous definitions
for $\vec{Q}_{\pm}$. The stationary trajectory follows a path along $q$, with no lateral displacement (\latin{i.e.}, $\vec{Q} = \vec{0}$). Since the combined action is
quadratic in the positions, integration over these variables can be
performed analytically to yield $\cfa(z_0, z_1, z)$.
Choosing the real parts of its arguments such that $\tau = \tfrac{\beta \hbar \kappaR}{\kappaR-\kappaP}$ and $\tau_0 = \tau_1 = 0$ makes the three-time correlation
function proportional to $\delta(t)$.
By analogy with $\margfb$ in \eqn{eq:J-def}, we then define
\begin{subequations}
\begin{align}
\label{eq:L4A-qm}
&\margfa(t_0, t_1) = \int_{-\infty}^{\infty} \cfa(\iu t_0,
\iu t_1, \tau + \iu t) \, \rmd t \\
& {} \sim \frac{\Delta^4}{\hbar^4}
 \sqrt{\frac{2 \pi m}{\beta (\kappaR-\kappaP)^2}} \,
 Z^{\ddagger} \mkern1mu \eu{-\beta V^{\ddagger}} \exp\!\left[
 \frac{\beta^3 \hbar^2 \kappa_0^2 \kappa_1^2}{96 m (\kappa_0 - \kappa_1)^2}
 \right] \nonumber \\
\label{eq:L4A-ht}
& \qquad \qquad \qquad {} \times  \exp \! \left[
 -\frac{d_0 t_0^2 + d_1 t_1^2 + d_{01} t_0^2 t_1^2}{2 \hbar}
 \right]
\end{align}
\end{subequations}
as $\beta \to 0$. Here $\Delta$ is evaluated at the MECP, and we define
a transition-state partition function
\begin{gather}
Z^{\ddagger} = \bigg\lvert \frac{\beta^2 \hbar^2 \widetilde{\mat{H}}}{m} \bigg\rvert^{-1/2} \ \ \text{with} \qquad
\widetilde{\mat{H}} = \frac{\kappa_0 \mat{H}_1 - \kappa_1 \mat{H}_0}%
{\kappa_0 - \kappa_1},
\end{gather}
as well as the coefficients
\begin{equation}
\label{eq:dn-ht}
d_n = \beta \hbar \, \frac{\kappa_n^2}{4 m}, \qquad
d_{01} = \frac{1}{\beta \hbar} \frac{(\kappa_0 - \kappa_1)^2}{m}.
\end{equation}

\begin{figure}[ht]
\includegraphics{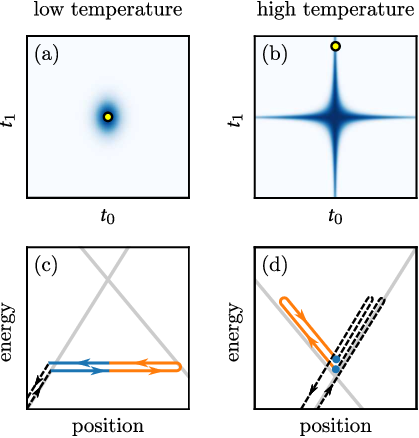}
\caption{\label{fig:L4A}
(a) Marginalised $\margfa(t_0, t_1)$ correlation function
for a system of linear diabats
with $\kappaR = 4,\, \kappaP=-3,\, m = 1$, $\hbar = 1$ and $\beta=5$.
(b) The same for $\beta = 0.1$. In both cases
the darkest blue corresponds to the maximum value,
white corresponds to zero, and
$-2 \leq t_0, t_1 \leq 2$. At low temperature the function
is approximately Gaussian and dominated by $t_0 = t_1 = 0$,
whereas at high temperature the function becomes star-shaped, developing rays
along the $t_0$ and $t_1$ axes.
In the top panels, one point of importance is picked out with a yellow circle and defines the parameters of the instanton shown in the lower panels.
(c) Shows the low-temperature half-instanton trajectory
associated with $t_0 = t_1 = 0$,
and (d) shows the high-temperature half-instanton for
a point on the ray along $t_1$.
Blue colour is used for the half-instanton segments that reside on $V_0$
and orange is used for $V_1$. The dashed black lines depict system dynamics
on $V_0$ before and after the scattering events associated with the half-instantons.}
\end{figure}

In general, the asymptotic relation in \eqn{eq:L4A-ht} only holds
at high temperatures and short times $t_0 \text{ and } t_1$. However
for one-dimensional linear diabats, $V_n(x) = V^{\ddagger} + \kappa_n x$, \eqn{eq:L4A-ht} is not an approximation and is
precisely equal to the quantum $\margfa(t_0, t_1)$.
It is therefore meaningful to make a digression into the low-temperature regime
($\beta \to \infty$). In this limit, the coefficient $d_{01}$ of the
quartic term in \eqn{eq:L4A-ht} becomes vanishingly small, and so the
marginalised correlation function $\margfa(t_0, t_1)$ becomes
well approximated by a Gaussian, as seen in \fig{fig:L4A}(a).
Neglecting the quartic term and integrating over
$t_n$ yields the low-temperature limit of $\kfreda$ for a linear system,
\begin{equation}
\label{eq:k4a-lt}
\kfredaltl = \sqrt{\frac{2 \pi m}{\beta (\kappaR - \kappaP)^2}} \,
\frac{4 \pi m \Delta^4 \mkern2mu \eu{-\beta V^{\ddagger}}}{ \beta \hbar^4
\lvert\kappaR \kappaP\rvert} \exp\!\left(\frac{\bred^3}{48}\right)\!,
\end{equation}
to be compared with the golden-rule rate\cite{GoldenGreens}
\begin{align}
\label{eq:k2-lin}
\kgrred^{\mathrm{lin}}  =
\sqrt{\frac{2 \pi m}{\beta (\kappaR - \kappaP)^2}}\,
\frac{\Delta^2 \mkern2mu \eu{-\beta V^{\ddagger}} }{\hbar^2} \, \exp\!\left(\frac{\bred^3}{12}\right),
\end{align}
where
\begin{equation}
\label{eq:bred}
\bred = 2 \left[\frac{d_0 d_1}{\hbar d_{01}}\right]^{\mathrlap{\!1/3}} \ = \beta \left[
   \frac{\hbar^2 \kappaR^2 \kappaP^2}{2 m (\kappaR-\kappaP)^2}
\right]^{\mathrlap{\!1/3}}.
\end{equation}
The similarity between \eqs{eq:k4a-lt}{eq:k2-lin} arises
because at low temperatures steepest-descent integration of $\cfa$
is a straightforward generalisation of the procedure
for the golden-rule correlation function $\cgr$.
One can therefore expect to see some similarities between the physical
interpretations of these two expressions.

In the golden-rule case, the value of the rate
constant $\kgrred$ is determined by the behaviour of
$\cgr(\tau + \iu t)$ in the small-$t$ limit.
Resolved in terms of position eigenstates,
the function at $t=0$ reads
\begin{equation}
\label{eq:gr-scatter}
c_2(\tau) = \iint_{-\infty}^{\infty} \!\!\rmd \vec{x}_\mathrm{i} \rmd \vec{x}_\mathrm{f}
 \left \lVert
\braket{\vec{x}_\mathrm{f} | \KP\! \left(\tfrac{\tauP}{2}\right)\! \tfrac{\DRP}{\hbar} \mkern1mu
\KR\!\left( \tfrac{\tauR}{2} \right)\! | \vec{x}_\mathrm{i} }
\right \rVert^2\!\!,
\end{equation}
which is dominated by $\vec{x}_\mathrm{i}$ and $\vec{x}_\mathrm{f}$ near
the GR instanton turning points [see \fig{fig:k4a-instanton}(c)]. Reading
the integrand from right to left, the dominant contribution
corresponds to the probability of the system starting near the reactant turning point, evolving in imaginary time for $\tauR/2 \equiv (\beta \hbar - \tau)/2$, ``switching'' from $V_0$ to $V_1$, propagating
for $\tauP/2 \equiv \tau/2$, and finally arriving at the product turning
point. Imaginary-time evolution indicates that the transition proceeds
via a tunnelling mechanism, with the dominant
tunnelling pathway following the half-instanton from
$\vec{x}_{\mathrm{i}}$ to $\vec{x}_{\mathrm{f}}$.
The dynamics before and after tunnelling
is not explicitly included in the half-instanton but may be deduced
from the locations of the turning points and the corresponding momenta
(which are zero). Given these boundary conditions we conclude that the
thermally activated system initially moves up $V_0$ with just enough energy to reach $\vec{x}_{\mathrm{i}}$,
at which point it tunnels to $\vec{x}_{\mathrm{f}}$ and continues moving
down $V_1$. This is precisely the kind of qualitative picture
we were looking for in order to describe the processes underlying
$\kgrred$, and it was possible to deduce from just the GR half-instanton.

A similar analysis can be conducted for $\kfreda$, whose value at low
temperatures is determined by $\cfa(z_0, z_1, z)$ around
$z_0 = z_1 = 0,$ $z = \tau$ [see \fig{fig:L4A}(a)].
Expanding $\cfa$ in terms of position eigenstates yields
\begin{align}
\label{eq:cfa-phi-res1}
& \cfa(0, 0, \tau) = \\
& \quad \iint_{-\infty}^{\infty} \!\!\rmd \vec{x}_{\mathrm{i}} \rmd \vec{x}_{\mathrm{f}}
 \left \lVert
\braket{\vec{x}_{\mathrm{f}} |  \KR\! \left(\tfrac{\tauR}{4}\right)\! \tfrac{\DRP}{\hbar} \mkern1mu
\KP\! \left(\tfrac{\tauP}{2}\right)\!
\tfrac{\DRP}{\hbar} \mkern1mu
\KR\!\left( \tfrac{\tauR}{4} \right)\! | \vec{x}_{\mathrm{i}} }
\right \rVert^2\!\!,
\nonumber
\end{align}
where the dominant contributions to the integral come from
$\vec{x}_{\mathrm{i}}$ and $\vec{x}_{\mathrm{f}}$ \emph{both} in the vicinity of the reactant turning
point. The integrand is once again in the form of a probability density
and describes a double scattering event in which the system
tunnels from the reactant state into the product state, then back to reactant.
This scattering process is associated with the half-instanton
derived from the double-loop trajectory in \fig{fig:k4a-instanton}(b),
and is shown schematically with solid lines in \fig{fig:L4A}(c). The real-time
dynamics immediately before and after the tunnelling event are indicated
on the same figure with dashed lines.

It is now clear why
$\kfreda$ decreases the full nonadiabatic rate (recall
that it appears in \eqn{eq:k4-therm} with a minus sign).
The golden-rule expression implicitly assumes that after every
reactant-to-product transition the system remains
in the product state. In reality there is a non-zero probability that
the system recrosses back to reactant, which the golden rule
entirely neglects, thus overestimating the full rate constant.
The $\kfreda$ term accounts for such recrossing events to leading order in $\Delta$,
correcting the overestimate.
According to our analysis, this correction is expected to
be small at low temperatures, since the underlying tunnelling mechanism
is relatively inefficient compared to the direct (reactive) transition.
For the linear case this can be deduced  from the exponents of
\eqs{eq:k4a-lt}{eq:k2-lin}, and we will see later that the same
applies to other potentials.

When considering the high-temperature ($\beta \to 0$) limit,
our approach has to be modified, since we can no longer neglect the quartic term in \eqn{eq:L4A-ht}. The change comes about because the coefficient
$d_{01}$ is proportional to $\beta^{-1}$, whereas the quadratic coefficients $d_n$ are linear in $\beta$ [\eqn{eq:dn-ht}]. This highlights
a curious feature of the $\cfa$ correlation function: unlike its
GR counterpart, it cannot be uniformly approximated as Gaussian in its arguments. Steepest-descent
integration over $t$ can always be done in the usual manner, giving us
$\margfa^{\SC}(t_0, t_1)$, but the remaining two variables show
some unexpected behaviour. At low temperatures, $\margfa^{\SC}$ has an (approximately) Gaussian shape that broadens with increasing temperature.
Eventually the broadening becomes noticeably anisotropic and largely
confined to the $t_0$ and $t_1$ axes. Along all other directions
the function shrinks rapidly, so that in
the $\beta \to 0$ limit it assumes the distinctly non-Gaussian four-pointed
star shape
in \fig{fig:L4A}(b). Hence we must modify the
steepest-descent procedure outlined in \sec{ssec:semiclassical} and take
the fourth-order term into account. This results in the asymptotic relation
\begin{align}
\label{eq:k4asc-ht}
\kfreda(\beta) & \sim \frac{\Delta^4}{\hbar^4}
 \sqrt{\frac{\pi m}{2 \beta (\kappaR-\kappaP)^2}} \,
 Z^{\ddagger} \mkern1mu \eu{-\beta V^{\ddagger}} \\
& \quad \quad {} \times
\sqrt{\frac{2 \pi \hbar}{d_{01}}} \mkern2mu
\exp\!\left(
 \frac{\bred^3}{48}
 \right) \mkern-2mu \rbessel{0}\!\mkern-1mu\left(
 \frac{\bred^3}{32}
 \right) \nonumber
\end{align}
as $\beta \to 0$, where $\rbessel{0}(y) \equiv \eu{y} \bessel{0}(y)$ and $\bessel{0}$ is a modified Bessel function of the second kind.\cite{Gradshteyn}
For one-dimensional linear diabats this relation
is exact and hence valid
at all temperatures.
From the
small-argument asymptotic behaviour of $\rbessel{0}$ it follows
that $\kfreda$ tends to
\begin{equation}
\label{eq:k4a-ht}
\kfredaht(\beta) =
\frac{ \pi m \Delta^4 Z^{\ddagger} \mkern1mu \eu{-\beta V^{\ddagger}}}{\hbar^3 (\kappa_0 - \kappa_1)^2} \!
 \left\{
\ln \! \left( \frac{64}{\bred^3}\right)  - \gamma
\right\}
\end{equation}
as $\beta \to 0$, where $\gamma = 0.577\ldots$ is the Euler--Mascheroni constant.
It is absolutely necessary to include the quartic $d_{01}$ term to arrive
at this result.

To understand the change in the asymptotics of $\kfreda$, we refer to
\fig{fig:L4A}(b), which shows that $\cfa(z_0, z_1, z)$ still has a stationary point
at $z_0 = z_1 = 0$ and $z=\tau$. The corresponding instanton
has the same shape as in the low-temperature regime, but is now contracted
almost to a point at the transition state, implying that
nuclear tunnelling does not play a major role.
From the same figure we see that significant contributions to $\kfreda$
also come from combinations of arguments where one of
$z_0$ or $z_1$ assumes a non-zero imaginary value.
Without loss of generality we focus on $z_0 = 0$ and $z_1 = \iu t_1$
(the alternative simply interchanges the roles of $V_0$ and $V_1$).
At this point the $\cfa$ correlation function can be written as
\begin{align}
\label{eq:cfa-phi-res2}
& \cfa(0, \iu t_1, \tau) = \\
& \iint_{-\infty}^{\infty} \!\!\rmd \vec{x}_{\mathrm{i}} \rmd \vec{x}_{\mathrm{f}}
 \left \lVert
\braket{\vec{x}_{\mathrm{f}} |  \KR\! \left(\tfrac{\tauR}{4}\right)\! \tfrac{\DRP}{\hbar} \mkern1mu
\KP\! \left(\tfrac{\tauP}{2} + \iu t_1\right)\!
\tfrac{\DRP}{\hbar} \mkern1mu
\KR\!\left( \tfrac{\tauR}{4} \right)\! | \vec{x}_{\mathrm{i}} }
\right \rVert^2\!\!,
\nonumber
\end{align}
closely resembling \eqn{eq:cfa-phi-res1}. As before, the integrand
describes the probability of an unreactive back-and-forth transition,
although now there is hardly any imaginary-time propagation, since
both $\tauR$ and $\tauP$ tend to 0 in the high-temperature limit.
The matrix element in \eqn{eq:cfa-phi-res2}
now corresponds to a system starting in the reactant state and almost
immediately getting scattered into the product state. The product then follows a real-time
trajectory of duration $t_1$, at the end of which it is scattered back into
the reactant state. The corresponding half-instanton
follows a path in the complex plane whose imaginary component
becomes vanishingly
small as $\beta \to 0$. The real part of the path is shown with solid
lines in \fig{fig:L4A}(d). On this occasion,
the momenta at the end-points of the half-instanton are non-zero,
from which it follows that before and after the
scattering event the system follows the trajectories indicated
with dashed lines.

The change in the asymptotics of $\kfreda$ is therefore a consequence of a  change in mechanism:
at low temperature the recrossing proceeds largely via
tunnelling, whereas at high temperature such transitions are mostly due to
high-energy ``over-the-barrier'' trajectories that overshoot
the hopping point at first and are only scattered on their way back.
We note that these trajectories (emerging
naturally from our theory) are precisely what motivates
the Holstein transmission coefficient\cite{Holstein1959,NikitinBook}
based on Landau--Zener theory,\cite{Landau1932LZ,Zener1932LZ}
which was previously used to construct uniform nonadiabatic rate expressions.\cite{Rips1995ET,Rips1996ET,UlstrupBook} It should then come
as no surprise that at high temperatures, rates based
on the Holstein transmission coefficient are consistent with
our \eqn{eq:k4a-ht}, as shown in \app{sec:high-t-limit}.
What our new theory brings to the table is
an accurate description of transitions at energies close to the MECP,
whereas LZ only applies in the high-energy limit. As a result,
the Holstein expression can only establish the dependence
in \eqn{eq:k4a-ht} up to an additive constant.
Furthermore, this expression only accounts for over-the-barrier transitions, hence it cannot be used to describe low-temperature
reaction rates, which are dominated by
tunnelling under the barrier. Our new approach can do so
rigorously, even for multidimensional nonlinear potentials, as shown below.

\paragraph{General case.}

The high-temperature regime requires additional care because
we need to account for the $\mathcal{O}(t_0^2 t_1^2)$ term in \eqn{eq:L4A-ht} in order to obtain the correct semiclassical limit.
This can be done with relative ease for a classical action in its short-time asymptotic form [\eqn{eq:ht-propa}].
However deriving the term without relying on the short-time/linear approximation
is not straightforward, since the derivation calls for high-order derivatives of
the action not normally available from standard instanton calculations.
At the same time, our previous discussion indicates
that such high-order terms are negligible at low temperatures,
implying that the standard steepest-descent prescription can be safely
followed in this regime. We therefore aim to derive a semiclassical
approximation to \eqn{eq:L4A-qm} correct to second order in $t_0$ and $t_1$,
expressing all the relevant coefficients in terms of second derivatives of the
classical action [\eqn{eq:action}]. The quartic term will be
based on an analytically tractable model, yielding a $d_{01}$ coefficient
that has the correct high-temperature limit. It will not necessarily be
accurate at low temperatures, but this should not be an issue provided
the coefficient vanishes sufficiently quickly as $\beta \to \infty$, since
in that regime $\kfredasc$ is to leading order
independent of $d_{01}$.

To begin, we locate the set of arguments $\cvec{x} = \big( \vec{x}_1,
\vec{x}_2, \vec{x}_3, \vec{x}_4 )$ and $\cvec{\tau} = \big( \tau_0, \tau_1,  \tau \big)$ that make the combined action in \eqn{eq:s4a-defn} stationary.
Generally this action will not be known in closed form, however it can be
calculated efficiently using numerical algorithms.\cite{GoldenRPI,InstReview,Ansari2021}
At the stationary point, the derivatives of the action satisfy
\begin{equation}
\label{eq:k4a-sd}
\pder[]{S_{\fa}}{\cvec{x}} = 0, \qquad \pder[]{S_{\fa}}{\gvec{\tau}} = 0.
\end{equation}
The trajectory that makes $S_{\fa}$ stationary at temperature~$\beta$
is related
to the trajectory that makes $S_2$ stationary at temperature $\beta/2$, as shown in
\fig{fig:k4a-instanton}. All the information required to calculate
$ \kfredasc $ can therefore be extracted from a GR instanton at twice the temperature
(that is $\beta/2$). For the remainder of this section all
quantities pertaining to the GR instanton, namely the hopping point
$\vec{x}^{\ddagger}$, the imaginary time $\tauGR$, the stationary action
$S_2$ and its Hessian $\mat{\Sigma_2}$ are quantities calculated at this higher temperature.

Explicitly, the conditions in \eqn{eq:k4a-sd} are satisfied by $\vec{x}_{1\ldots4} = \vec{x}^{\ddagger}$, $\tau_0 = \tau_1 = 0$ and $\tau = 2 \tauGR$.
It follows that the stationary action $S_{\fa} =
2 S_2$. To relate the derivatives of the two actions,
let us denote $S_0 \equiv S_0(\vec{x}'', \vec{x}', \tfrac{\beta \hbar}{2} - \tauGR)$ and
$S_1 \equiv S_1(\vec{x}'', \vec{x}', \tauGR)$. We then define
\begin{equation}
\label{eq:S4A-blocks}
\begin{gathered}
   \mat{\Theta} = \pders{S_0}{\vec{x}'}{\vec{x}'} + \pders{S_1}{\vec{x}''}{\vec{x}''},  \qquad
   \mat{\Theta}_n = \pders{S_n}{\vec{x}'}{\vec{x}''}, \\
   \vec{w}_n = \pders{S_n}{\vec{x}'}{\tauGR} = \pders{S_n}{\vec{x}''}{\tauGR}, \qquad
   \zeta_n = \pder{^2 S_n}{\tauGR^2},
\end{gathered}
\end{equation}
where $\mat{\Theta}, \, \mat{\Theta}_n$ are symmetric $f \by f$ matrices,
$\vec{w}_n$ are $f$-dimensional column vectors, $\zeta_n$ are scalars, and
the derivatives are all evaluated at the GR stationary point. Thus
all of the above quantities are directly available from a GR instanton
calculation and can be evaluated numerically within the ring-polymer formulation as described in \Refx{GoldenRPI}.

Given this information one can derive the general steepest-descent approximation to \eqn{eq:L4A-qm} by following the
procedure in \app{sec:dn}. Defining
\begin{equation}
\label{eq:theta-pm}
\mat{\Theta}_{\pm} = \mat{\Theta} \pm (\mat{\Theta}_0 - \mat{\Theta}_1),
\end{equation}
we find that the coefficients $d_n$ in
\eqn{eq:L4A-ht} are given by
\begin{subequations}
\label{eq:d0d1-defn}
\begin{align}
d_0 & = 4 \vec{w}_0\trsp \mat{\Theta}_{+}^{-1} \vec{w}_0 - 2 \zeta_0, \\
d_1 & = 4 \vec{w}_1\trsp \mat{\Theta}_{-}^{-1} \vec{w}_1 - 2 \zeta_1.
\end{align}
\end{subequations}
An analogous expression for $d_{01}$ would require high-order derivatives
of $S_2$ that are not readily available from standard GR instanton calculations.
We can circumvent this by noting that, based on \eqn{eq:dn-ht} and on general
physical considerations, the ratio $\hbar d_{01}/ d_0 d_1 $ tends to zero
as $\beta \to \infty$. In this limit
the factor arising from integrating over $t_0$ and $t_1$ behaves as
\begin{equation}
\sqrt{\frac{2 \pi \hbar}{d_{01}}} \ \widetilde{K}_0\!\left(
\frac{d_0 d_1}{4 d_{01} \hbar}
\right) \sim \frac{2 \pi \hbar}{\sqrt{d_0 d_1}} \left(
1 - \frac{\hbar}{2} \frac{ d_{01}}{d_0 d_1}
\right)\!,
\end{equation}
\latin{i.e.}, to leading order it is independent of $d_{01}$. Therefore any reasonable approximation
to $d_{01}$ should lead to a good estimate of $\kfredasc$, provided
the expression satisfies $\hbar d_{01} / d_0 d_1 \ll 1$ as $\beta \to
\infty$ and tends to \eqn{eq:dn-ht} in the high-temperature limit.
We suggest the following expression, which is exact for a one-dimensional
spin--boson model [\eqn{eq:sb-pes} with $f=1$]:
\begin{align}
\label{eq:d01-sb}
& d_{01} = \mathrm{max} \bigg[0, \, \tfrac{\omega}{2 m} \bigg \{
 (\kappaR-\kappaP)^2 \csch\! \left(
\tfrac{\beta \hbar \omega}{2}
\right) \\
& \quad {} +  \kappaR \kappaP
 \sech\!\left(
  \tfrac{\beta \hbar \omega}{4}
  \right)^{\!2} \left[
   \left(\tfrac{\kappaR+\kappaP}{\kappaR-\kappaP}\right)^{\mathrlap{\!2}} +
   \csch\!\left(
     \tfrac{\beta \hbar \omega}{4}
     \right)^{\!2}
\right]^{-\frac{1}{2}} \!
\bigg\}\bigg].  \nonumber
\end{align}
Here $\kappa_n$ are the signed norms of $\grad V_n$ at the MECP and $\omega > 0$
is a parameter ensuring that $d_{01}$ vanishes sufficiently quickly as $\beta \to \infty$. A suitable value can be obtained from the curvature along the
direction of steepest descent at the MECP, $\omega^2 = m^{-1} \grad^2_q \, V_n(\vec{x}^{\ddagger})$, calculated for whichever diabatic PES yields
the larger value. Unlike \eqn{eq:dn-ht}, this expression for $d_{01}$
can become negative at sufficiently low temperatures, tending
to zero from below as $\beta \to \infty$. In such cases the most straightforward course of action is to set $d_{01} = 0$, which is equivalent to
neglecting contributions from the quartic term.\footnote{%
This is also the more consistent treatment, since at temperatures low enough that $d_{01} < 0$, the corresponding contribution to the steepest-descent integral is small and no longer dominates the other quartic terms ($t^4$ and $t_n^4$), which are neglected. There is therefore no reason to give the
$d_{01}$ contribution privileged treatment and include it
into the final result.}
More sophisticated approximations to $d_{01}$
could be constructed, but
any modification is expected to only have a significant effect at temperatures where $\kfredasc$ is subdominant to the other components,
namely $\kfredbsc$, as will be shown later.

Combining \eqss{eq:d0d1-defn}{eq:d01-sb}{eq:k4a-mats},
we arrive at the final expression for the rate constant,
\begin{equation}
\label{eq:k4asc}
\begin{aligned}
\hspace*{-1ex}\kfredasc(\beta) = \left[
       \kgrredsc \mkern-1mu\big(\tfrac{\beta}{2}\big)
   \right]^2 \!\! \sqrt{
      \frac{\mathllap{-}\lvert \mat{\Sigma}_{2} \rvert}{
      2 d_{01}
      \lvert \mat{\Theta}_{+} \rvert
      \lvert \mat{\Theta}_{-} \rvert}
      } \,
      \rbessel{0} \! \left(
          \frac{d_0 d_1}{4 \mkern1mu d_{01} \hbar}
      \right)\!,
\end{aligned}
\end{equation}
where $\kgrredsc$ is defined in \eqn{eq:kgrred-sc}. The expression
is a uniform approximation that is expected to be valid at any temperature.
Additionally, the $\omega \to 0$ limit of \eqn{eq:d01-sb} is precisely equal to \eqn{eq:dn-ht}, and so our current prescription for $d_{01}$
exactly recovers the quantum $\kfreda$ for two-state systems of one-dimensional linear diabats.

\subsection{\label{ssec:kb}%
Semiclassical \emph{B}-type contribution}

\begin{figure}[ht]
\includegraphics{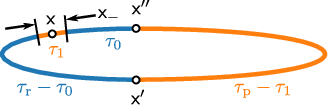}
\caption{
\label{fig:k4b-blip}%
An instanton trajectory that contributes significantly
to the $\kfredbsc$ rate [\eqn{eq:kfb-ht}]. Here $\tauR = \beta \hbar - \tauGR$, $\tauP = \tauGR$
and $\tauGR$ is the stationary imaginary time of a GR instanton
at temperature $\beta$.
}
\end{figure}

The considerations in \sec{ssec:exact-kb} allow us to write
\begin{align}
\label{eq:k4b-sc}
\kfredb \sim  \kfredasc  + 2 \Real & \Bigg\{
\bigg[
    \iint_{\mathcal{A}_{\mathrlap{\tau \tau}}} +
    \iint_{\mathcal{A}_{\mathrlap{\tau t}}} + {} \\
& \quad
    \iint_{\mathcal{A}_{\mathrlap{t \tau}}} +
    \iint_{\mathcal{A}_{t t}} \bigg]
    \margfb^{\SC}(z_0, z_1) \, \rmd z_0 \rmd z_1
    \Bigg\}, \nonumber
\end{align}
where $\margfb^{\SC}$ is obtained by substituting semiclassical propagators
into \eqn{eq:J-def} and integrating over positions and time $t$ by steepest descent. As before, we identify the combined action,
\begin{align}
\label{eq:s4b-defn}
 S_{\fb} & \equiv S_0\big(\vec{x}_1, \vec{x}_4, \zR - z_0\big)
   + S_1\big(\vec{x}_4, \vec{x}_3, z_1\big) \nonumber \\
    & {} + S_0\big(\vec{x}_3, \vec{x}_2, z_0\big)
   + S_1\big(\vec{x}_2, \vec{x}_1, \zP - z_1\big),
\end{align}
which we expect to be minimised by a two-bounce (single-loop)
trajectory like the one shown in \fig{fig:k4b-blip}.
To prove this we need to determine
the optimal integration domains (\latin{i.e.}, the optimal $\tau_n^{*}$)
which is most easily done in the high-temperature limit, as we now show.

\paragraph{High-temperature limit.}

Once again, we expand the diabatic PESs in a Taylor series about the MECP and consider the corresponding propagators defined by \eqn{eq:ht-propa}.
Integration over position and time $t$ gives
\begin{equation}
\label{eq:JSC-ht}
\begin{gathered}
\margfb^{\SC}(z_0, z_1) \sim
\frac{\Delta^4}{\hbar^4}
\sqrt{\frac{2 \pi m}{\beta (\kappaR-\kappaP)^2}}
Z^{\ddagger} \mkern2mu
\eu{-\beta V^{\ddagger} + \bred^3\!/12}
 \\
 {} \times \exp\bigg\{
 {- \frac{1}{\hbar}} \bigg[
\frac{(\kappa_0 - \kappa_1)^2}{2 m \beta \hbar} z_0^2 z_1^2
 - \frac{\beta \hbar \mkern1mu \kappa_0 \kappa_1}{2 m} z_0 z_1 \\
 {} -  \frac{\kappa_0 - \kappa_1}{2 m} (\kappa_0 z_0^2 z_1 - \kappa_1 z_1^2 z_0)
\bigg]
\bigg\},
\end{gathered}
\end{equation}
where all the terms are defined as in \sec{ssec:ka}.
From here it can be shown that the remaining integrals
over $z_n$ in \eqn{eq:k4b-sc} can be performed
most easily by setting
\begin{equation}
\label{eq:tau-opt}
\tau_0^{*} = \frac{\beta \hbar}{2} \frac{\kappaP}{\kappaP - \kappaR},
\qquad
\tau_1^{*} = \frac{\beta \hbar}{2} \frac{\kappaR}{\kappaR - \kappaP}.
\end{equation}
This choice removes any oscillations in the integrand and allows for
further simplifications, which will be discussed shortly.
Note that these values of $\tau_n^{*}$ are related to the stationary imaginary time
of a golden-rule instanton at temperature $\beta$ in a linear system or
the high-temperature limit,
namely $\tau_1^{*} = \tauGR/2$ and
$\tau_0^{*} = (\beta \hbar - \tauGR)/2$.
Under this definition, contributions from $\mathcal{A}_{\tau t}$
and $\mathcal{A}_{t \tau}$ are purely imaginary and are therefore discarded.
Furthermore, the contribution from $\mathcal{A}_{t t}$ cancels exactly with
$\kfredasc$ in \eqn{eq:k4b-sc}, meaning that the semiclassical rate
is
\begin{equation}
\label{eq:kfb-ht}
\kfredbsc \sim
 2
 \int_{0}^{\tau_0^{*}} \!\!\!
 \int_{0}^{\tau_1^{*}} \margfb^{\SC}(\tau_0, \tau_1)
 \, \rmd \tau_0 \rmd \tau_1
\end{equation}
as $\beta \to 0$. We observe that similarly to $\kfreda$,
the $\kfredb$ rate cannot be approximated with a Gaussian integral.
Also like with $\kfreda$, the high-temperature
asymptotic relation given for $\kfredb$ by \eqt{eq:JSC-ht}{eq:kfb-ht}
is exact across all temperatures for the linear system. Hence we may
use the results of this section to inspire the derivation in the general case.

\begin{figure}[ht]%
\includegraphics{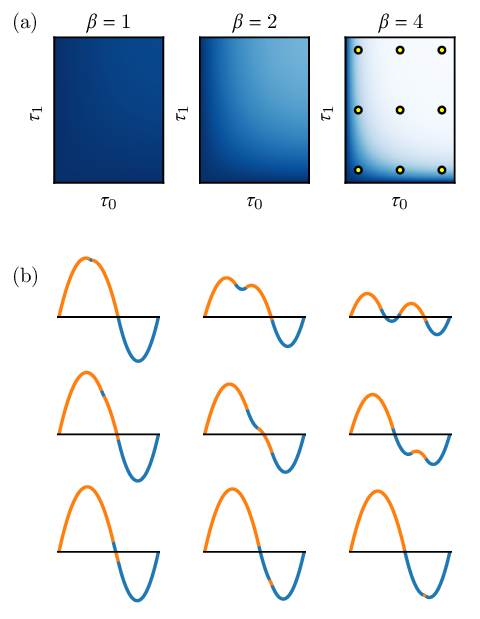}
\caption{\label{fig:k4b-t0t1}%
Panel~(a) shows the marginalised $\margfb(\tau_0, \tau_1)$
correlation function for a system of linear diabats
with $\kappaR = 4$, $\kappaP = -3$, $m=1$, $\hbar=1$.
The plotting ranges are $\tau_n \in [0, \tau_n^{*}]$ as defined in
\eqn{eq:tau-opt}. The darkest blue corresponds to the maximum value,
and white corresponds to $0$.
For $\beta = 4$ we highlight a representative set of $(\tau_0, \tau_1)$ values in yellow and plot
the corresponding minimal action instanton trajectories in panel~(b). The $x$-axis marks the imaginary time along the instanton, and the $y$-axis shows the position relative to the hopping point $\vec{x}^{\ddagger}$. Blue and orange trajectory segments
reside on reactant and product diabats respectively.
}
\end{figure}

In \fig{fig:k4b-t0t1}(a) we plot $\margfb$ at different
temperatures for $(\tau_0, \tau_1) \in \mathcal{A}_{\tau \tau}$. As $\beta \to 0$, the function becomes flat across the
entire domain of integration; as $\beta \to \infty$, the function becomes
dominated by the edges for which $\tau_0$ or $\tau_1$ is $0$.
In \fig{fig:k4b-t0t1}(b) we show the instanton trajectories that correspond to
different points inside the domain of integration. As anticipated
at the beginning of \sec{sec:new-theory}, trajectories that dominate the integral at low temperatures have two consecutive segments that bounce once, and
two consecutive segments that travel directly between the endpoints.
Overall, these follow the path of a golden-rule
instanton, with either a short reactant segment along a product trajectory or vice versa. As we move to trajectories for
which neither $\tau_n$ is small, the segments become elongated and the
entire instanton more localised around the barrier region.
In the most extreme case (top right trajectory), we arrive at
the four-bounce instanton that corresponds to the $\kfredasc$ term
in \sec{ssec:ka}. For large $\beta$ these trajectories have considerably
larger action and do not contribute much to the integral in \eqn{eq:kfb-ht}. Hence we expect that the general semiclassical expression for $\kfredb$ will be dominated by trajectories for which at least one of the direct segments is vanishingly short
($\tau_0 \to 0$ or $\tau_1 \to 0$). All such trajectories reduce exactly to a standard (two-bounce) GR instanton corresponding to temperature~$\beta$.

\paragraph{General case}

In order to derive a general expression for $\kfredbsc$ we need to express the
coefficients in
\begin{align}
\label{eq:margfbsc}
\hspace*{-1ex}
\margfb^{\SC}(\tau_0, \tau_1) \sim \margfb^{\SC}(\tau_0, 0) \exp\!\left\{ \!
{-\frac{1}{\hbar}} \! \left[
\alpha_1 \tau_1 - \frac{ \gamma_1 \tau_1^2}{2}
\right]\!\right\}
\end{align}
(valid as $\tau_1 \to 0$) in terms of derivatives of the GR action. From \eqs{eq:k4b-exact}{eq:J-def} it follows that
\begin{equation}
\label{eq:margfb-tnull}
\margfb^{\SC}(\tau_0, 0) = \kgrredsc(\beta)
\frac{\Delta[\vec{x}(\tau_0)]^2}{\hbar^2},
\end{equation}
where $\Delta[\vec{x}(\tau_0)]$ is the diabatic coupling along
the reactant segment of a GR instanton after travelling for an
imaginary time $\tau_0$ away from the hopping point. Deriving
expressions for $\alpha_1$ and $\gamma_1$ (both functions
of $\tau_0$) is somewhat laborious, but the underlying approach is
similar to the one used in \sec{ssec:ka} to derive $d_n$.
Treating $\tau_0$ as a parameter, we consider a series expansion of $S_{\fb}$ in the
remaining six variables and integrate over $\left(t,\,\vec{x}_1,\,\vec{x}_2,\,
\vec{x}_3,\,\vec{x}_4\right)$ by steepest descent. The resulting effective
action is in the form of the bracketed expression in
\eqn{eq:margfbsc}, and so yields the desired coefficients.
To begin, we transform to a more convenient set of coordinates,
\begin{equation}
\label{eq:k4b-diffx}
\begin{aligned}
\vec{x}' & \equiv \vec{x}_1, & \quad
\vec{x}'' & \equiv \vec{x}_2, \\
\vec{x} & \equiv (\vec{x}_3+\vec{x}_4)/2, & \quad
\vec{x}_{-} & \equiv \vec{x}_4-\vec{x}_3,
\end{aligned}
\end{equation}
as indicated in \fig{fig:k4b-blip}. We then define
\begin{subequations}
\label{eq:s4b-parts}
\begin{align}
S_0' & \equiv S_0(\vec{x}\mathrlap{'}, \ \vec{x},\ \beta \hbar - \tau - \tau_0), \\
S_0''& \equiv S_0(\vec{x}, \ \vec{x}\mathrlap{''},\ \tau_0), \\
S_1  &\equiv S_1(\vec{x}\mathrlap{''}, \ \vec{x}\mathrlap{'}, \ \tau),
\end{align}
\end{subequations}
and  introduce the notation $\cvec{v} \equiv \left(\tau \ \vec{x}' \ \vec{x} \ \, \vec{x}''  \right)$. Treating $\tau_0$ as a parameter,
we expand $S_{\fb}$ to second order about $\tau_1 = 0$, $\vec{x}_{-} = \vec{0}$,
and a $\cvec{v}$ such that
\begin{equation}
\label{eq:k4b-stat}
\pder{}{\cvec{v}} \left(
S_0' + S_0'' + S_1
\right) = \cvec{0}.
\end{equation}

We then integrate (the exponential of) the resulting expression over all variables
except for $\tau_0$ and $\tau_1$, as discussed in \app{sec:k4b-ag}. This leads to the
desired Taylor series expansion of the effective action, with coefficients
\begin{subequations}
\label{eq:k4b-ag}
\begin{align}
\rule[-0.75em]{0pt}{1em}%
\alpha_1 & = V_1(\vec{x}) - V_0(\vec{x}), \\
\gamma_1 & = \pder{(V_1-V_0)}{\vec{x}} \cdot \left[
\mat{\Phi}_{0}^{-1}
\right]_{\vec{xx}} \! \cdot \pder{(V_1-V_0)}{\vec{x}} \\
{} & + \pder{(V_1-V_0)}{\vec{x}} \cdot \left[
\mat{\Phi}_{0}^{-1}
\right]_{\vec{x}\cvec{v}} \! \cdot \pders{(S_0' + S_0'')}{\cvec{v}}{\tau_0},
\nonumber
\end{align}
\end{subequations}
where $\vec{x}$ is the coordinate along the reactant segment of the GR instanton at time $\tau_0$ and
\begin{equation}
\label{eq:k4b-Phi0}
\mat{\Phi}_{0} = \pders{(S_0' + S_0'' + S_1)}{\cvec{v}}{\cvec{v}}.
\end{equation}
We use $\left[
\mat{\Phi}_{0}^{-1}
\right]_{\vec{x}\cvec{v}}$ to denote the submatrix formed from $\mat{\Phi}_0^{-1}$
by taking the rows corresponding to
$\vec{x}$ and all of the columns; $\left[
\mat{\Phi}_{0}^{-1}
\right]_{\vec{x}\vec{x}}$ is formed from the rows \emph{and} columns corresponding to $\vec{x}$. The same considerations apply to the asymptotic form of $\margfb^{\SC}(\tau_0, \tau_1)$ as $\tau_0 \to 0$, and expressions for $\alpha_0$ and $\gamma_0$ can
be obtained by swapping the indices $0 \leftrightarrow 1$ in the preceding derivation.

It remains to evaluate the double integral in \eqn{eq:kfb-ht} over the
rectangular domain $\mathcal{A}_{\tau \tau}$, defined by
\mbox{$\tau^{*}_1 = \tfrac{1}{2} \tauGR(\beta)$},
\mbox{$\tau^{*}_0 = \tfrac{1}{2}[\beta \hbar - \tauGR(\beta)]$}.
To do this, we split the domain along the diagonal connecting the bottom left
and top right corners of the rectangles in \fig{fig:k4b-t0t1}(a),
\begin{align}
\label{eq:t0t1-intsplit}
&\int_0^{\tau^{*}_0} \rmd \tau_0
\int_0^{\tau^{*}_1} \rmd \tau_1 = {} \\
&\qquad \qquad
\int_0^{\tau^{*}_0} \rmd \tau_0
\int_0^{\frac{\tau^{*}_1}{\tau^{*}_0}\mathrlap{\tau_0}} \rmd \tau_1
+
\int_0^{\tau^{*}_1} \rmd \tau_1
\int_0^{\frac{\tau^{*}_0}{\tau^{*}_1}\mathrlap{\tau_1}} \rmd \tau_0.%
\nonumber
\end{align}
We shall focus on the first of the two terms; the result for the second
term follows upon exchanging the labels $0 \leftrightarrow 1$.
The outer integral over $\tau_0$ is not amenable to
steepest-descent integration but can readily be evaluated numerically
using information directly available from standard ring-polymer
instanton optimisation.
The inner integral is of the form
\begin{equation}
\label{eq:j0}
j_0^{\SC}(\tau_0) = \int_0^{\frac{\tau^{*}_1}{\tau^{*}_0}\mathrlap{\tau_0}}
\eu{
{-\frac{1}{\hbar}} \! \left[
\alpha_{1}(\tau_0) \tau_1 - \tfrac{1}{2} \gamma_1(\tau_0)  \tau_1^2
\right]
} \ud \tau_1,
\end{equation}
where $\alpha_1$ and $\gamma_1$ are defined in \eqn{eq:k4b-ag}.
The standard procedure (\onlinecite[see Chapter~6.4 of Ref.~][]{BenderBook}) for calculating this integral by steepest descent
is to replace the integrand with
\begin{equation}
\label{eq:j0-int}
\exp \! \left[
-\frac{\alpha_{1} \tau_1}{\hbar} + \frac{\gamma_1 \tau_1^2}{2 \hbar}
\right] \sim \eu{-\alpha_{1} \tau_1 / \hbar} \left[
1 + \frac{\gamma_1 \tau_1^2}{2 \hbar}
\right]
\end{equation}
and integrate the resulting expression analytically. This
yields
\begin{equation}
\label{eq:k4b-taylor}
\begin{aligned}
j_0^{\SC}(\tau_0) & = \frac{\hbar}{\alpha_1}
\left[
1 - \eu{-\phi_1}
\right] \\
& \quad {} + \frac{\hbar^2 \gamma_1}{\alpha^3_1}
\left[
1 - \eu{-\phi_1} \! \left(
\tfrac{\phi_1^2}{2}  + \phi_1 + 1
\right)
\right],
\end{aligned}
\end{equation}
with $\phi_1 = \tfrac{\alpha_1}{\hbar} \mkern-1mu \tfrac{\tau_1^{*}}{\tau_0^{*}} \tau_0$, and hence the semiclassical rate
\begin{equation}
\label{eq:k4b-sc-uni}
\begin{aligned}
 \kfredbsc(\beta) = \frac{2 \kgrredsc(\beta)}{\hbar^2} \bigg \{
& \int_0^{\tau^{*}_0}\!\! \Delta[\vec{x}_{\mathrm{r}}(\tau_0)]^2 \,
j_0^{\SC}(\tau_0) \, \rmd \tau_0 \\
{} + & \int_0^{\tau^{*}_1}\!\! \Delta[\vec{x}_{\mathrm{p}}(\tau_1)]^2
\, j_1^{\SC}(\tau_1) \, \rmd \tau_1
\bigg\},
\end{aligned}
\end{equation}
where $\Delta[\vec{x}_{\mathrm{r}}(\tau_0)]$ is the diabatic coupling
at time $\tau_0$ along the reactant segment of the GR instanton, $\vec{x}_{\mathrm{r}}(0)$ is the hopping point, $\vec{x}_{\mathrm{r}}(\tau_0^{*})$ is the turning point, and
analogous definitions hold for $\vec{x}_{\mathrm{p}}$.
Despite the lengthy derivation, \eqn{eq:k4b-sc-uni} is easy to evaluate
in practice. All one needs are the potentials $V_n$,
their gradient $\grad V_n$ and the diabatic coupling
$\Delta$ along the GR instanton trajectory at temperature $\beta$,
together with the derivatives of the stationary action directly
available from a standard GR calculation. No reoptimisation is required,
and the additional computational overhead is minimal.

There are three points remaining to be addressed. First, \eqn{eq:margfbsc}
with $\alpha_n,\,\gamma_n$ defined as in \eqn{eq:k4b-ag}
exactly recovers the quantum $\margfb$ (at all $\tau_n$)
for linear diabats $V_n = V^{\ddagger} + \kappa_n x$.
Therefore for that system we need not perform the expansion
in \eqn{eq:j0-int} and can instead evaluate \eqn{eq:j0}
directly, as described in \app{sec:kfb-lin}. This leads to an
analytic expression for the quantum $\kfredb(\beta)$ for a system of
linear diabats, which we
can compare with the asymptotic approximation in \eqn{eq:k4b-sc-uni}.
It can be shown that for any combination of parameters the latter
tends to the quantum result in both the high- and low-temperature limits,
and is never in error by more than~$2.8\%$.

Second, we should stress that the rectangular domains in \fig{fig:k4b-t0t1},
which we have adopted throughout the derivation, are only strictly optimal in the
high-temperature limit (unless the system is linear, in which case the domains are
optimal at any temperature). It is also only in those limits that
integrals over $\mathcal{A}_{\tau t}$ and $\mathcal{A}_{t \tau}$ are purely
imaginary, and therefore do not contribute,
and only here that the integral over $\mathcal{A}_{t t}$ cancels exactly with $\kfredasc$. Nevertheless, any significant
deviations are expected to occur at low temperatures, for which the integral is
entirely dominated by regions of small $\tau_0$ or $\tau_1$.
Deformation of the optimal $\mathcal{A}_{\tau \tau}$ from the
rectangular shape would only become prominent in parts of the $(\tau_0, \tau_1)$
plane that contribute little to the overall integral.
Furthermore, we expect terms derived from integration over optimal
$\mathcal{A}_{\tau t}$, $\mathcal{A}_{t \tau}$ and $\mathcal{A}_{t t}$,
as well as the rate constant $\kfredasc$, to become subdominant to
$\kfredbsc$ in the low-temperature limit. Therefore any real part
acquired by the first two terms, or any imperfect cancellation between
the latter are expected to be negligible, and thus \eqn{eq:k4b-sc-uni}
remains the correct asymptotic limit of the $\kfredb$ rate constant.

Finally, we come to the question of physical interpretation. Mirroring the discussion
in \sec{ssec:ka}, we consider $\cfb(z_0, z_1, z)$ at a point
that makes a significant contribution to $\kfredb$, expressing the
function in the position basis.
In what follows we consider the contribution that
corresponds to the instanton in \fig{fig:k4b-blip}, for which
$z_0 = \tau_0 \in (0, \tfrac{\tauR}{2})$, $z_1 \to 0$ and $z = \tau$.
This can be written as
\begin{align}
\label{eq:cfb-phi-res}
& \cfb(\tau_0, 0, \tau) =
\iint_{-\infty}^{\infty} \!\!\rmd \vec{x}_{\mathrm{i}} \rmd \vec{x}_{\mathrm{f}}
\braket{\vec{x}_{\mathrm{i}}|\KR\! \left(\tfrac{\tauR}{2}\right)\!
\tfrac{\DRP}{\hbar} \mkern1mu
\KP\! \left(\tfrac{\tauP}{2}\right)\!  | \vec{x}_{\mathrm{f}} } \times {} \nonumber \\
&
\braket{ \vec{x}_{\mathrm{f}} | \KP\! \left(\tfrac{\tauP}{2}\right)\! \tfrac{\DRP}{\hbar} \mkern1mu
\KR(\tau_0)
\tfrac{\DRP}{\hbar} \mkern1mu
\KP(0)
\tfrac{\DRP}{\hbar} \mkern1mu
\KR\!\left( \tfrac{\tauR}{2} - \tau_0 \right) \! | \vec{x}_{\mathrm{i}} }.
\end{align}
Unlike in \eqs{eq:cfa-phi-res1}{eq:cfa-phi-res2}, here the integrand is
not the square of a single matrix element and hence
cannot be directly identified with the probability
of a scattering event. Instead, the integrand takes the form of a quantum
interference term. The first of the two interfering alternatives is
described by the matrix element on the first line of \eqn{eq:cfb-phi-res}
and corresponds to a golden-rule transition, as follows from
comparing to \eqn{eq:gr-scatter}. The second alternative, corresponding to the
second line of \eqn{eq:cfb-phi-res}, is a new type of scattering process that involves an excitation into a \emph{virtual} state.

\begin{figure}[ht]
\includegraphics{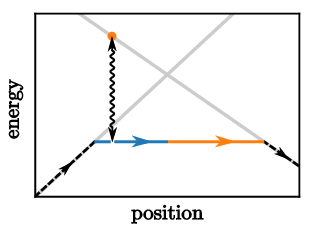}
\caption{
\label{fig:L4B-traj}%
Schematic depiction of the half-instanton that corresponds
to $\cfb(\tau_0, 0, \tau)$ in \eqn{eq:cfb-phi-res}.
The half-instanton segments residing on $V_0$ are shown in blue
and those on $V_1$ are shown in orange.
The wavy line indicates an excitation into a ``virtual'' state,
which is infinitesimally short and not energy-conserving.
The dashed black lines depict the system dynamics before and after the scattering event associated with the half-instanton.
}
\end{figure}

The half-instanton trajectory corresponding to this process is shown in \fig{fig:L4B-traj} and is identical to the GR case, except for
the infinitesimally short excitation onto~$V_1$. We call this excitation
virtual because, firstly, the amount of time spent on $V_1$ is vanishingly small
and, secondly, the excitation does not conserve the energy along the half-instanton. As $\tau_0$ is varied (\latin{e.g.}, in the first integral in
\eqn{eq:k4b-sc-uni}), the excitation moves from the turning point
to the hopping point, ``sampling'' $V_1$ all along the
reactant segment of the half-instanton. When $\tau_0 \to 0$ and
$\tau_1 \in (0, \tfrac{\tauP}{2})$, as in the second integral in
\eqn{eq:k4b-sc-uni}, the excitation moves along the product segment and
samples $V_0$ instead.

It seems reasonable that this type of scattering process should describe how the \emph{upper} diabatic state influences the probability of
tunnelling from reactant to product. Casting the system in the adiabatic representation,
it becomes evident that the shape of the lower
Born--Oppenheimer PES depends on $\Delta$ and \emph{both} of the diabats $V_n$
all along the tunnelling pathway (not just at the hopping point). $\kfredb$ takes
this into account and serves to increase the overall rate. This is consistent
with the notion that a larger $\Delta$ will tend to lower the (adiabatic)
activation energy barrier, increasing the probability of tunnelling and hence speeding up
the reaction.

\subsection{Semiclassical partition-function correction%
\label{ssec:sc-partf}}

It is often the case that the diabatic coupling $\Delta(\vec{x})$ becomes negligible
as $\vec{x}$ approaches the minimum of $V_n$. The term
considered in this section is then itself negligible and can be omitted altogether. If the coupling remains considerable around the
reactant/product minimum, it can be accounted for as follows.
We begin by rewriting \eqn{eq:zrred2} as
\begin{align}
\label{eq:z2red-ints}
\ZRredn{2} & = \int_{0}^{\frac{\beta \hbar}{2}} \! \!
                                [\beta \hbar - 2 u] \, c_2(u) \, \rmd u  \\
                     {} & - \bigg[
                     \int_{0}^{\frac{\beta \hbar}{2}} \! \!
                                 u \, c_2(u) \, \rmd u +
                     \int_{0}^{\frac{\beta \hbar}{2}} \! \!
                                 u \, c_2(\beta \hbar - u) \, \rmd u
                     \bigg] \frac{\ZRn{0}}{\ZPn{0}}. \nonumber
\end{align}
For imaginary times $u \in [0, \beta \hbar]$, the only stationary point of
$c_2(u)$ is a minimum at $u = \tauGR(\beta)$, which corresponds to the GR
instanton. In the absence of maxima, the asymptotic expansion of \eqn{eq:z2red-ints} is determined by how $c_2(u)$ behaves near the edges of the integration domains, namely
$u=0$ and $u=\beta \hbar$. These correspond to instantons collapsed at the bottom of the reactant and the product well respectively.

\begin{figure}[ht]
\includegraphics{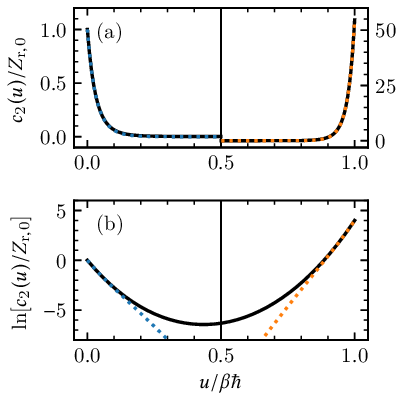}
\caption{%
\label{fig:z2red}%
(a)~The GR correlation function for a 1D spin--boson model
[\eqn{eq:sb-pes}] at $\beta = 2$, $\epsilon=2$, $d=3$ and all other
parameters equal to $1$. The function is normalised by $\ZRn{0}$. Note the change of scale either side of the vertical
line at $u = \beta \hbar/2$. The blue dotted line
overlaying the black curve is the asymptotic approximation as
$u \to 0$ [\eqn{eq:c2-asympt}]. The orange dotted line
is the analogous approximation as $u \to \beta \hbar$.
(b)~The logarithms of these functions. Note
the minimum at $\tauGR/\beta \hbar \approx 0.435$.
}
\end{figure}

In principle we should also consider the integrand around $u = \beta \hbar/2$,
but the corresponding contribution is small and can usually be neglected. It only becomes important
if the system is in the near-activationless regime
or the activation energy is appreciable but the system is asymmetric and the temperature is extremely low. In both these cases the value of $\cgr(\beta \hbar /2)$ can
become comparable to (or exceeds) one of $\cgr(0)$ or $\cgr(\beta \hbar)$. Other than in such cases,
our approximation holds because the instanton trajectory
associated with $\cgr(\beta \hbar/2)$ is localised near the barrier region.
It will thus resemble a GR instanton and give rise to a relatively large stationary action, making its contribution negligible compared
to those from the reactant and product wells.
This is shown in \fig{fig:z2red}, where on panel~(a) we plot a representative
$c_2(u)$, normalised by $\ZRredn{0}$. Panel~(b) shows the logarithm of $c_2(u)$---in this case
simply the stationary action at a given $u$, multiplied by $-1/\hbar$.

Hence to obtain an asymptotic approximation to \eqn{eq:z2red-ints}
we require the asymptotic form of $c_2(u)$ as $u \to 0$,
\begin{align}
\label{eq:c2-asympt}
c_2(u) & \sim \ZRn{0}^{\SC}
 \tfrac{\Delta_{\mathrm{r}}^2}{\hbar^2}  \mkern2mu \eu{-\frac{1}{\hbar}(\alpha_{\mathrm{r}} u -
\frac{1}{2} \gamma_{\mathrm{r}} u^2)}  \\
{} & \sim \ZRn{0}^{\SC}
 \tfrac{\Delta_{\mathrm{r}}^2}{\hbar^2}  \mkern2mu
\eu{-\alpha_{\mathrm{r}} u/\hbar}
\left(1 + \tfrac{\gamma_{\mathrm{r}} u^2}{2 \hbar} \right), \nonumber
\end{align}
where $\Delta_{\mathrm{r}} \equiv \Delta(\vec{x}_{\mathrm{r}})$, and
$\vec{x}_{\mathrm{r}}$ is the minimum of $V_0$.
The coefficients $\alpha_{\mathrm{r}}$ and $\gamma_{\mathrm{r}}$
are derived analogously to the $\alpha_n$ and $\gamma_n$ in \sec{ssec:kb},
\begin{subequations}
\label{eq:Zagr}
\begin{align}
\alpha_{\mathrm{r}} & = V_1\big( \vec{x}_{\mathrm{r}} \big) -
V_0\big( \vec{x}_{\mathrm{r}} \big) \\
\gamma_{\mathrm{r}} & = \sum_{j=1}^{f} \frac{
g_{\mathrm{p},j}^2 \coth \! \Big( \frac{\beta \hbar \omega_{\mathrm{r},\mkern-1mu j}}{2} \Big)
}{ 2 m \omega_{\mathrm{r},j} },
\end{align}
\end{subequations}
and $\ZRn{0}^{\SC}$ is defined in \eqn{eq:zreact-sc}.
As before, $ \omega_{\mathrm{r},j} $ is the frequency of the $j$-th reactant normal
mode, and $g_{\mathrm{p},j}$ is the derivative
of the product diabat at $\vec{x}_{\mathrm{r}}$ with respect to
that mode. An analogous expansion can be obtained for $c_2(\beta \hbar-u)$
as $u \to 0$ by exchanging the subscripts
$\mathrm{r} \leftrightarrow \mathrm{p}$ and potentials
$V_0 \leftrightarrow V_1$ in \eqs{eq:c2-asympt}{eq:Zagr}. These expressions
are plotted as dotted lines in \fig{fig:z2red}, alongside the quantum $c_2(u)$,
which they are seen to approximate accurately, at least in regions that
contribute significantly to the integrals over $u$.

Following standard procedure, we substitute the asymptotic expansion of
$c_2(u)$ on the second line of \eqn{eq:c2-asympt} into the original integrals,
which yields
\begin{subequations}
\label{eq:zrred2-bound}
\begin{align}
&
\begin{aligned}
& \frac{1}{\ZRn{0}}\int_{0}^{\frac{\beta \hbar}{2}} \! \!
                                [\beta \hbar - 2 u] \, c_2(u) \, \rmd u \sim
\frac{\Delta_{\mathrm{r}}^2}{\alpha_{\mathrm{r}}} \left[
\beta - \frac{2 \theta_{\mathrm{r}}}{\alpha_{\mathrm{r}}}
\right] \\
& {} + \Delta_{\mathrm{r}}^2 \frac{\hbar \gamma_{\mathrm{r}}}{\alpha_{\mathrm{r}}^2}
\left[
-\frac{6 \theta_{\mathrm{r}}}{\alpha_{\mathrm{r}}^2} + \frac{\beta (3 - 2\theta_{\mathrm{r}})}{\alpha_{\mathrm{r}}} +
\frac{\beta^2 \eu{-\beta \alpha_{\mathrm{r}}/2} }{4}
\right]
\end{aligned} \\
&
\begin{aligned}
& \frac{1}{Z_{\mathrm{s},0}}
\int_{0}^{\frac{\beta \hbar}{2}} \! \!
u \, c_2(u_{\mathrm{s}}) \, \rmd u \sim \frac{\Delta_{\mathrm{s}}^2}{\alpha_{\mathrm{s}}} \left[
\frac{\theta_{\mathrm{s}}}{\alpha_{\mathrm{s}}} -
\frac{\beta \eu{-\beta \alpha_{\mathrm{s}}/2}}{2}
\right] \\
& \! {} + \Delta_{\mathrm{s}}^2 \frac{\hbar \gamma_{\mathrm{s}}}{\alpha_{\mathrm{s}}}
\left[
\frac{3 \theta_{\mathrm{s}}}{\alpha_{\mathrm{s}}^3} -
\left(
\frac{3 \beta}{2 \alpha_{\mathrm{s}}^2} + \frac{3 \beta^2}{8 \alpha_{\mathrm{s}}}
+\frac{\beta^3}{16}
\right) \! \eu{-\beta \alpha_{\mathrm{s}}/2}
\right]\mathrlap{\!,}
\end{aligned}
\end{align}
\end{subequations}
where $\theta_{\mathrm{s}} = 1 - \eu{-\beta \alpha_{\mathrm{s}}/2}$,
the subscript `s' is either `r' or `p',
$u_\mathrm{r} \equiv u$ and $u_\mathrm{p} \equiv \beta \hbar-u$.
Substituting \eqn{eq:zrred2-bound} into \eqn{eq:z2red-ints} then gives
the semiclassical approximation to $\ZRredn{2}$
that applies when both $V_0$ and $V_1$ have
stable minima. When $V_1$ is unbound, only the first integral
in \eqn{eq:z2red-ints} needs to be calculated, as in this case
$\ZRn{0}/\ZPn{0}$ evaluates to zero, and so the second line of
\eqn{eq:z2red-ints} makes no contribution.

The correction discussed here predominantly accounts for
changes in the potential energy at the bottom of the reactant well. In the adiabatic
representation, an increase in $\Delta$ lowers the energy of
the well, thus increasing the activation energy barrier.
The term in \eqn{eq:z2red-ints} serves therefore to decrease the overall reaction rate
[see \eqn{eq:k4-therm}]---an effect that we correctly predict to be
negligible if $\Delta(\vec{x})$ decays to zero in the vicinity of the PES minima.

\begin{figure*}[t]
\includegraphics{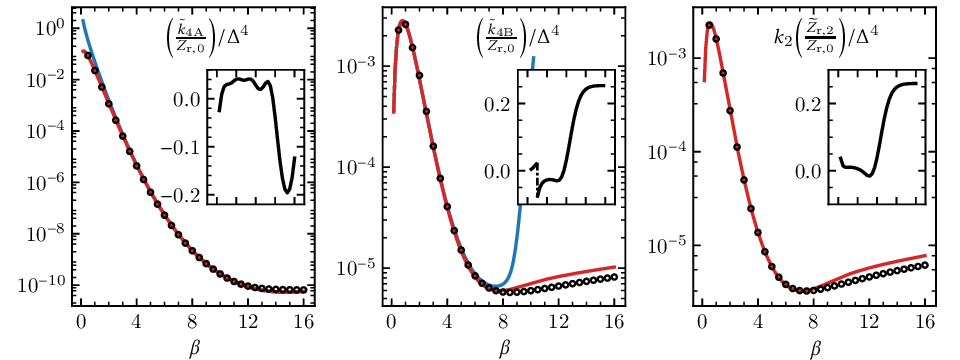}
\caption{\label{fig:ho-morse4}%
The three components of the $k_4$ thermal rate, calculated for the potentials in \eqn{eq:ho-morse} at constant diabatic coupling~$\Delta$. The black circles
are the quantum results. The semiclassical approximations to $\kfreda$ on the leftmost panel are calculated
according to \eqn{eq:k4asc}, with $d_{01}$ given by \eqn{eq:d01-sb} (red line)
or $d_{01} = 0$ (blue line). On the central panel, the semiclassical $\kfredb$ are
calculated according to \eqn{eq:k4b-sc-uni}, in combination with either
\eqn{eq:k4b-taylor} (red line) or \eqn{eq:k4b-lin} (blue line). On the rightmost
panel, the semiclassical approximation to $\ZRredn{2}/\ZRn{0}$, in red, is calculated as described in \sec{ssec:sc-partf}. Each panel includes an inset showing the
relative error in the red semiclassical estimate. The dash-dotted line on the middle inset indicates
the point at which the reference quantum result is changed as described in the main text.
 }
\end{figure*}

\section{\label{sec:models}%
Application to model systems}

\Eqn{eq:k4asc} for the $\kfreda$ component of the rate constant
is identical to the quantum result for
one-dimensional linear systems. \Eqn{eq:k4b-sc-uni} for $\kfredb$
is a highly accurate approximation of
the quantum result for the same type of system, reproducing it to within
$2.8\%$ or better. It remains to see how accurate these expressions are for nonlinear
or anharmonic potentials and in multiple dimensions.
To ensure high numerical accuracy of our tests,
in this section we consider systems for which the stationary action
can either be found using a one-dimensional root-finding algorithm
(\sec{ssec:prediss}) or is known analytically (\sec{ssec:spin-boson}).
Since the semiclassical expressions being tested are based on the
golden-rule instanton, we anticipate no particular difficulties
in extending the approach to more general potentials.
We have discussed in other work how to calculate the GR instanton
in such cases using the ring-polymer
formalism,\cite{GoldenRPI,InstReview,Ansari2021,Heller2021}
and our fourth-order components
can be obtained from the output of these calculations during post-processing.

\subsection{\label{ssec:prediss}%
Predissociation model}

In this example we show that the new theory performs well in anharmonic potentials,
using as our test case the predissociation model considered in \Refs{nonoscillatory,Lawrence2018Wolynes,inverted},
\begin{subequations}
\label{eq:ho-morse}
\begin{align}
\label{eq:pre-V0}
V_0(x) & = \tfrac{1}{2} m \omega^2 x^2 \\
\label{eq:pre-V1}
V_1(x) & = D_{\mathrm{e}} \eu{-2 \alpha (x - \zeta)} - \epsilon,
\end{align}
\end{subequations}
where $m = 1$,  $\hbar = 1$, $\omega = 1$, $D_{\mathrm{e}} = 2$,
$\alpha = 0.2$, $\zeta = 5$, $\epsilon = 2$. The quantum results were calculated
by expanding Eqs.~\eqref{eq:k4a-exact}, \eqref{eq:k4b-exact}, and the first line
of \eqref{eq:z2red-ints}, in terms of the eigenstates\cite{nonoscillatory} of $\Hop_{0,1}$ and performing the time integrals analytically. Where applicable, the infinite integration limits for $t$ and $t_1$ were
replaced with $t_{\mathrm{max}} = 10$, and with $\pi/\omega$ for $t_0$.
Energies up to $E_{\mathrm{max}}=75$ were included in the calculation
and only those eigenfunctions of $\HP$ were considered that had a node at $x_{\mathrm{max}}=35$,
which is equivalent to truncating the space at that point. $E_{\mathrm{max}}$, $t_{\mathrm{max}}$ and $x_{\mathrm{max}}$
are all convergence parameters that were confirmed to be sufficiently
large for the range of temperatures considered in this section.

All semiclassical results were derived from
the stationary action, which is directly available\cite{feynman2010quantum} for the
harmonic reactant potential in \eqn{eq:pre-V0}, and can be related to the abbreviated action for the repulsive Morse potential in \eqn{eq:pre-V1},
as discussed in \app{sec:morse0}.

We have taken the diabatic coupling to be constant, $\Delta(x) = \Delta$,
and calculated the quantum and semiclassical GR thermal rates along with
their fourth-order corrections. The three components of the latter are plotted
in \fig{fig:ho-morse4} across a set of temperatures ranging
from the classical to the deep-tunnelling regime. For the $\kfreda$ contribution
plotted on the leftmost panel, we show semiclassical results calculated
according to \eqn{eq:k4asc} using two different approximations for $d_{01}$:
the red curve uses the expression in \eqn{eq:d01-sb}, whereas the blue
curve sets $d_{01} = 0$. The latter was included to show
why it is important to account for the $t_0^2 t_1^2$  term in \eqn{eq:L4A-ht}, as
failing to do so causes the semiclassical estimate to diverge from the
correct result in the high-temperature limit.
In this example neglecting the quartic term leads to
an overestimation of the $\kfreda$ component by a factor of 3 at
the highest temperature considered in our quantum calculations
($\beta = 0.5$).

\begin{figure}
\includegraphics{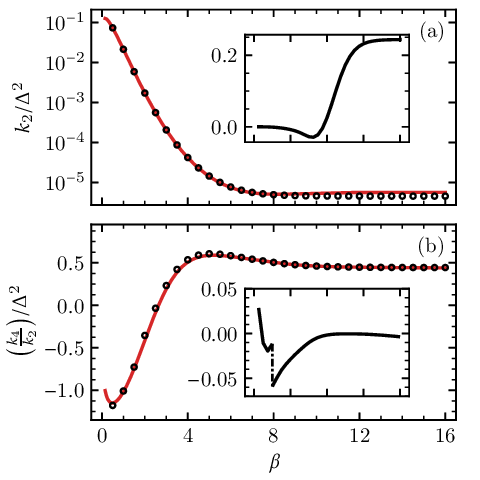}
\caption{(a)~The golden-rule thermal rate constant $k_2$ and
(b)~the estimate of its relative deviation from the full rate,
\mbox{$(k - k_2)/k \sim k_4/k_2$}, plotted for
the systems in \eqn{eq:ho-morse} at constant
diabatic coupling $\Delta$. The black circles represent
the quantum results, and the red lines are the corresponding
semiclassical approximations. Panel (a) includes an inset
showing the error in the semiclassical GR rate relative to the
quantum result, \mbox{$\kgrsc/\kgr^{}-1$}, and panel~(b) includes an
inset with the absolute error, $\Delta^{\!-2}\left(\kfsc/\kgrsc - \kf/\kgr\right)$.
The dash-dotted line indicates
the point at which the reference quantum $\kfredb$ is changed as described in the main text.
\label{fig:ho-morse-all}}
\end{figure}

Including the $d_{01}$ term results in a value
that is within a few per cent from the quantum
expression down to $\beta=12$. At lower temperatures,
the magnitude of the
relative error increases to approximately $20\%$, which is comparable to the $\approx25\%$ error developed by the GR instanton rate
in the $\beta \to \infty$ limit [\fig{fig:ho-morse-all}(a)]. Further down,
the relative error appears to decrease---an artefact that can be
traced down to our quantum results. Their calculation
involves truncating the integration bounds for $t_0$ at
$\pm \pi/\omega$ because of recurrences in the $\cfa$ correlation
function.  At sufficiently low temperatures, this function
no longer decays to zero by the cut-off points,
and so the quantum $\kfreda$ becomes ill-defined.

The $\kfredb$ contribution, plotted on the middle panel of \fig{fig:ho-morse4} is also
approximated using two different semiclassical expressions: the red curve follows
the general prescription in \eqs{eq:k4b-taylor}{eq:k4b-sc-uni}, and the blue curve
replaces \eqn{eq:k4b-taylor} with \eqn{eq:k4b-lin}, which would recover the quantum result if the
system were linear. Quantum results are calculated according to
\eqn{eq:k4b-exact}, with the integration contours for $z_n$ deformed as in \eqn{eq:z0z1-int}. We use the
$\tauGR$ from the GR instanton to define the deformed contours and to
set the real part of $z$ in \eqn{eq:J-def}. Ideally, instead of using a fixed
$\Re(z)$ for all $z_n$, one would choose an optimal value for each combination
$(z_0, z_1)$. This would minimise the oscillations of the integrand
in \eqn{eq:J-def} and reduce the error due to truncating the integration
range to $(-t_{\mathrm{max}}, t_{\mathrm{max}})$. However we found this computationally demanding and
used a fixed $\Real(z) = \tauGR$ instead. We can verify that this is a reasonable approximation by comparing the output of \eqn{eq:k4b-exact} to the value obtained
from only the $\mathcal{A}_{\tau \tau}$ term. The two are expected to be
close across the entire temperature range, and indeed for $\beta \geq 2$
the values are within $10\%$ of each other. For smaller $\beta$
the magnitudes of $\kfreda$ and $\mathcal{A}_{tt}$ are large enough that accurately
calculating \eqn{eq:k4b-exact} poses numerical difficulties.
For this reason the $\kfredb$ plotted at the three highest temperatures includes only the contribution from $\mathcal{A}_{\tau\tau}$.
This change is the cause of the
small discontinuity in the relative error shown on the inset.

As anticipated,
the two semiclassical approximations are very close at high temperatures
(small $\beta$), but
only \eqn{eq:k4b-sc-uni} is well-behaved for large $\beta$. \eqn{eq:k4b-lin} predicts
a rate constant that rapidly diverges from the quantum result as $\beta \to \infty$. In the same regime, the ``good'' approximation develops a constant
relative error of around $25\%$. This is not an error in the fourth-order expression
\latin{per~se}, but is rather a consequence of $\kfredbsc$ being proportional
to $\kgrredsc$, which itself overestimates the GR rate in the
$\beta \to \infty$ limit, as shown in \fig{fig:ho-morse-all}(a).
The same observation applies to the partition-function component plotted on the rightmost panel of \fig{fig:ho-morse4}. Like $\kfredbsc$, it is
proportional to the semiclassical golden-rule rate constant and hence
develops a (nearly) constant relative error in the
low-temperature regime, inherited from $\kgrsc$. Otherwise, the quantum and
semiclassical results for this component are in good agreement.

One of the key applications of a fourth-order rate theory
is estimating the error
introduced by approximating the full thermal rate with just the second-order
(golden-rule) term.
Specifically, we are interested in the relative error, whose estimate
in the small-$\Delta$ limit is $k_4/\kgr$,
provided \mbox{$k_4 \neq 0$}. The latter is important to bear in mind,
since $k_4$ contains both positive and negative components. Their relative magnitudes are temperature-dependent, such that
$-\kfreda/\kgrred < 0$ dominates as $\beta \to 0$ and
$(\kfredb/\kgrred - \ZRredn{2}/\ZRn{0}) > 0$ dominates as $\beta \to \infty$.
Hence there will be a temperature at which these expressions cancel exactly.
At that point, strictly speaking, one has to consider the
next term in the series, $k_6$, in order to estimate the error.

With this in mind, we plot the golden-rule rate constant $k_2$ and the ratio $k_4/k_2$ in panels (a) and (b) of \fig{fig:ho-morse-all} respectively, alongside
the corresponding semiclassical approximations. As mentioned previously,
$\kgrsc$ develops a constant relative error in the low-temperature
regime, where the rate plateaus. However, because the terms
that dominate $\kfsc$ at such temperatures are themselves
proportional to $\kgrsc$, the ratio $\kfsc / \kgrsc$
benefits from near-complete error cancellation and
is a remarkably accurate approximation to the quantum result
across the entire temperature range. Like the quantum result, it (approximately) plateaus
at low temperatures, when only the reactant vibrational ground state
contributes to the process. The semiclassical expression deviates
from quantum by no more than 5\% of its maximum absolute value,
with the largest deviations found at
$ \beta \leq 2.5$. The errors at these high temperatures
are likely overestimated because of numerical artefacts in the corresponding quantum calculations,
but even the apparent level of accuracy is more than sufficient
to gauge the validity of the GR approximation. Under
favourable circumstances these results should even enable us
to correct the GR approximation, which we explore further in \sec{ssec:spin-boson}.

\subsection{\label{ssec:spin-boson}%
Spin--boson model}

\begin{table*}[t]
\caption{
Thermal rate constants for the underdamped spin--boson system in
\eqn{eq:sb-pes}, calculated for $\beta=1$, $\hbar=1$, $\Lambda = 60$,
and $\gamma = \Omega$. The quantum results were calculated by numerical
quadrature of the corresponding correlation functions.
All results are divided by $\Delta^2$.%
\label{tab:low-gamma}
}
\begin{ruledtabular}
\begin{tabular}{>{\rule[-0.2em]{0pt}{1.6em}$}r<{$}*8{>{$}c<{$}}}
\multicolumn{1}{c}{\rule[-0.4em]{0pt}{1.4em}$\epsilon$} & \multicolumn{4}{c}{$0$} & \multicolumn{4}{c}{$15$} \\
\cline{1-1} \cline{2-5} \cline{6-9}
\multicolumn{1}{c}{\rule[-0.45em]{0pt}{1.7em}$\Omega$}  & 0.5  & 2  & 4  & 6  & 0.5  & 2  & 4  & 6  \\
 \hline
\kgr &\num{7.56e-08} & \num{1.80e-07} & \num{1.00e-06} & \num{5.25e-06} & \num{5.30e-05} & \num{1.12e-04} & \num{4.59e-04} & \num{1.57e-03} \\
\kgrsc &\num{7.56e-08} & \num{1.80e-07} & \num{9.91e-07} & \num{5.12e-06} & \num{5.30e-05} & \num{1.12e-04} & \num{4.56e-04} & \num{1.56e-03} \\
\frac{\kfreda}{\kgrred}  & 0.947 & 0.0744 & 0.0122 & 0.0042  & 0.992 & 0.0866 & 0.0181 & 0.0088  \\ 
\frac{\kfredasc}{\kgrredsc} & 0.859 & 0.0679 & 0.0115 & 0.0042  & 0.891 & 0.0778 & 0.0164 & 0.0079  \\ 
\frac{\kfredb}{\kgrred} &0.122 & 0.0942 & 0.0637 & 0.0485 & 0.114 & 0.0900 & 0.0626 & 0.0487 \\
\frac{\kfredbsc}{\kgrredsc} &0.122 & 0.0956 & 0.0649 & 0.0492 & 0.115 & 0.0914 & 0.0639 & 0.0494 \\
\frac{\ZRredn{2}}{\ZRn{0}} &0.0161 & 0.0162 & 0.0164 & 0.0168 & 0.0225 & 0.0229 & 0.0239 & 0.0255 \\
\frac{\ZRredn{2}^{\SC}}{\ZRn{0}^{\SC}} &0.0160 & 0.0161 & 0.0164 & 0.0167 & 0.0222 & 0.0225 & 0.0231 & 0.0239 \\
 \hline
\frac{\kf}{\kgr} &-0.8417 & 0.0036 & 0.0350 & 0.0274 & -0.9000 & -0.0194 & 0.0206 & 0.0143 \\
\frac{\kfsc}{\kgrsc} &-0.7527 & 0.0117 & 0.0370 & 0.0284 & -0.7986 & -0.0088 & 0.0244 & 0.0177 \\
\end{tabular}
\end{ruledtabular}
\end{table*}

Here our attention turns to the multidimensional spin--boson model. As mentioned
in the introduction, this type of model potential is not
our main target. Even so, it provides a good test system as the corresponding stationary action
is known exactly and can be evaluated with relative ease (see \app{sec:sb-tcfs}).
Furthermore, because the model has been extensively studied in the past,
benchmark non-perturbative quantum rates
are available for a broad parameter range.\cite{Lawrence2019ET}

The diabatic PESs for the spin--boson model in ``reaction coordinate'' form are
given by
\begin{subequations}
\begin{align}
V_n(q, \vec{Q}) & = \frac{1}{2}\Omega^2\! \left(
	q \pm \sqrt{\frac{\Lambda}{2 \Omega^2}}
\right)^{\mathrlap{\!\!2}} + V_{\mathrm{sb}}(q, \vec{Q}) - \epsilon_n \\
V_{\mathrm{sb}}(q, \vec{Q}) & =  \sum_{j = 1}^{f-1} \frac{1}{2} \omega_{j}^2 \!
\left(
    Q_j - \frac{c_{j} q}{\omega_{j}^2}
\right)^{\!\!2},
\end{align}
\end{subequations}
where $\vec{Q}$ are the bath modes, $q$ is the reaction coordinate
and $\Lambda$ is the Marcus reorganisation energy. The plus and minus signs
are taken for $n = 0$ and $1$ respectively,
and $\epsilon_1 - \epsilon_0 = \epsilon$. The coupling
coefficients~$c_{j}$ and normal-mode frequencies~$\omega_{j}$
are all encoded in the spectral density
\begin{equation}
J(\omega) =  \frac{\pi}{2} \sum_{j = 1}^{f}
\frac{c_{j}^2}{\omega_{j}} \delta(\omega - \omega_{j}).
\end{equation}
The diabatic coupling, $\Delta$, is taken to be a constant.
Following \Refx{Lawrence2019ET}, we go to the continuum limit ($f \to \infty$)
and consider a purely Ohmic spectral density
\begin{equation}
J(\omega) = \gamma \omega,
\end{equation}
where $\gamma$ is the friction coefficient along the reaction coordinate. We then
re-express the diabatic potentials in the conventional
spin--boson form,
\begin{equation}
\label{eq:sb-pes}
V_n(\vec{x}) = \sum_{j = 1}^{f} \frac{1}{2} \tilde{\omega}_{j}^2 \!
\left(
    x_j \pm \frac{\tilde{c}_{j}}{\tilde{\omega}_{j}^2}
\right)^{\!\!2} - \epsilon_n,
\end{equation}
where the new coordinates $\vec{x}$ are related to $(q, \vec{Q})$ by an
orthogonal transformation, and $\tilde{\omega}_j, \tilde{c}_j$ derive
from the Brownian oscillator spectral density,\cite{Garg1985spinboson,Weiss}
\begin{equation}
J(\omega) = \frac{\Lambda}{2} \frac{\gamma \Omega^2 \omega}{
  (\omega^2 - \Omega^2)^2 + \gamma^2 \omega^2
}.
\end{equation}
The two-state system in \eqn{eq:sb-pes} has been studied extensively,
and some of the previously derived analytical results are
used by us below.

For harmonic potentials, the semiclassical propagator is exact,
the corresponding action $S_n$ quadratic in the end-points,
and the prefactor $C_n$ independent of positions. Therefore,
semiclassical three-time correlation functions for the
spin--boson model (obtained following the prescriptions in this paper)
are identical to their quantum counterparts.
As shown, for example, by Weiss,\cite{Weiss} these are generally of the form
$\eu{-\phi_4(z, z_0, z_1)/\hbar}$. It follows that the
time derivatives of $\phi_4$ agree exactly with the coefficients
$d_{n}$ in \eqn{eq:d0d1-defn}, and $\alpha_{n}, \gamma_{n}$ in \eqn{eq:k4b-ag}.
The same considerations
apply to the single-time correlation function from
which the GR rate constant and the partition-function correction
are derived. In \app{sec:sb-tcfs} we give the general form
of $\phi_{2\nu}(\cvec{z})$, where $\nu = 1,\,2,\,\ldots,$
and $\cvec{z}$ is a set of $2 \nu - 1 $ complex time variables.
These are used to calculate the quantum
$k_2$ and $k_4$ rate constants, and it is also from these
that we derive all of the coefficients required for the
steepest-descent approximation. The only exception is $d_{01}$,
whose rigorous calculation would in general require high-order derivatives
of the action that are not available from standard GR
instanton calculations. For simplicity, we
use the approximate expression for $d_{01}$ in \eqn{eq:d01-sb} with $\omega = \Omega$, even though in this
case the exact value can in principle be obtained.

In \tab{tab:low-gamma} we list the second- and fourth-order rate
constants for a set of underdamped
symmetric ($\epsilon=0$) and asymmetric ($\epsilon=15$)
spin--boson models, with \mbox{$\beta=1$}, $\hbar=1$, $\Lambda = 60$,
and $\gamma = \Omega$ ranging from $0.5$ to~$6$. With temperature,
reorganisation energy and bias kept constant within each set of
systems, rates predicted by classical theories such as
Marcus\cite{Marcus1985review,Marcus1964review,Nitzan}
or Zusman\cite{Zusman1980,Garg1985spinboson,Gladkikh2005}
are all of the same order of magnitude within a given set.
Hence any major variation seen in practice is due to contributions from quantum tunnelling, which becomes more important with increasing $\Omega$.

All quantum results are calculated by numerical integration of the
corresponding correlation functions, and $\kfb$ only includes
the contribution from the imaginary-time integral over
$\mathcal{A}_{\tau \tau}$. As expected, $\kgrsc$ is in close agreement
with the quantum GR rate for the entire parameter range.
Semiclassical approximations of the fourth-order rate components
are also accurate across the board, lying within about 10\% of quantum.

Notwithstanding the accuracy of its individual components,
at $\Omega = 2$ the semiclassical approximation to the \emph{total}
$\kf$ rate constant [\eqn{eq:k4-therm}] deviates quite significantly
from the quantum result. This should come as no
surprise, since the three components contribute to the total
fourth-order rate constant with different signs.
One expects therefore to encounter a combination of parameters
for which these contributions cancel exactly. Unless there is some fortuitous cancellation of errors, semiclassical predictions around that point are necessarily expected to show a large \emph{relative} error. However the
key point is that the \emph{absolute} error remains small.
What the semiclassical theory
predicts correctly is that here the fourth-order dependence of the
total rate on $\Delta$ is weak, and that the GR expression remains accurate
up to greater diabatic coupling strengths than would normally be expected. For
additional information one could consider the next term
in the perturbation series, $k_6$, which we do not pursue in this work.
With this proviso, the semiclassical results
in \tab{tab:low-gamma} are of sufficient accuracy to establish
whether a system is in the GR regime and to estimate the sign and magnitude of the error introduced by making the GR approximation.

\begin{table}[b]
\caption{
Thermal rate constants for the overdamped spin--boson model in
\eqn{eq:sb-pes} with $\hbar=1$, $\Lambda = 60$, $\epsilon = 0$,
$\gamma = 32 \Omega$ and $\Omega = 2$.
All results are divided by $\Delta^2$.%
\label{tab:high-gamma}
}
\begin{ruledtabular}
\begin{tabular}{>{\rule[-0.2em]{0pt}{1.6em}$}r<{$}*3{>{$}c<{$}}}
\multicolumn{1}{c}{\rule[-0.45em]{0pt}{1.7em}$\beta$} & 1 & 2 & 4 \\
 \hline
\kgr & \num{7.895e-08} & \num{4.929e-14} & \num{2.781e-26} \\
\kgrsc & \num{7.893e-08} & \num{4.928e-14} & \num{2.781e-26} \\
\frac{\kfreda}{\kgrred} & 2.309 & 1.701 & 0.789 \\
\frac{\kfredasc}{\kgrredsc} & 0.399 & 0.557 & 0.762 \\
\frac{\kfredb}{\kgrred} & 0.121 & 0.443 & 1.289 \\
\frac{\kfredbsc}{\kgrredsc} & 0.123 & 0.469 & 1.424 \\
\frac{\ZRredn{2}}{\ZRn{0}} & 0.016 & 0.033 & 0.066 \\
\frac{\ZRredn{2}^{\SC}}{\ZRn{0}^{\SC}} &0.016 & 0.033 & 0.066 \\
 \hline
\frac{\kf}{\kgr} &-2.204 & -1.291 & 0.433 \\
\frac{\kfsc}{\kgrsc} &-0.292 & -0.121 & 0.595 \\
\end{tabular}
\end{ruledtabular}
\end{table}

\begin{figure}[b]
\includegraphics{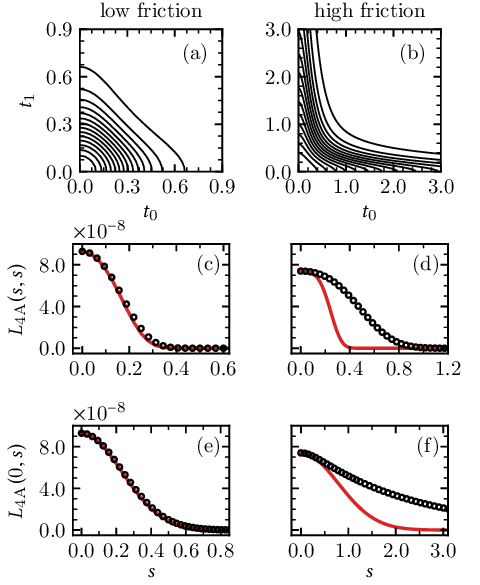}
\caption{
The marginalised $\margfa(t_0, t_1)$ quantum correlation
function (black) and its semiclassical approximation (red) for
a spin--boson model with $\Delta = 1$, $\epsilon = 0$,  $\beta = 1$, $\Omega = 2$, and $\Lambda = 60$. On the left-hand side the
friction is $\gamma = \Omega$, and on the right-hand side
$\gamma = 32 \Omega$.
Panels~(a--b) show contour plots of the quantum
$\margfa(t_0, t_1)$, with contours
drawn at evenly spaced values between 0 and the maximum value at the origin.
Panels~(c--f) compare the quantum and the semiclassical $\margfa(t_0, t_1)$ along
selected cuts through the $(t_0, t_1)$ plane.%
\label{fig:sb-margfa}
}
\end{figure}

It was mentioned in the introduction that the methods developed in this
paper are not generally aimed at reactions in solution and do not
attempt to describe rate processes involving diffusive motion
along the reaction coordinate.
Hence it was reasonable to base the preceding discussion
on an underdamped spin--boson model ($\gamma = \Omega$).
Even so, it is instructive to also analyse the predictions
for a strongly overdamped system,
which may shed light on the limitations of the method and
give insight into the underlying physics.
In \tab{tab:high-gamma} we list the results for a symmetric spin--boson model
at a range of temperatures, with $\Omega = 2$, $\gamma = 32 \Omega$, and the
remaining parameters kept the same as in \tab{tab:low-gamma}.
At $\beta = 1$, $\kgrredsc$, $\kfredbsc$ and $\ZRredn{2}^{\SC}$ are all in excellent agreement with their quantum counterparts, whereas $\kfredasc$ underestimates the corresponding quantum value by about an order of magnitude,
completely compromising the total rate, $\kf$. The same pattern (not shown)
is found at other values of $\epsilon$ and $\Omega$ listed in \tab{tab:low-gamma}\@.
This suggests that the assumptions underlying our derivation of $\kfredasc$ do not hold in the
high-temperature, high-friction regime.
We can narrow the issue down to the behaviour of $\margfa(t_0, t_1)$,
previously defined in \eqn{eq:L4A-qm}. In~\fig{fig:sb-margfa} we compare this function
to its semiclassical approximation
\begin{align}
\label{eq:SB-L4ASC}
\margfa^{\SC}(t_0, t_1) & = \sqrt{2 \pi \hbar} \,
\frac{\Delta^4}{\hbar^4} \!
\left(
\pder[2]{\phi_{\fa}}{t}
\right)^{\!\!-1/2} \!\! \eu{-\phi_{\fa}/\hbar} \\
& \qquad {} \times \exp\left(
-\frac{d_0 t_0^2 + d_1 t_1^2 + d_{01} t_0^2 t_1^2}{2 \hbar}
\right), \nonumber
\end{align}
with $\phi_{\fa}$ defined as in \eqn{eq:phi4A}, and both it and its derivative
evaluated at the stationary point.

The comparison allows us to identify two distinct factors that
cause the accuracy of the semiclassical approximation
to deteriorate. Firstly, from the plots of $\margfa(s,s)$ in \fig{fig:sb-margfa}(c--d) we see that the approximation in \eqn{eq:d01-sb} significantly overestimates the true value of $d_{01}$ at high friction.
In principle this can be fixed by basing the approximation
on a more sophisticated analytically soluble model. Alternatively
one could evaluate $d_{01}$ numerically, which requires
implementing the necessary high-order action derivatives\footnote{%
In practice evaluating the exact $d_{01}$ for anharmonic potentials would involve calculating
terms containing third- and  fourth-order derivatives of the PESs, which may be impractical in \latin{ab initio} simulations.
However we expect that in most cases a very accurate approximation
to $d_{01}$ would result even if such terms were omitted.} in addition to those calculated during GR instanton optimisation.\cite{GoldenRPI}
If necessary, we may pursue one or both of these strategies in future work.

Nevertheless a refined estimate of $d_{01}$ would still not bring the
semiclassical $\kfredasc$ into alignment with the quantum results
due to a second feature, illustrated
in \fig{fig:sb-margfa}(e--f). There we plot $\margfa(0, s)$, whose semiclassical approximation is proportional to $\exp(-d_1 s^2 / 2 \hbar)$
and contains no contributions from the quartic $d_{01}$ term.
It is apparent that the quantum and semiclassical values agree at
short times (indicating that $d_1$ is calculated correctly), but
whereas the semiclassical function is shaped like a Gaussian,
at long times the quantum expression behaves like a decaying exponential. The emergence
of this ``fat'' exponential tail is almost certainly due to the motion along the reaction coordinate becoming diffusive.\cite{Garg1985spinboson}
This implies that the system spends more time in the barrier region
and is more likely to undergo an unreactive transition
of the kind described in \sec{ssec:ka}, resulting in a larger
$\kfreda$ and a diminished overall rate constant.

That the semiclassical approximation fails to capture the long-time behaviour
of $\margfa(t_0, t_1)$, and hence the true magnitude of $\kfreda$, is not
a problem of the instanton method \latin{per~se}, but rather a breakdown
of steepest-descent in time that affects all transition-state theories (TST).\cite{Nitzan} To avoid this, one
has to define a more suitable dividing surface/projection operator $\hat{h}$, leading to an altogether different set of instantons. Thankfully we need
not implement so radical a change, since the primary target of our semiclassical rate theory is gas-phase reactions in the deep-tunnelling regime, where
instanton theories are typically most useful. In this regime, the reaction
rate is unlikely to have significant contributions from diffusive over-the-barrier
motion, even if it happens to be a feature at higher temperatures. Instead
the rate is dominated by nuclear tunnelling contributions, which are
captured accurately by our theory, as shown by the low-temperature results in \tab{tab:high-gamma}.
 As $\beta$ increases,
the $\kfreda$ term decreases relative to the other fourth-order contributions and becomes better approximated by the semiclassical expression. At the
lowest temperature considered ($\beta = 4$), the overall agreement between
the semiclassical and the quantum rates is comparable to that in \tab{tab:low-gamma}, even though the system is heavily overdamped.
We expect that a similar level of accuracy will be seen for more
chemically realistic potentials.

Our results so far imply that
\eqn{eq:k4-therm} evaluated semiclassically provides a reliable
indicator of whether a reaction is in the golden-rule limit
(as long as the rate is not dominated by diffusive motion through the barrier
region).
The GR rate, $\kgr$, should be a reliable approximation to the full (non-perturbative) rate only if $\lvert \kf / \kgr \rvert \ll 1$.
When this condition is not satisfied,
we may attempt to correct the GR approximation using the Pad\'{e}-summed expression, as discussed in \sec{sec:pade}.

\begin{figure}
\includegraphics{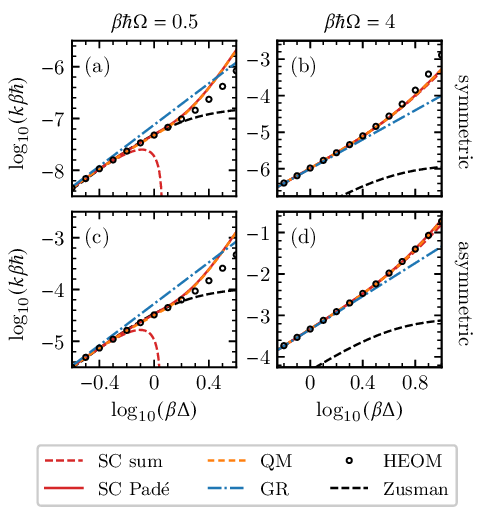}
\caption{Exact (non-perturbative) rates calculated with HEOM\cite{Lawrence2019ET}
(black circles),
compared against three semiclassical approximations:
golden-rule $\kgrsc$ on its own (GR),
the sum $\kgrsc + \kfsc$ (SC sum), and the Pad\'{e} approximation in
\eqn{eq:k24-pade} (SC Pad\'{e}). Also shown are \eqn{eq:k24-pade} evaluated with quantum rate constants (QM)
and rates calculated using the Zusman equation [Eq.~(70) in \Refx{Lawrence2019ET}].
The semiclassical approximation is almost exactly
on top of the analogous Pad\'{e}-summed quantum result. Top-row panels (a--b) correspond
to symmetric systems, $\epsilon = 0$,
and bottom-row panels (c--d) to asymmetric systems, $\epsilon = 15\beta^{-1}$.
Plots on the left are for the high-temperature regime
($\beta \hbar \Omega = 0.5$), and plots on the right are for low
 temperature ($\beta \hbar \Omega = 4$).
In all cases $\hbar = 1$, $\Omega = \gamma$ and $\beta \Lambda = 60$.
\label{fig:sb-rates}
}
\end{figure}
In \fig{fig:sb-rates} we compare
the semiclassical results to the exact (non-perturbative) rate constants  from \Refx{Lawrence2019ET}.
Plotted alongside is the same Pad\'e expression calculated
using the quantum GR and fourth-order rates
from \tab{tab:low-gamma}\@. The semiclassical
results overlap their quantum counterparts
almost exactly. Hence deviations between our theory and the full nonadiabatic rate that emerge at large $\Delta$ are due to the truncation of the perturbation series, and
not the result of making a semiclassical
approximation.
Also included are the plain GR rate $\kgrsc$,
the partial sum $\kgrsc + \kfsc$, and the output of the Zusman equation.\cite{Zusman1980,Garg1985spinboson,Gladkikh2005} The latter
is accurate at high temperature and weak coupling, and is a classical reference that illustrates the magnitude of nuclear quantum effects.

In all cases the corrected semiclassical expression (be it the sum or Pad\'{e})
is to within graphical accuracy coincident with the exact non-perturbative quantum results
up to larger $\Delta$ than $\kgrsc$ on its own. Granted, at even larger $\Delta$
the partial sum may diverge
quite dramatically from the reference result, as in
panels (a) and (c) of \fig{fig:sb-rates}, whereas here the golden rule continues to give results within an order of magnitude of HEOM\@. This
should, however, not be seen as a failure of our approach or a sign that the golden-rule approximation is somehow better.
The divergence occurs because the $\kf$ term
becomes comparable in magnitude to $\kgr$, giving a clear indication that
the rate no longer
scales as $\Delta^2$. That alone
provides valuable mechanistic insight and invalidates the golden-rule approximation.

At high temperatures [panels (a) and (c) of \fig{fig:sb-rates}],
where the correction is dominated by the $\kfa$ term,
there is a clear benefit to using the Pad\'{e} approximant over the partial sum.
At low temperature [panels (b) and (d)] the difference between the two corrected
expressions becomes marginal, as $\kfa$ is relatively small. Overall,
\eqn{eq:k24-pade} is at least as good or better than the more na\"{i}ve
expression for the corrected rate, and so we recommend its use at
all temperatures.

Our goal was to
derive a correction to GR in the weak-coupling limit, and consequently we have not incorporated any information from the strong-coupling (adiabatic) limit into our theory.
Because of this, our approach
cannot rival the global accuracy of the interpolation formula by Lawrence and
co-workers at large $\Delta$.\cite{Lawrence2019ET} It does, however, go a long way towards fixing the
rate at intermediate coupling. Furthermore, the new theory predicts the correct
high-temperature dependence of $k_4$ [\eqn{eq:k4a-ht}] for reactions
whose rate is not solvent controlled (\latin{i.e.}, away from the high-friction regime). As mentioned in \sec{ssec:ka},  \app{sec:high-t-limit} shows that classical rates based on the Landau--Zener transmission probability\cite{Landau1932LZ,Zener1932LZ} are consistent with our theory in the high-temperature limit. In contrast, the Zusman equation, and hence the interpolation formula, are not---a discrepancy due to  Zusman's underlying assumption of strong solvent friction.\cite{Zusman1980,Garg1985spinboson,Gladkikh2005}
Our approach thus offers
insight into nonadiabatic rate processes at
temperatures and frictions that could not be accurately described by pre-existing (semi)classical theories.

\section{\label{sec:discuss}%
Discussion}

In this paper we have derived the second term, $k_4$,
in the perturbation series expansion of the nonadiabatic rate constant $k$
in powers of the diabatic coupling $\Delta$ (the first term being the golden-rule rate constant $k_2$). Our expression consists of three components,
each of which is well approximated by semiclassical
instanton methods. The first component, $\kfredasc$, corresponds to a
four-bounce instanton path. It dominates $\kf$ at high temperatures,
decreasing the overall rate, and accounts for recrossing
transitions between the two diabats, which can either take
place via tunnelling or during the passage of the system
through the barrier region.
The second component, $\kfredbsc$, corresponds to a two-bounce instanton path.
It dominates $\kf$ at low temperatures and accounts for the enhanced
tunnelling probability at stronger diabatic coupling, increasing the overall rate.
The final component, $\ZRredn{2}$, derives from the minima of
the reactant and product wells. Along with $\kfredbsc$,
it dominates at low temperature, but decreases the rate instead.
This term largely accounts for changes in the potential energy around the
reactant minimum.

Like the semiclassical golden-rule rate, the $\kfredasc$ and $\kfredbsc$ components
are identical to their corresponding quantum rate constants
for systems comprised of two linear
diabats (at least if $\kfredbsc$ is defined as in \app{sec:kfb-lin}). As far as we are aware, this is
the first time that these expressions are reported in the literature, and
hence the first time that the high-temperature asymptotics of $\kf$ are
rigorously derived for a system of one-dimensional diabatic potentials.
By extension, we expect
that the high-temperature rates for multidimensional systems
with low to moderate friction along the reaction coordinate are more
accurately described by our theory
than by, \latin{e.g.}, the Zusman equation\cite{Zusman1980,Garg1985spinboson,Gladkikh2005} or uniform rate expressions\cite{Rips1995ET,Rips1996ET,Nitzan}
derived from the Holstein nonadiabatic transmission coefficient\cite{Holstein1959}
(based in turn on Landau--Zener theory).\cite{Landau1932LZ,Zener1932LZ} It should, however,
be noted that the latter is
consistent with our new theory,
as discussed in \app{sec:high-t-limit}, although the assumptions
underlying the Holstein expression mean it can only give the
asymptotic form of $\kf(\beta \to 0)$ up to a
temperature-independent constant. Our theory goes
beyond LZ and Holstein expressions: not only does it fully establish
the high-temperature asymptotics of $\kf$, it also captures the
mechanistic transition from over-the-barrier to tunnelling
processes, and can therefore be applied at low temperatures.

Calculations of $k_4$ in a one-dimensional predissociation
model have shown that our semiclassical expressions readily generalise to
anharmonic systems, with the model in \sec{ssec:prediss} displaying deviations
from the quantum ratio $\kf/\kgr$ that are no greater than 5\% of its maximum absolute value.
Analogous calculations for a
spin--boson model have shown that the semiclassical rates are
also accurate in multidimensional systems. Problems only
arise when both the temperature and friction are high,
at which point the motion of the nuclei through the barrier region becomes diffusive\cite{Garg1985spinboson} and recrossing events quite likely.
In line with our physical interpretation, this affects the $\kfredasc$ term, causing it to
considerably underestimate the value of its quantum counterpart.
One could attempt
to fix the problem either by redefining the product projection operator
(which would lead to a different semiclassical theory) or by employing
our knowledge of the real-time dependence of the quantum
$\cfa$ time correlation function for the spin--boson
model. Neither is pursued in this work, as our intended domain of application
is low-temperature reactions in the gas phase, which are not likely to be
rate-limited by diffusion along the reaction coordinate.
Nevertheless, in future work it could be interesting to
study the spin--boson $\cfa$ correlation function more closely, as such
investigations could lead to a generalisation of the
Zusman formula\cite{Zusman1980,Garg1985spinboson,Gladkikh2005}
that would rigorously account for nuclear quantum effects.

For spin--boson models at moderate friction our theory captures the
behaviour of the full non-perturbative rate for a wider range of diabatic coupling strengths
than the GR rate on its own, and correctly predicts the sign and
magnitude of the error introduced by making the golden-rule approximation.
Under favourable conditions, one
can use the fourth-order terms to obtain significantly improved estimates of the full nonadiabatic rate.
Crucially, few additional calculations are
necessary to evaluate the new expression. Its $\kfa$ component derives from
a GR instanton at $\beta/2$, \latin{i.e.}, at twice the temperature of the simulation. Since instanton optimisation typically involves starting
at high temperature and cooling the system down,\cite{Andersson2009Hmethane,Rommel2011locating,InstReview} it is easy to arrange
for $\beta/2$ to be visited along the way. All the quantities
needed to evaluate $\kfredasc$ are in that case directly available from the
output of the optimisation.

The $\kfb$ component derives from a GR instanton at the simulation
temperature, $\beta$. In addition to the output of the GR optimisation, the calculation
of $\kfredbsc$ requires the potential energy and gradient of the upper diabatic
state ($V_1$ on the reactant side and $V_0$ on the product side), as well as
the diabatic coupling $\Delta$ along the instanton trajectory. However no instanton re-optimisation is required, and the additional quantities
need only be evaluated on a relatively sparse grid.
The computational overhead thus remains marginal compared to the initial
GR optimisation.

The final (partition-function) component, rather than deriving from a
delocalised instanton structure, comes from just two points---the
minima of the reactant and product diabats.
Consequently this term should also add little
to the overall computation time. Moreover, this term
is expected to often be insignificant in practice,
as it is proportional to the square of the diabatic coupling
at the reactant/product minimum, which may be negligible.

On the whole, lengthy derivation notwithstanding,
one can obtain $\kfsc$ in a simple post-processing step to
a GR instanton optimisation, involving only a few
additional electronic-structure calculations.
It has all the makings of a practical method
that, like GR instanton theory,
can be applied to \latin{ab~initio} simulations of
real molecules.\cite{Heller2021}
Our theoretical approach should also help address the long-standing issue of tackling
the Marcus inverted regime beyond the golden-rule limit. %
Although here we have initially assumed that the two-state system
under consideration is in the normal regime,
the analytical formulae we have derived for linear diabats are in fact
general.\footnote{The only difference is that in the inverted regime
both $\kfreda$ and $\kfredb$ reduce the total rate, so the sign of $\kfredb$ must be changed accordingly.}
This suggests that $\kfsc$ can be reformulated to encompass the inverted regime
for arbitrary potentials, much like has been done for golden-rule instantons.\cite{inverted,Ansari2021} We intend to pursue a similar generalisation of
our fourth-order semiclassical rates in future work.

Furthermore, the derivations in this paper
suggest how one might obtain further high-order terms: $k_{6}^{\SC},\,k_8^{\SC},$
and so on.
Based on the Green's function formalism we expect
$k_{6}^{\SC}$ to comprise contributions from three instantons.
The first two are likely simple generalisations of the terms
encountered so far: a six-bounce 6A term and a two-bounce 6C term, now with \emph{two} points sampling the upper diabat.
The remaining 6B term is expected to combine these elements, resulting
in a four-bounce path with a single point sampling the upper diabat.
In addition to that, $k_{6}^{\SC}$ is expected to contain contributions derived from
the minima of the diabatic PESs, likely to be insignificant
in practice due to the small magnitude of the diabatic coupling at these points.
We see no reason why the methods presented in this paper cannot
be applied to the derivation of all these components.

Of course it is likely that, beyond a certain point, there is little merit
to calculating increasingly high-order terms. Since all of them
are associated with golden-rule, rather than adiabatic, instantons,
this may not be the most accurate or efficient way of describing nonadiabatic
rates close to the Born--Oppenheimer limit. An alternative would be to
apply the ideas developed in this paper to deriving a correction
to the \emph{adiabatic} rate constant, and to then
connect the two limits using a suitable interpolation formula
(\latin{e.g.}, a two-point Pad\'{e} approximation),\cite{BenderBook} not unlike
\Refx{Lawrence2019ET}. Apart from encompassing the full range of coupling
strengths, this approach would offer new mechanistic insight into
nonadiabatic reactions, rigorously derived from the instantons that underlie
the correction terms.

Looking beyond semiclassical instantons, our theory may
inspire the search for fourth-order analogues to Wolynes theory\cite{Wolynes1987nonadiabatic,Bader1990golden}
or GR-QTST.\cite{GRQTST,GRQTST2}
In turn this is expected to
help in the development of more rigorous path-integral sampling and dynamics methods, including nonadiabatic extensions of RPMD\@.

 \begin{acknowledgments}
 The authors acknowledge financial support by the Swiss National Science Foundation
 through SNSF Project 207772 and thank Joseph E.~Lawrence for helpful discussions.
 \end{acknowledgments}

\appendix

\renewcommand{\theequation}{\thesection.\arabic{equation}}

\section{\label{sec:dn}%
Asymptotic expansion of L\textsubscript{4A}}

Given the quantities in \eqn{eq:S4A-blocks}, the Hessian of
$S_{\fa}$ [\eqn{eq:s4a-defn}] at its stationary point can be
written as the symmetric matrix\footnote{For clarity we only show the lower triangle and indicate how the matrix can be divided into blocks.}
\begin{equation}
\hspace*{-0.9ex}%
\mat{\Sigma}_{\fa} =
\pders{S_{\fa}}{\cvec{v}\mkern1mu}{\cvec{v}} =
\setlength{\mycolw}{1.6em}
\!\left[
\begin{array}{*{2}{>{\rule[-0.25em]{0pt}{1.5em}\!\!}L{\mycolw}}|%
*{1}{>{\rule[-0.25em]{0pt}{1.5em}}L{0.9em}}
*{3}{>{\rule[-0.25em]{0pt}{1.5em}}L{1.1em}}
>{\rule[-0.25em]{0pt}{1.5em}}L{0.7em}}
\ \, 2\zeta_0 &   &   &   &   &   &   \\
\phantom{-}0  &\ \, 2\zeta_1  &   &    &     &    &     \\
\hline
\phantom{-}0  & \phantom{-} 0  &  \bar{\zeta}  &   &    &    &     \\
{-\vec{w}_0}  & {-\vec{w}_1} & \bar{\vec{w}}  & \mat{\Theta} &
   &   &   \\
\phantom{-}\vec{w}_0  &  {-\vec{w}_1} & \bar{\vec{w}} &
\mat{\Theta}_{\mathrlap{1}} &  \mat{\Theta} &    &    \\
\phantom{-}\vec{w}_0  &  \phantom{-}\vec{w}_1 & \bar{\vec{w}}
&  \, \mathbb{0}   &  \mat{\Theta}_{\mathrlap{0}}   &  \mat{\Theta}   &   \\
{-\vec{w}_0}  & \phantom{-}\vec{w}_1  & \bar{\vec{w}}
&  \mat{\Theta}_{\mathrlap{0}} & \, \mathbb{0} & \mat{\Theta}_{\mathrlap{1}} & \mat{\Theta} \\
\end{array}
\right]\!,
\end{equation}
where $\cvec{v}$ is a vector formed by concatenating $\cvec{\tau}$ and
$ \cvec{x} $, $\bar{\vec{w}} = \tfrac{1}{2}(\vec{w}_0 + \vec{w}_1)$ and
$\bar{\zeta} = \tfrac{1}{2}(\zeta_0 + \zeta_1)$. To integrate
\eqn{eq:L4A-qm} by steepest descent we rewrite the Hessian as
\begin{equation}
\label{eq:schur}
\begin{aligned}
 \mat{\Sigma}_{\fa} & \equiv
\left[
    \begin{array}{>{\rule[-0.4em]{0pt}{1.6em}}l|l}
    \mat{M}_{11} & \mat{M}_{12} \\
    \hline
    \mat{M}_{21} & \mat{M}_{22}
    \end{array}
\right]\\
& = \left[
    \begin{array}{*{2}{>{\rule[-0.4em]{0pt}{1.6em}}R{2.2ex}}}
    \mathbb{1}_2 & \mat{B}\trsp \\
    \mathbb{0}\mkern6mu & \mathbb{1}_j
    \end{array}
\right]
\left[
    \begin{array}{>{\rule[-0.4em]{0pt}{1.6em}}C{1em}%
    L{1.5em}}
    \mat{A} &  \ \mathbb{0} \\
    \mathbb{0} & \mat{M}_{22} \\
    \end{array}
\right]
\left[
    \begin{array}{*{2}{>{\rule[-0.4em]{0pt}{1.6em}}L{2.2ex}}}
    \mathbb{1}_2 & \mathbb{0} \\
    \mat{B} & \mathbb{1}_j
    \end{array}
\right]\!,
\end{aligned}
\end{equation}
where $\mat{B} = \mat{M}_{22}^{-1} \mat{M}_{21}$,
 $\mat{A} \equiv
 \mat{M}_{11} - \mat{M}_{12} \mat{M}_{22}^{-1} \mat{M}_{21}  $,
$\mat{M}_{11}$ is a $2 \by 2$ submatrix and the other blocks are
shaped accordingly. Additionally $\mathbb{1}_j$ denotes
a $j\by j$ identity matrix and here $j = 4f + 1$.
\eqn{eq:schur} defines a transformation that casts $\mat{\Sigma}_{\fa}$ into block-diagonal form, decoupling $\tau_{n}$ from the other variables. Specifically,
\begin{equation}
\begin{gathered}
\cvec{v}\trsp  \mat{\Sigma}_{\fa} \cvec{v} =
\binom{\tau_0}{\tau_1}\trsp[-8] \mat{A}  \, \binom{\tau_0}{\tau_1}
+ \cvec{y}\trsp \mat{M}_{22} \mkern2mu \cvec{y} \\
\cvec{y} =
\left[\, \mat{B} \quad \mathbb{1}_j \right] \cvec{v},
\end{gathered}
\end{equation}
allowing one to integrate over $\cvec{y}$ to get a prefactor
$\sqrt{-(2 \pi \hbar)^{4f+1}/\lvert \mat{M}_{22} \rvert}$ and
an effective action
\begin{equation}
S_{\fa}^{\mathrm{eff}} = 2 S_2 + \frac{1}{2}
\binom{z_0}{z_1}\trsp[-8] \mat{A}  \ \binom{z_0}{z_1}
\qquad (\lvert z_n \rvert \to 0),
\end{equation}
where $S_2$ is evaluated at the stationary point.  The second term, multiplied
by $-1/\hbar$, corresponds to the quadratic part of the argument of the exponential
on the second line of \eqn{eq:L4A-ht}.
Subsequent calculations are simplified by the
orthogonal transformation given by
\begin{equation}
\mat{U} = \left[
\begin{array}{R{1.2ex}R{1.3em}*{3}{>{\rule[-0.5em]{0pt}{1.7em}}R{2em}}}
1 & \vec{0}\trsp[-3] & \vec{0}\trsp[-3] & \vec{0}\trsp[-3] & \vec{0}\trsp[-3] \\
\vec{0} & \frac{1}{2} \mathbb{1}  & \frac{1}{2} \mathbb{1}  & \frac{1}{2} \mathbb{1}  & \frac{1}{2} \mathbb{1}  \\
\vec{0} & \frac{1}{2} \mathbb{1}  & -\frac{1}{2} \mathbb{1}  & \frac{1}{2} \mathbb{1}  & -\frac{1}{2} \mathbb{1}  \\
\vec{0} & \frac{1}{2} \mathbb{1}  & \frac{1}{2} \mathbb{1}  & -\frac{1}{2} \mathbb{1}  & -\frac{1}{2} \mathbb{1}  \\
\vec{0} & \frac{1}{2} \mathbb{1}  & -\frac{1}{2} \mathbb{1}  & -\frac{1}{2} \mathbb{1}  & \frac{1}{2} \mathbb{1}
\end{array}
\right]
\end{equation}
(here $\mathbb{1} \equiv \mathbb{1}_{f}$), which can be used to show that
\begin{equation}
\label{eq:k4a-mats}
\lvert \mat{M}_{22} \rvert =
\lvert \mat{U}\trsp \mat{M}_{22} \mat{U} \rvert =
\frac{1}{2}\lvert \mat{\Sigma}_{2}\rvert \lvert \mat{\Theta_{+}} \rvert
\lvert \mat{\Theta_{-}} \rvert,
\end{equation}
with $\mat{\Sigma}_2$ and $\mat{\Theta}_{\pm}$ defined in \eqs{eq:sigma2}{eq:theta-pm} respectively. Using the same
transformation it can be shown that
\begin{equation}
\mat{A} = \mat{M}_{11} - \mat{M}_{12} \mat{U} (\mat{U}\trsp\mkern-1mu \mat{M}_{22} \mat{U})^{-1} \mat{U}\trsp\mkern-1mu \mat{M}_{21}
 =  -\!\left[
\begin{array}{*{2}{C{0.8em}}}
 d_0 & 0 \\
 0 & d_1
\end{array}
\right]\!,
\end{equation}
where we identify the diagonal elements with the
coefficients $d_n$, whose explicit form in terms of GR action  derivatives is given in \eqn{eq:d0d1-defn}.

\section{
\label{sec:k4b-ag}%
Asymptotic expansion of \textit{L}\textsubscript{4B}
}

Here we derive the expressions in \eqn{eq:k4b-ag}. To this end,
it is convenient to perform steepest-descent integration of $S_{\fb}$ in stages,
treating the difference coordinate $\vec{x}_{-}$ first, followed by all the remaining
variables. We begin by expressing \eqn{eq:s4b-defn} in terms of the new position
variables in \eqn{eq:k4b-diffx} and use the short-time form of the action for the
segment of length $\tau_1$. The remaining terms are Taylor expanded to second
order about $\tau_1 = 0$ and $\vec{x}_{-} = \vec{0}$, yielding
\begin{gather}
\label{eq:s4b-asympt}
 S_{\fb} \sim S'_0 + S''_0 + S_1 + \frac{m \lVert \vec{x}_{-} \rVert^2}%
 {2 \tau_1} + \tau_1 V_1(\vec{x}) \nonumber \\
 {} +  \bigg(
 \pder{S'_0}{\vec{x}} -  \pder{S''_0}{\vec{x}}
 \bigg) \! \cdot \! \bigg( \frac{\vec{x}_{-}\!}{2} \bigg)
 - \pder{S_1}{\tau} \tau_1  \\
 {} + \frac{1}{2}
  \bigg( \frac{\vec{x}_{-}\!}{2} \bigg) \! \cdot \!
  \left[
  \pders{(S'_0+S''_0)}{\vec{x}\mkern1mu} {\vec{x}}
  \right]
  \! \cdot \! \bigg( \frac{\vec{x}_{-}\!}{2} \bigg) + \frac{1}{2} \pder[2]{S_1}{\tau} \tau_1^2,
  \nonumber
\end{gather}
with $S_0', S_0'',$ and $S_1$ defined in \eqn{eq:s4b-parts}. Substituting
this expansion for $S_{\fb}$, we evaluate the integral
\begin{equation}
\int_{-\infty}^{\infty} \! \eu{-S_{\fb} / \hbar} \, \rmd \vec{x}_{-} \sim
A \, \eu{-S_{\fb}^{\mathrm{eff}}/\hbar},
\end{equation}
where
\begin{equation}
\label{eq:s4b-eff}
\begin{aligned}
S_{\fb}^{\mathrm{eff}} & \sim S_0' + S_0'' + S_1 \\
& + \tau_1 \left\{
V_1(\vec{x}) - \frac{\lVert \bar{\vec{p}} \rVert^2}{2 m} - \pder{S_1}{\tau}
\right\} \\
& + \frac{1}{2} \tau_1^2 \left\{
\frac{\bar{\vec{p}}}{2 m} \cdot \pders{(S_0' + S_0'')}{\vec{x}}{\vec{x}}
\cdot \frac{\bar{\vec{p}}}{2 m} +
\pder[2]{S_1}{\tau_1}
\right\}
\end{aligned}
\end{equation}
and we define
\begin{equation}
\bar{\vec{p}} = \frac{1}{2} \pder{(S_0''-S_0')}{\vec{x}}.
\end{equation}
The symbol $A$ denotes a prefactor whose derivation can be skipped here,
since it can be deduced by simpler means and has already been given in \eqn{eq:margfb-tnull}.

The new,
effective action is now expanded to second order in the fluctuations $\delta \cvec{v}$
about the point $\cvec{v}$ that satisfies \eqn{eq:k4b-stat}. At this point
$\vec{x}',\, \vec{x}''\!,$ and $\tau$ assume the values that minimise the GR
instanton action; $\vec{x}$ is the coordinate along the GR instanton after travelling for imaginary time $\tau_0$ on the reactant diabat, away from $\vec{x}''$.
A few simplifications can now be made. \Eqn{eq:dsdstuff} relates the derivative
of the stationary action to the energy of the corresponding classical imaginary-time
trajectory, which we write as
\begin{subequations}
\label{eq:s4b-s0dt}
\begin{align}
\label{eq:s4b-s01dt}
-\pder{S_0'}{\tau_0} & = -\frac{1}{2 m} \left \lVert \pder{S_0'}{\vec{x}} \right\rVert^2 + V_0(\vec{x}) \\
\label{eq:s4b-s02dt}
\pder{S_0''}{\tau_0} & = -\frac{1}{2 m} \left \lVert \pder{S_0''}{\vec{x}} \right\rVert^2
+ V_0(\vec{x}).
\end{align}
\end{subequations}
Differentiating the difference of these two equations with respect to either
$\vec{x}$ or $\tau_0$ and using the fact that
at the stationary $\cvec{v}$
\begin{equation}
\label{eq:s4b-dx}
\bar{\vec{p}} = \pder{S_0''}{\vec{x}} = -\pder{S_0'}{\vec{x}},
\end{equation}
we can show that
\begin{equation}
\frac{\bar{\vec{p}}}{2 m} \cdot \pders{(S_0' + S_0'')}{\vec{x}}{\vec{x}}
\cdot \frac{\bar{\vec{p}}}{2 m} = \frac{1}{4} \pder[2]{(S_0' + S_0'')}{\tau_0}.
\end{equation}
Furthermore, considering the factor multiplying $\tau_1$ in \eqn{eq:s4b-eff} together
with \eqs{eq:s4b-s01dt}{eq:s4b-dx}, we can write
\begin{equation}
\begin{aligned}
- \frac{\lVert \bar{\vec{p}} \rVert^2}{2 m} - \pder{S_1}{\tau} & =
V_1(\vec{x}) - V_0(\vec{x}) - \pder{S_0'}{\tau_0} - \pder{S_1}{\tau} \\
{} & = V_1(\vec{x}) - V_0(\vec{x}),
\end{aligned}
\end{equation}
where the last equality follows from the relation
\begin{equation}
\pder{S_0'}{\tau_0} = \pder{S_0'}{\tau} = -\pder{S_1}{\tau}
\end{equation}
at the stationary $\cvec{v}$. Last, to calculate the $\mathcal{O}(\tau_1 \delta \cvec{v})$ coefficient, we need the derivative of the same factor with respect to
$\cvec{v}$. In particular, consider
\begin{equation}
\frac{1}{2m} \pder{\lVert \bar{\vec{p}} \rVert^2}{\cvec{v}} =
\frac{1}{2} \pders{(S_0'' - S_0')}{\cvec{v}}{\vec{x}} \cdot \left[
\frac{1}{2 m} \pder{(S_0'' - S_0')}{\vec{x}}
\right].
\end{equation}
According to \eqn{eq:s4b-dx} this can be rewritten as
\begin{equation}
\begin{aligned}
& \frac{1}{2} \left\{
\pders{S_0''}{\cvec{v}}{\vec{x}} \cdot \left[
\frac{1}{m} \pder{S_0''}{\vec{x}}
\right] +
\pders{S_0'}{\cvec{v}}{\vec{x}} \cdot \left[
\frac{1}{m} \pder{S_0'}{\vec{x}}
\right]
\right\} \\
& \qquad {} = \frac{1}{2} \pder{}{\cvec{v}} \left\{
\frac{1}{2m} \left\lVert
  \pder{S_0''}{\vec{x}}
\right \rVert^2 +
\frac{1}{2m} \left\lVert
  \pder{S_0'}{\vec{x}}
\right \rVert^2
\right\}.
\end{aligned}
\end{equation}
From \eqn{eq:s4b-s0dt} it then follows that
\begin{equation}
\frac{1}{2m} \pder{\lVert \bar{\vec{p}} \rVert^2}{\cvec{v}} =
\pder{V_0(\vec{x})}{\cvec{v}} - \frac{1}{2} \pders{(S_0''-S_0')}{\cvec{v}}{\tau_0}.
\end{equation}
Combining all these results we have the following terms in the Taylor series expansion
of $S_{\fb}^{\mathrm{eff}}$:
\begin{subequations}
\label{eq:s4b-taylor}
\begin{align}
\mathcal{O}(1) \ : & \ S_0' + S_0'' + S_1 \equiv S_2(\beta), \\
\label{eq:s4b-O1}
\mathcal{O}(\tau_1) \ : & \ \tau_1 \left[
V_1( \vec{x} ) - V_0( \vec{x} )
\right], \\
\mathcal{O}(\tau_1^2)  \ : & \
\frac{1}{2} \tau_1^2
 \left[
\frac{1}{4}
\pder[2]{(S_0'+S_0'')}{\tau_0} + \pder[2]{S_1}{\tau}
\right], \\
\mathcal{O}(\delta \cvec{v} \delta \cvec{v})  \ : & \
\frac{1}{2} \delta \cvec{v}  \cdot
\mat{\Phi}_{0}  \! \cdot  \delta \cvec{v}, \\
\mathcal{O}(\tau_1 \delta \cvec{v})  \ : & \
 \tau_1 \delta \cvec{v} \cdot \cvec{a},
\end{align}
\end{subequations}
where $\mat{\Phi}_0$ is defined in \eqn{eq:k4b-Phi0},
$\left[ \mat{\Phi}_{0} \right]_{\cvec{v}\tau}$ refers to its column
of derivatives with respect to $\tau$, and
\begin{equation}
\cvec{a} = \pder{[V_1(\vec{x}) - V_0(\vec{x})]}{\cvec{v}}
+ \frac{1}{2} \pders{(S_0' + S_0'')}{\cvec{v}}{\tau_0}
- \left[ \mat{\Phi}_{0} \right]_{\cvec{v} \tau}.
\end{equation}
The exponential of this truncated Taylor series can now be integrated over $\delta
\cvec{v}$ to get another effective action, now depending
on $\tau_n$ only. This has the same $\mathcal{O}(1)$ and
$\mathcal{O}(\tau_1)$ terms as in \eqn{eq:s4b-taylor}, and a quadratic
term
\begin{equation}
\label{eq:s4b-O2}
\frac{1}{2} \tau_1^2
 \left[
\frac{1}{4}
\pder[2]{(S_0'+S_0'')}{\tau_0} + \pder[2]{S_1}{\tau}
- \cvec{a} \cdot \mat{\Phi}_0^{-1} \!\! \cdot \cvec{a}
\right].
\end{equation}
The latter can be simplified by noting that
\begin{equation}
\mat{\Phi}_0^{-1} \cdot \left[ \mat{\Phi}_{0} \right]_{\cvec{v} \tau}
= \left( 1,\, 0,\, \ldots,\, 0 \right)\trsp
\end{equation}
and
\begin{gather}
\pder[2]{(S_0'+S_0'')}{\tau_0} -
\pders{(S_0' + S_0'')}{\tau_0}{\cvec{v}} \cdot
\mat{\Phi}_0^{-1} \cdot
\pders{(S_0' + S_0'')}{\cvec{v}}{\tau_0} \nonumber \\
{} = \der[2]{\left(
S_0' + S_0'' + S_1
\right)}{\tau_0} = 0.
\end{gather}
The final relation follows because  $S_0' + S_0'' + S_1$
is equal to the GR instanton stationary action regardless of the value of
$\tau_0$. Therefore a full derivative of this expression with respect to $\tau_0$
necessarily evaluates to zero. After this simplification, the bracketed terms in \eqs{eq:s4b-O1}{eq:s4b-O2}
can be identified with $\alpha_1$ and $-\gamma_1$ respectively, leading to
the final result in \eqn{eq:k4b-ag}.

\section{\label{sec:kfb-lin}%
Exact \emph{\~{k}}\textsubscript{4B} for linear diabats}

For a general potential, evaluating \eqn{eq:j0} exactly
is not mathematically justified,\cite{BenderBook} since the integrand is an
asymptotic approximation to the quantum $\margfb$
that is only valid for $\tau_1 \to 0$. Linear diabats are
a notable exception, since in this case the quantum $\margfb$
is precisely of the form in \eqn{eq:margfbsc}. Hence for
these systems it is meaningful to evaluate the expression
in \eqn{eq:j0} as is, yielding
\begin{align}
\label{eq:k4b-lin}
& j^{\mathrm{lin}}_0(\tau_0) = \sqrt{\frac{2 \hbar}{\gamma_1}}
\bigg[
\dawson( \lambda_1 ) \\
& \qquad \qquad {} - \exp \! \left(
 \frac{\phi_1^2}{4 \lambda_1^2} -\phi_1
\right)
  \dawson\!\left(\lambda_1 - \frac{\phi_1}{2 \lambda_1}
\right)
\bigg], \nonumber
\end{align}
where $\phi_1 = \tfrac{\alpha_1}{\hbar} \mkern-1mu \tfrac{\tau_1^{*}}{\tau_0^{*}} \tau_0$,
$\lambda_1 = \alpha_1 / \sqrt{2 \hbar \gamma_1}$, and
\begin{equation}
\dawson(x) \equiv \eu{-x^2} \! \int_0^{x} \eu{t^2} \, \rmd t
\end{equation}
is the Dawson function. Substituting this and the analogous
$j^{\mathrm{lin}}_1$ into \eqn{eq:k4b-sc-uni}
instead of $j_n^{\SC}$ defines the quantum $\kfredb$ for
linear diabats with constant coupling $\Delta$. Alternatively
we can evaluate \eqn{eq:kfb-ht} directly, without splitting the
$\mathcal{A}_{\tau \tau}$ domain along the diagonal. Both approaches
lead to the same expression,
\begin{align}
\label{eq:k4blin}
\kfredb^{\mathrm{lin}}(\beta) = \kgrred^{\mathrm{lin}}(\beta)
\frac{\Delta^{\!2} \mkern-1mu \sqrt{\rule{0pt}{0.9em}%
8 \beta m} }{\hbar \lvert\kappaR-\kappaP\rvert}
\int_0^{\frac{\pi}{2}} \!\!\! \dawson  \bigg( \mkern-1mu
    \frac{\bred^{\frac{3}{2}} \sin \theta}{4} \mkern-1mu
\bigg) \rmd \theta,
\end{align}
with $\kgrred^{\mathrm{lin}}$ and $\bred$ defined in \eqs{eq:k2-lin}{eq:bred}.

It can be shown that the rate calculated using the approximate
$j_n^{\SC}$ tends to the quantum
result for a system of linear diabats both as $\beta \to 0$ and $\beta \to \infty$.
Furthermore, for any combination of parameters the semiclassical
rate, $\kfbsc$, is never in error by more than $2.8\%$, with the
largest deviation seen at $\widetilde{\beta} \approx 3.5$ (in all
instances the approximation is a lower bound).

\section{\label{sec:morse0}%
Stationary action for the scattering\\
Morse potential}

The abbreviated action is defined as
\begin{equation}
W(x'', x', E) = \int_{x'}^{x''} \vec{p}(x, E) \cdot
\rmd \vec{x},
\end{equation}
where $\vec{p}$ is the imaginary-time momentum whose magnitude is
$p(x, E) = \sqrt{2 m [V(x) - E]}$. For the scattering
Morse potential in \eqn{eq:pre-V1}, this integral can be evaluated analytically.
One must distinguish two types of trajectories: those going directly
from $x'$ to $x''$, with action
\begin{equation}
\hspace*{-0.75ex}W_{\mathrm{d}}(x''\mkern-2mu, x', E) =
\left \lvert
\frac{p' - p''\!}{\alpha} -
\frac{\mu}{\alpha} \tan^{-1} \! \left[
\frac{\mu p' - \mu p''}{\mu^2 + p' p''}
\right] \right \rvert,
\end{equation}
and those passing through a turning point, with
\begin{equation}
W_{\mathrm{b}}(x''\mkern-2mu, x', E) =
\frac{p' + p''\!}{\lvert \alpha \rvert} -
\frac{\mu}{\lvert \alpha \rvert} \tan^{-1} \! \left[
\frac{\mu p' + \mu p''}{\mu^2 - p' p''}
\right].
\end{equation}
Here $p' \equiv p(x', E)$, $p'' \equiv p(x'', E)$,
$\mu = \sqrt{2 m (E - \epsilon)}$ and $\tan^{-1}(y/x)$ refers to
the principal value of the argument of $x + \iu y$. The abbreviated
action satisfies
\begin{equation}
S(x'', x', \tau) = W(x'', x', E) + E \tau,
\end{equation}
where for a given $\tau$ the energy is such that
\begin{equation}
\pder{W}{E} = -\tau.
\end{equation}
The first-order derivatives of $S$ are given by
\begin{equation}
\pder{S}{x'} = \pder{W}{x'}, \qquad
\pder{S}{x''} = \pder{W}{x''}, \qquad
\pder{S}{\tau} = E,
\end{equation}
and the second-order derivatives are
\begin{subequations}
\begin{align}
\pder[2]{S}{\tau} & = -\left(\pder[2]{W}{E}\right)^{\!\!-1}, \\
\pders{S}{x'}{\tau} & = -\pders{W}{x'}{E}
\left(\pder[2]{W}{E}\right)^{\!\!-1}, \\
\pders{S}{x'}{x''} & = \pders{W}{x'}{x''} -
\pders{W}{x'}{E} \left(\pder[2]{W}{E}\right)^{\!\!-1} \!\!\pders{W}{E}{x''},
\end{align}
\end{subequations}
along with all possible variations with $x'$ and $x''$ variously
interchanged. This gives all the information necessary to calculate
$\kgrredsc,\,\kfredasc$ and $\kfredbsc$. The only other, minor modification
is the change of integration variable in \eqn{eq:k4b-sc-uni} from imaginary time
to position,
\begin{equation}
\int_0^{\tau_n^{*}} j_n (\tau_n) \, \rmd \tau_n \to
\int_{x_{<}}^{x_{>}} \!\! \frac{m j_n (\tau_n(x,E))}{\sqrt{2 m [V_n(x)-E]}} \, \rmd x,
\end{equation}
where $x_{<}$ and $x_{>}$ are respectively the leftmost and the rightmost of $x^{\ddagger}$ (hopping point)
and $x_{\mathrm{b}}$ (turning point), and
\begin{equation}
\tau_n(x,E) = -\pder{W(x,\, x_{<}, E)}{E}.
\end{equation}
The integral over position gives the same result as the integral
over imaginary time, but is the simpler alternative to implement for a one-dimensional system.

\section{\label{sec:sb-tcfs}%
Quantum correlation functions
for the spin--boson model}

In \eqn{eq:partf-series} we give the perturbation series expansion of
$\Tr[\eu{-\beta \Hop} \ketbra{0}{0}]$ for the Hamiltonian in \eqn{eq:htot}.
The terms in this series are nested imaginary-time integrals,
whose integrands are precisely the multi-time correlation functions
that define the rate constants $k_{2\nu}$ in the perturbation
expansion of the total rate. Weiss, in Chapter~19.1 of \Refx{Weiss}, gives
the analytic form of these correlation functions for the spin--boson model
described in \sec{ssec:spin-boson}, namely
\begin{subequations}
\label{eq:gen-sb-tcf}
\begin{align}
& \frac{c_{2 \nu}(u_{1\ldots2\nu})}{\ZRn{0}} =
\frac{\Delta^{\!2\nu}}{\hbar^{2\nu}} \eu{-\phi_{2\nu}(u_{1\ldots2\nu}) / \hbar} \\
& \phi_{2\nu} = {-\epsilon} \sum_{j=1}^{\nu} s_j +  \sum_{j=2}^{2\nu} \sum_{i=1}^{j-1}
(-1)^{i+j} Q(u_j - u_i),
\end{align}
\end{subequations}
where
\begin{align}
Q(u) = \int_{0}^{\infty} \frac{4 J(\omega)}{\pi \omega^2} \left[
\frac{1 - \cosh(\omega u)}{\tanh \big( \frac{\beta \hbar \omega}{2} \big) }
+ \sinh(\omega u)
\right] \rmd \omega,
\end{align}
and $s_j = u_{2j} - u_{2j-1}$.
Obtaining the correlation function that corresponds to a particular component of the $k_{2 \nu}$ rate constants amounts to finding the appropriate variable
transformation from $u_{1\ldots2\nu}$ to $\{z, z_{0\ldots2\nu-2}\}$. Note that
because the exponent in \eqn{eq:gen-sb-tcf} depends only on the differences
of $u_{j}$, one of the original variables can be eliminated prior to the transformation.
For $\nu = 1$ this procedure yields the familiar expression used to define the GR rate,\cite{Lawrence2018Wolynes,inverted,UlstrupBook}
\begin{equation}
\phi_2 =  Q(z) - \epsilon z.
\end{equation}
For $\nu=2$ we get
\begin{subequations}
\begin{align}
\label{eq:phi4A}
\phi_{\fa} & =
      Q\Big(\tfrac{\beta \hbar-z}{2} + z_0 \Big)
    + Q\Big(\tfrac{z}{2} + z_1 \Big)  \\
&  + Q\Big(\tfrac{\beta \hbar-z}{2} - z_0 \Big)
    + Q\Big(\tfrac{z}{2} - z_1\Big) \nonumber \\
& - Q\Big(\tfrac{\beta \hbar}{2} - z_0 - z_1 \Big)
       - Q\Big(\tfrac{\beta \hbar}{2} - z_0 + z_1 \Big)
       - \epsilon z, \nonumber \\
\label{eq:phi4B}
\phi_{\fb} & = Q\left(\beta \hbar - z - z_0\right)
        + Q\left(z_1\right) \vphantom{\Big(\Big)} \\
& + Q\left(z_0\right)
        + Q\left(z - z_1\right) \vphantom{\Big(\Big)} \nonumber \\
& - Q\left(z + z_0 - z_1\right)
  - Q\left(z_0 + z_1\right)
       - \epsilon z. \nonumber \vphantom{\Big(\Big)}
\end{align}
\end{subequations}
Continuing in the same fashion, one can
generate further multi-time correlation functions,
although the expressions quickly become cumbersome,
as $\phi_{2\nu}+\epsilon z$ contains a total of  $\nu (2\nu - 1)$
terms.

\section{\label{sec:high-t-limit}%
Nonadiabatic rates in the high-temperature limit}

As $\beta \to 0$, the $\mathcal{O}(\Delta^4)$ correction to the GR rate
becomes dominated by the $\kfreda$ contribution, whose high-temperature
form is given in \eqn{eq:k4a-ht}. Combining this with the high-temperature
limit of the GR rate,
\begin{equation}
\label{eq:k2-cl}
\kgrht(\beta) = \sqrt{\frac{2 \pi m}{\beta \hbar^2}}
\frac{\Delta^2}{\hbar \lvert \kappaR - \kappaP \rvert}
\frac{Z^{\ddagger}}{\ZRn{0}} \, \eu{-\beta V^{\ddagger}},
\end{equation}
gives
\begin{align}
\label{eq:k24-ht}
k(\beta) & \sim  \frac{Z^{\ddagger} \mkern1mu \eu{-\beta V^{\ddagger}}}{\ZRn{0}}  \Bigg[ \sqrt{\frac{2 \pi m}{\beta \hbar^2}}
\frac{\Delta^2}{\hbar \lvert \kappaR - \kappaP \rvert}  - {} \\
& \quad  \frac{\pi m \Delta^4}{\hbar^3 (\kappa_0 - \kappa_1)^2} \!
 \left\{
\ln \! \left[
\frac{128 m (\kappa_0 - \kappa_1)^2}{\beta^3 \hbar^2 \kappa_0^2 \kappa_1^2}
\right]  - \gamma
\right\}
 \Bigg]. \nonumber
\end{align}
We know that this expansion is correct for a linear
model, and we have confirmed that it gives accurate rates
in nonlinear and multidimensional systems, except at high friction,
where the steepest-descent approximation to the integral over
$\cfa(t_0, t_1, z)$ breaks down. We especially want to emphasise this,
since the logarithmic dependence on $\beta$ at $\mathcal{O}(\Delta^4)$  seems unusual when compared, for example,
to the Zusman equation
\begin{equation}
\label{eq:ZUS}
\kZUS = \frac{\kgrht \mkern1mu \kad{,0}}{\kgrht + \kad{,0}},
\end{equation}
which gives the classical limit of the spin--boson rate
at small $\Delta$ and large friction. Here $\kad{,0}$ is the classical rate
on the lower (cusped) adiabatic surface in the limit as $\Delta \to 0$,
\begin{equation}
\kad{,\Delta} = \frac{1}{2 \pi \beta \hbar}
\frac{Z^{\ddagger} \mkern1mu  \eu{-\beta V^{\ddagger}}}{\ZRn{0}}
\exp\!\bigg[
2 \beta \Delta
\frac{\sqrt{-\kappaR \kappaP}}{\kappaR-\kappaP} \!
\bigg],
\end{equation}
and hence
\begin{align}
\label{eq:zus-ht}
\kZUS(\beta)  & \sim \kgrht(\beta) -  \frac{Z^{\ddagger} \mkern1mu \eu{-\beta V^{\ddagger}}\!}{\ZRn{0}}
 \frac{ 4 \pi^2 m  \Delta^4}{\hbar^3 (\kappa_0 - \kappa_1)^2},
\end{align}
which is clearly different from \eqn{eq:k24-ht}. Ultimately this should
not come as a surprise, since $\kZUS$ is derived assuming high friction,\cite{Zusman1980,Garg1985spinboson}
and is therefore valid in a different limit to our semiclassical rate theory.

The interpolation formula of \Refx{Lawrence2019ET}
also fails to recover \eqn{eq:k24-ht} as
$\beta \to 0$. The key change here compared to  \eqn{eq:ZUS} is that the $\kad{,0}$ in the denominator is replaced with $\kad{,\Delta}$. This
results in a modification to the $\mathcal{O}(\Delta^4)$ term in \eqn{eq:zus-ht}
that is of order $\beta$ and therefore negligible in this limit.
Additionally, the interpolation formula introduces an
unphysical term of order $\Delta^3$, although this is
proportional to  $\beta\Delta\kgrht$ and hence also
negligible in this limit. In short, the Zusman equation
and the interpolation formula agree with each other to leading order in $\beta \Delta$ but
predict the wrong high-temperature behaviour for the one-dimensional linear system.

The temperature dependence of our semiclassical rate constants
can also be compared to that of thermal rates derived from
the cumulative reaction probability
\begin{equation}
\label{eq:ph}
\PH(v) = \frac{2\PLZ(v)}{1 + \PLZ(v)},
\end{equation}
given in the form suggested by Holstein.\cite{Holstein1959}
Here $v$ is the velocity of the system along the reaction coordinate
in the vicinity of the hopping point, and
\begin{equation}
\label{eq:plz}
\PLZ(v) = 1 - \exp\left[-\frac{2 \pi \Delta^2}{\hbar v \lvert \kappaR-\kappaP \rvert}\right]
\end{equation}
is the Landau--Zener transmission coefficient.\cite{Landau1932LZ,Zener1932LZ} The latter approximates the probability of
transmission between two linear diabatic states
during a single passage through the hopping point, and is valid at all coupling strengths $\Delta$, but only
for sufficiently large $v$. $\PH(v)$ has been proposed as the starting point
for uniform approximations to classical nonadiabatic rates, via
\begin{equation}
\label{eq:k-uni}
k \ZRn{0} = \frac{\eu{-\beta V^{\ddagger}}}{2 \pi \hbar}  \int_0^{\infty} m v \PH(v) \mkern1mu
\eu{-\beta m v^2 / 2} \, \rmd v
\end{equation}
or similar expressions,\cite{PetersBook,Rips1995ET,Nitzan,UlstrupBook,Lykhin2016NATST} where for simplicity we have taken the system to be one-dimensional. \Eqn{eq:k-uni} is in part motivated by the observation that replacing $\PH(v)$ with its $\Delta \to 0$ limit correctly recovers the classical GR rate. However the next term in the weak-coupling expansion of $\PH(v)$,
which should give the $\mathcal{O}(\Delta^4)$ behaviour, causes the integral to
diverge. This happens because $\PLZ(v)$ is only
valid at high velocities. At $\mathcal{O}(\Delta^2)$, the (spurious) contribution from low velocities as
 $\beta \to 0$ is small enough that \eqn{eq:k-uni} can be integrated all the way from 0. At $\mathcal{O}(\Delta^4)$ this is no longer the case, and a non-zero bound has to be introduced.
Hence the correct $\mathcal{O}(\Delta^4)$ behaviour cannot be obtained
using $\PH(v)$ on its own---for this we would also require the
transmission coefficient at
low velocities, which is not available in closed form.\cite{NikitinBook}
Nevertheless, \eqn{eq:k-uni} gives us qualitative insight. First, we rewrite it as
\begin{align}
k & = \kgrht - \frac{\eu{-\beta V^{\ddagger}}}{\ZRn{0}} \frac{3 \pi m \Delta^4}{\hbar^3
 (\kappaR - \kappaP)^2} \int_{w_0}^{\infty} \frac{\eu{-w}}{w} \, \rmd w + \mathcal{O}(\Delta^6)
 \nonumber \\
 & \sim \kgrht - \frac{\eu{-\beta V^{\ddagger}}}{\ZRn{0}} \frac{\pi m \Delta^4}{\hbar^3
  (\kappaR - \kappaP)^2} \left\{
  \ln \! \left[ \frac{1}{w_0^3}  \right] - 3 \gamma
  \right\}, \label{eq:holstein-k4}
\end{align}
where $w = \beta m v^2 / 2$ and we have introduced a lower bound $0 < w_0 \ll 1$.
Next we note that the exact transmission coefficient for any linear system
can be expressed
in terms of just two parameters.\cite{Zhu1992a} One can then use
dimensional arguments to show that $w_0$ has to be proportional to
the reduced temperature $\bred$ that we defined previously in \eqn{eq:bred}.
It is no coincidence that the same parameter also appears in the
quantum $\kgrred$, $\kfreda$ and $\kfredb$ for the linear system [\eqss{eq:k2-lin}{eq:k4asc-ht}{eq:k4blin}].

Landau--Zener theory cannot give us the proportionality constant in
$w_0 = \bred / \alpha $,
as this requires knowing the small-$v$ behaviour of $P(v)$,
which falls outside the range of validity of \eqn{eq:plz}.
For the same reason, this approach fails to account for
transitions at energies  below $w_0/\beta + \Vdd$. However
the latter are expected to be small for $\beta \to 0$,
and indeed \eqn{eq:holstein-k4} reproduces the leading behaviour of $\kf$ in this limit, correctly predicting the
$\ln(1/\bred^{3})$ term inside the curly brackets. The
asymptotic behaviour of $\kf$ is therefore determined
up to an additive constant, and setting
$\alpha = 4 \, \eu{2 \gamma/3}$ happens to recover
\eqn{eq:k24-ht} exactly, bringing LZ into agreement
with our theory.
This suggests that expressions like \eqs{eq:ph}{eq:k-uni}, when combined
with our analysis,
can inform future work on rigorous uniform nonadiabatic rate theories,
valid across the entire spectrum of diabatic coupling strengths.

\bibliography{paper_refs}%

\end{document}